\documentclass[12pt]{article}
\usepackage[a4paper, total={6in, 8.6in}]{geometry}
\usepackage{graphicx}
\usepackage{float}
\newfloat{scheme}{htbp}{los}
\floatname{scheme}{Scheme}
\usepackage{amsmath}
\usepackage{amssymb}
\usepackage{amsthm}
\usepackage{times}
\usepackage{algorithm}
\usepackage{algpseudocode}
\usepackage[english]{babel}
\usepackage{tikz, graphicx}
\usepackage{colortbl}
\usepackage{xcolor}

\usepackage{cite}
\usepackage{blindtext}
\usepackage{multicol}
\usepackage{multirow}
\makeatletter
\def\NAT@def@citea{\def\@citea{\NAT@separator}}
\makeatother
\usepackage{booktabs}
\newtheorem{assumption}{Assumption}
\newtheorem{example}{Example}
\newtheorem{corollary}{Corollary}
\newtheorem{remark}{Remark}
\newtheorem{definition}{Definition}
\newtheorem{proposition}{Proposition}

\newtheorem{theorem}{Theorem}
\newtheorem{notation}{Notation}
\newtheorem{aim}{Aim}
\usepackage{blindtext}
\usepackage{hyperref}
\DeclareMathOperator*{\argmax}{argmax}
\DeclareMathOperator*{\argmin}{argmin}
\usepackage{indentfirst}
\usepackage{hyperref}
\hypersetup{
    colorlinks=true, 
    linkcolor=blue,  
    urlcolor=red,    
    citecolor=green  
}
\usepackage{xcolor}
\usepackage[
backend=biber,
style=authoryear,
sorting=ynt
]{}
\def\twogroup{\hskip\arraycolsep0 \hskip\arraycolsep\dots\hskip\arraycolsep 0}

\def\oneGroup{\hskip\arraycolsep\underbrace{0 \hskip\arraycolsep\dots \hskip\arraycolsep 0}_{\text{$n-n_j$ Ignored Securities}}}
\usepackage{blindtext}
\usepackage{hyperref}
\usepackage{indentfirst}
\usepackage{times}

\hypersetup{
    colorlinks=true,
    linkcolor=blue,
    urlcolor=cyan,
    pdftitle={Overleaf Example},
    pdfpagemode=FullScreen,
    citecolor=red,
    }
    
\providecommand{\keywords}[1]
{
  \small
  \textbf{\textbf{Keywords}} #1
}
\author{Hafid Lalioui\thanks{Africa Business School, University Mohammed $6$ Polytechnic. Email: \href{mailto: hafid.lalioui@um6p.ma}{\textcolor{black}{hafid.lalioui@um6p.ma}}.}\,, Amine Ben Amar\thanks{Africa Business School, University Mohammed $6$ Polytechnic. Email: \href{mailto: amine.benamar@um6p.ma}{\textcolor{black}{amine.benamar@um6p.ma}}.}\, and Makram Bellalah\thanks{Universit\'e Jules Verne Picardie. Email: \href{mailto: amine.benamar@um6p.ma}{\textcolor{black}{makram.bellalah@u-picardie.fr}}.}\\
}
\date{}
\begin{document}
\title{\LARGE Asset Pricing Model in Markets of Imperfect Information and Subjective Views}
\maketitle
\begin{abstract}
We provide closed-form market equilibrium formula consolidating informational imperfections and investors’ beliefs. Based on Merton's model, we characterize the equilibrium expected excess returns vector with incomplete information. We then derive the corresponding market portfolio as the solution to a non-linear system of equations and analyze the sensitivities of extra excess returns to shadow-costs and market weights. We derive the market reference model for excess returns under random shadow-costs. The conditional posterior distribution of excess returns integrates the pick-matrix and pick-vector of views and the vector of shadow-costs into a multivariate distribution with mean and covariance dependent on the reference model.
\end{abstract}
\section{Introduction}\label{Sec1}
Since capital markets are generally considered inefficient, a more realistic asset allocation model must consider the investor's experience, accumulated over years of dealing with these markets, as well as the available information at any given time and how this information is shared among investors. Our paper derives an original market equilibrium model that combines both aspects by integrating investors' subjective beliefs into markets with incomplete information.
\par Sophisticated asset allocation models in capital markets have evolved over several decades. The building block theories for these models range from the mean-variance (MV) approach of modern portfolio theory (MPT) by Markowitz \cite{mrk52,mrk91}, to continuous-time portfolio models by Merton \cite{mert69,mert75} and Samuelson \cite{sam70,sam75}, and the most recent generative machine learning and differential game-based models, see for example Wiese \cite{Wiese20} and Guo et al\cite{G2022}. Each time an approach is introduced, several researchers present empirical anomalies an theoretical limitations of the models generally due to unrealistic adopted assumptions or market structure, see e.g., Best and Grauer \cite{BRG91} and Green and Hollifield \cite{GR92}. Practitioners obtain weird portfolios when only quantitative models are employed within a MV optimizer, and the resulting optimal portfolios are highly sensitive to the model parameters such as expected returns of securities. Researchers have then criticized the MV approach \cite{BRG91,GR92} and shown limitations for the early derived traditional equilibrium methods such as capital asset pricing model (CAPM) of Sharpe-Linter-Mossin \cite{Sh64,Lin65,Mos66}. Portfolio managers tend to explore alternative ways and have always expressed interest in models combining quantitative aspects with the non-quantitative properties of either markets or investors. The factor model by Fama and French \cite{FF92,FF93} and the arbitrage pricing theory (APT) by Ross \cite{Ross76,Ross13} gave an extension to the CAPM to include additional factors such as size and value for the three-factor model, or many factors in APT. The analysis of the efficient market hypothesis (EMH) and topics on behavioral finance emerged to deal with different market structures and subjective perspectives of investors. Indeed, incomplete information equilibrium models show that assumptions on the observability of asset parameters, variables and their risk estimation significantly impact equilibrium models of asset prices, see e.g., papers by Klein and Bawa \cite{CB77} and Barry and Brown \cite{BB84,BB84',BB85} about differential information, and Detemple \cite{Det86} and Gennotte \cite{Gen86}. Additionally, other approaches for capital markets equilibrium consider psychological aspect such as feelings of regret and herding behavior, see e.g., Kahneman and Tversky \cite{KT76}, Loomes and Sugden \cite{LR82}, Bikhchandani and Sharma \cite{BS2000}, and Barberis and Thaler \cite{BT03} for more information.
\par The topic of incomplete information in economics, games and finance has attracted the attention of the scientific community for several decades \cite{Har67,Wil77,ddd,R94,Ra95,Berr06,Bre06,JaZ09,PjD10,FW12,BerMo13,EgCl15,BDR18}. Previous capital market models of incomplete information range from those where the reference model of the market is such that the parameters related to the securities returns do not have the same estimate for all investors \cite{BB84,BB84',BB85,Det86,Gen86}, to the models where each investor has only access to a subset of market's securities such as the model by Merton \cite{Mert87}. In $1987$, Merton \cite{Mert87} developed the capital market equilibrium model of equation (\ref{eq2}) below for non-perfect information markets where the random variable
\begin{equation}\label{eq01}
\tilde{R}_k:=\bar{R}_k+b_k\tilde{Y}+\sigma_k\tilde{\varepsilon}_k
\end{equation}
describes each risky security's return, for $\bar{R}_k:=\mathbb{E}[\tilde{R}_k]$ the mathematical expectation of the return, $b_k$ the function related to a common factor $\tilde{Y}$ and $\sigma_k$ the function related to the specific noise $\tilde{\varepsilon}_k$. Models of differential information \cite{BB84,BB84',BB85,Det86,Gen86} assume that all investors have different estimates for each security parameters $\bar{R}_k,b_k$ and $\sigma_k$. To the contrary, in Merton's model and the formulation we shall consider in this study, it is assumed that all investors have the same quality of information across securities with the reference model of equation (\ref{eq01}). That is, the same precision of the estimates of each security's parameters, but different distributions of that information across investors. Thus, each investor knows only the parameters related to some securities and not all of the market's risky securities and updates his/her portfolio to the information he/she acquired. In the equilibrium of this market of incomplete information with $N$ investor, each security $k$ has a dynamical expected return which depends on the dispersion of the information about $k$ among all investors, and which is characterized by
\begin{equation}\label{eq2}
\mathbb{E}\bigl[\tilde{R}_k\bigr]=r_f+\lambda_k+\beta_k\Bigl(\mathbb{E}\bigl[\tilde{R}_M\bigr]-r_f-\lambda_M\Bigr),
\end{equation}
where
\begin{equation*}
    \lambda_k:=\frac{1}{N}\sum_{j=1}^{N}\lambda^j_k\;\text{and}\;\lambda_M:=\sum_{k=1}^{N}x_k\lambda_k
\end{equation*}
are the asset-specific and market-wide premiums that account for the information incompleteness, respectively. Here, the classical $k$-th asset's systematic risk measure is $\beta_k:=\text{Cov}(\tilde{R}_k,\tilde{R}_M)/\text{Var}\bigl(\tilde{R}_M\bigr)$, and its corresponding market portfolio's weight $x_k$ is computed as the proportion of its market capitalization to the total market capitalization. This equation shows that any asset $k$ enjoys the additional asset-specific return $\lambda_k$, also known as the \textit{shadow-cost} of asset $k$, resulting from the fact that not all investors know about security $k$. That is, any investor $j$ who ignores security $k$, contributes the positive fraction $(1/N)\lambda^j_k$ of his/her shadow-cost in the expected return of security $k$ which is not allocated in his/her portfolio, i.e., $w^j_k=0$. In addition, the equilibrium expected return of equation (\ref{eq2}) is penalized by the term $\beta_k\lambda_M$, where $\lambda_M$ represents the extra return earned by the market with incomplete information. At the same time, $\lambda_k-\beta_k\lambda_M$ is the additional premium enjoyed by asset $k$. In other words, the value $\mathbb{E}[\tilde{R}_M]-(r_f+\lambda_M)$ represents the new \textit{Market Risk Premium} and $r_f+\lambda_k$ is the minimum modified risk-free return that an informed investor who is aware about security $k$ has to guarantee. Investors take advantage of the fact that some of them ignore security $k$, meaning that in the case where all investors are informed about the security $k$, both $\lambda_k$ and $\lambda_M$ vanish and equation (\ref{eq2}) reduces to the classical CAPM-equilibrium model. In this model of incomplete information, the market is not mean-variance efficient for all securities. In fact, for any security $k$ and since any investor's portfolio does not contain all securities, $\lambda_k-\beta_k\lambda_M$ is different from zero. Moreover, the optimal allocation $x_k^*$ for an informed investor is given by the market portfolio allocation $x_k$ plus an additional term that depends on the shadow-cost $\lambda_k$ and compensates for the information held by the investor concerning security $k$.
\par In the early 1990s, the Black-Litterman (BL) model \cite{blaLit90,blaLit91,blaLit92} presented a new formulation for the equilibrium in capital markets by combining modern quantitative general equilibrium theory aspects with the traditional approach to portfolio selection based on investors' subjective views. This formulation takes the CAPM model of perfect information markets as the neutral equilibrium and adjusts its output vector of expected excess returns $\pi^c$ to incorporate investors' subjective views. Each view $k$ is characterized by $k$-th row of a view's pick-matrix $P$, $k$-th diagonal element of a view's confidence matrix $\Omega$, and $k$-th element of a view's pick-vector $q$. Black and Litterman used a reverse optimization approach in which the market portfolio weights vector
\begin{equation}\label{eq04}
w_M^c:=(\delta\Sigma)^{-1}\pi^c,\;\text{for}\;\delta:=\frac{\mathbb{E}\bigl[\tilde{R}_M\bigr]-r_f}{\sigma_M^2}\;\text{being the market price of risk},
\end{equation}
related to the CAPM model, are tilted towards the investor's subjective views when applying the mean-variance optimizer to the derived excess returns vector mean. Here $\delta$ is the risk-aversion coefficient related to the market return $\tilde{R}_M$ whose standard deviation is $\sigma_M$, $\Sigma$ the covariance matrix of securities' returns, and $r_f$ the risk-free rate. The authors began by confirming that neither historical means, equal means, nor risk-adjusted equal means are suitable to model a neutral equilibrium portfolio for the prior estimate of excess returns. Before considering the general CAPM-equilibrium for risk-averse investors with unconstrained quadratic utility mean-variance optimization problem, however, they stated, \textit{"Our model does not assume that the world is always at CAPM-equilibrium, but rather that when expected returns move away from their equilibrium values, imbalances in markets will tend to push them back."} Therefore, our study aims at examining the impact of the "imperfection" of information in a more realistic market with incomplete information on the distribution of the equilibrium excess returns, capturing the subjective views of investors. In their inception work, Black and Litterman considered the reference model of the market such that the securities' random vector of excess returns
$$\tilde{R}:=[\tilde{R}_1-r_f,\dots,\tilde{R}_n-r_f]^\top\sim\mathcal{N}(\tilde\mu;\Sigma)$$
follows a multivariate normal distribution, with the random mean $\tilde\mu\sim\mathcal{N}(\pi^c;\Sigma_{\tilde\mu})$, where $\pi^c$ is the CAPM estimate of the mean of the excess returns, and $\Sigma_{\tilde\mu}$ represents the variance of this estimate. In other words, the random mean of $\tilde{R}$ is viewed as $\tilde\mu:=\pi^c+\tilde{\epsilon}$, it is then considered to be normally distributed around $\pi^c$ with a random disturbance value $\tilde\epsilon$ assumed to be normally distributed with mean zero and variance $\Sigma_{\tilde\mu}$, and uncorrelated with the random mean $\tilde{\mu}$. The reference model for the random vector of excess returns is then given, as in the classical BL model and related papers, such that
\begin{equation}\label{equ5}
\tilde{R}\sim\mathcal{N}(\pi^c;\tau\Sigma),\;\text{for}\;\tau\Sigma:=\Sigma+\Sigma_{\tilde{\mu}}\;\text{where}\;\tau\;\text{is a non-negative real scaling factor}.
\end{equation}
The posterior distribution of $\tilde{R}$ at the equilibrium of perfect information markets, which incorporates subjective views ${\tilde{\nu}}\sim\mathcal{N}(q;\Omega)$ of investors, is derived in the BL model using "Theil's Mixed Estimation" model or "Bayesian Inference" approach as follows:
\begin{equation}\label{equ4}
\tilde{R}_{\tilde{\nu}}\sim\mathcal{N}\Bigl(\mathbb{E}\bigl(\tilde{R}_{\tilde{\nu}}\bigr):=\text{Var}\bigl(\tilde{R}_{\tilde{\nu}}\bigr)\Bigl[(\tau\Sigma)^{-1}\pi^c+P^\top\Omega^{-1} q\Bigr];\text{Var}\bigl(\tilde{R}_{\tilde{\nu}}\bigr)\Bigr),\;\text{for}\;\tilde{R}_{\tilde{\nu}}=\tilde{R}|\tilde{\nu},
\end{equation}
where
$$\text{Var}\bigl(\tilde{R}_{\tilde{\nu}}\bigr)=\bigl[(\tau\Sigma)^{-1}+P^\top\Omega^{-1}P\bigr]^{-1}.$$
The BL model characterizes the mean of the distribution of excess returns at equilibrium as a scaled weighted sum of the vector of implied excess returns, $\pi^c$, weighted by $(\tau\Sigma)^{-1}$, and the vector of subjective views, $q$, weighted by $P^\top\Omega^{-1}$. The scaling factor is the variance of the equilibrium excess returns vector. An extensive literature on the BL model has been developed during $90$'s and $20$'s in \cite{HeLitter02,Litt04,Idz07,SatScow07,AlexSil09,Win10}, and recently in \cite{Ber12,Waltet13,Walt14,PJP19,SA22,Z_al22,FuHo23} for new directions. Unlike the BL model, for which beliefs are assigned means and variances a priori and no role for sample information about expected returns is defined, the approach by P\'astor and Stambaugh \cite{PS} investigates the portfolio choices of investors who use sample evidence to update prior beliefs centered on either risk-based or characteristic-based pricing models. This approach relies on the strength of
the sample's information about violations of the pricing model and the investor's prior confidence in the model.
\par In contrast to previous works, our study aims to model a posterior distribution of the random vector of excess returns for markets with incomplete information, while capturing investors' beliefs. Thus, we derive a generalization of the formula of equation (\ref{equ4}) to the more realistic markets with incomplete information. Such a market, also known as an imperfect information market, assumes that each investor only has access to a subset of securities they are informed about. Our approach first involves a reverse optimization of the prior equilibrium value of expected excess returns $\pi$ from the incomplete information model in equation (\ref{eq2}), providing a comprehensive study of the sensitivity of each asset's extra excess returns to its corresponding shadow-cost of information and market portfolio weight, and characterizing the related market equilibrium portfolio. Moreover, assuming that the vector of shadow-costs, $\lambda$, is the mean of a random vector $\tilde\lambda$, we use the Bayesian inference to derive the posterior distribution of the market excess returns equilibrium conditioned on both $\tilde{\lambda}$ and the random vector of subjective views $\tilde{\nu}$. It is assumed that the distributions of $\tilde{\lambda}$ and $\tilde{\nu}$ are multivariate normal with given mean and variance values, $\tilde{\lambda}\sim\mathcal{N}(\lambda;\Lambda)$ and $\tilde{\nu}\sim\mathcal{N}(q;\Omega)$, respectively. In the context of optimizing portfolio allocations, optimal strategies are assumed to be dependent on a market parameter vector $\theta$ and are defined by the allocation function
\begin{equation}\label{eq7_}
    \alpha(\theta)=\argmax_{\alpha\in\mathcal{C}_\theta}\bigl\{\mathcal{S}_\theta(\alpha)\bigr\},\;\text{for the satisfaction function}\;\mathcal{S}_\theta(\alpha):=\mathbb{E}\Bigl[U\bigl(\Psi_\alpha^\theta\bigr)\Bigr],
\end{equation}
these allocation are denoted by $\alpha$ and belongs to a set of admissible allocations $\mathcal{C}_\theta$, where the investor's preferences are modeled by the expected utility framework of Von Neumann and Morgenstern \cite{VNM44}, for the objective $\Psi_\alpha^\theta$ being the wealth, net profits or other specifications, and $U$ a given utility function. Considering $\Psi_\alpha^\theta\equiv\alpha^\top M_\theta$, for a simple affine function market vector $M_\theta$ which depends on the market prices, and whose distribution can be represented by a probability density function $f_\theta(m)$ which is fully determined by the market parameter vector $\theta$. The problem in (\ref{eq7_}) reduces to the estimation of the allocations $\alpha(\theta)$, for $\mathcal{M}$ being the probability support of the market vector $m$, as
\begin{equation}
    \alpha(\theta)=\argmax_{\alpha\in\mathcal{C}_\theta}\biggl\{\int_{\mathcal{M}}U(\alpha^\top m)f_\theta(m)dm\biggr\},\;\text{for a fixed}\;\theta,
\end{equation}
such an allocation is extremely sensitive to the random parameters $\theta$. Additionally, the Bayesian approach consists first in describing the possible outcomes of $\theta$ by a posterior probability density function $f(\theta)$ satisfying
\begin{equation*}
    \bar{\alpha}=\argmax_{\alpha\in\mathcal{C}}\biggl\{\int_\Theta\mathbb{E}\bigl[U(\Psi_\alpha^\theta)\bigr]f(\theta)d\theta\biggr\},
\end{equation*}
for $\Theta$ being the probability support of $\theta$, respectively. The smooth optimal allocation function $\bar{\alpha}$ considers the weighted average over all possible outcomes of the market parameter to reduce the sensitivity effect of the parameter $\theta$. The main advantage of the Bayesian estimation for optimal allocations is that it accounts for estimation risk by letting $f_\theta$ be dependent on the market information, $i_T$, at time horizon $T$, and the investor's experience, or knowledge, $e_C$, with confidence level $C$. Thus, as extensively discussed in Attilio \cite[Chapter $9$]{Attilio05}, this approach of allocating portfolios, depending on the market parameter $\theta$ and market vector $M_\theta$, allows to adopt the posterior density
\begin{equation*}
    f_p(m;i_T,e_C)=\int_\Theta f_\theta(m)f(\theta;i_T,e_C)d\theta,
\end{equation*}
 and then express the Bayesian allocation as
\begin{equation}\label{eq11}
\alpha_B[i_T,e_C]=\argmax_{\alpha\in\mathcal{C}}\int_{\mathcal{M}}U(\alpha^\top m)f_p(m;i_T,e_C)dm.
\end{equation}
This optimal allocation turns out to be $\argmax_{\alpha\in\mathcal{C}}\mathbb{E}\bigl[U(\Psi_\alpha^{i_T,e_C})\bigr]$, which is dependent on the market information and investor's knowledge, and maximizes the expected utility of the objective for an expectation computed according to the posterior distribution of the market $f_p$. Based on the above discussion, a combination of investors' experience and market information in a single portfolio allocation model, such as the Bayesian allocation decision of equation (\ref{eq11}), is of great interest, and a well-determined posterior probability density function $f_p$ is the key for such a combination. Therefore, our study seeks to provide an equilibrium model that encapsulates each investor's density function $f_p$ in the equilibrium vector of excess return, where the investment decision criterion is based on incorporating subjective views in the subset of securities about which the investor holds information. As a starting neutral point for our modeling, such an investor considers Merton's equilibrium excess returns vector $\pi$ of the imperfect information market as an intuitive equilibrium that accounts for the incompleteness of information, and adjusts it based on his/her subjective beliefs about the performance of securities in their subset of information. Hence, a posterior probability density function $f_p$ is implicitly obtained for the market vector of excess returns by exploring Merton's model and Black-Litterman's approach. An optimal allocation that considers the investor's experience and market information is then deduced.
\par The remainder of the paper is organized as follows: In Section \ref{sec2}, we first use Merton's equilibrium model of incomplete information to derive the implied equilibrium expected excess returns vector $\pi$ in markets with incomplete information. Then we discuss the sensitivities, to shadow-costs and market weights, of the extra excess returns resulting from the imperfection of information among investors. Next, we carefully derive a system of non-linear equations characterizing the incomplete information market's portfolio. Market structure is then described, and investors' subjective views are formulated to be incorporated within this market equilibrium in Section \ref{sec.3}. Finally, by exploring the Bayesian approach of statistical inference, we provide the closed-form formula for pricing securities in the imperfect information markets with subjective views of investors in Section \ref{sec.4}. The last Section \ref{sec.6} gives some numerical simulations for markets with imperfect information and subjective beliefs.
\section{Imperfect Information Market Equilibrium}\label{sec2}
The developments in this Section relate to the incomplete information market equilibrium model described in equation (\ref{eq2}). A characterization of the related market portfolio and the excess returns vector is derived to serve as the implied market equilibrium for further development. This is particularly relevant for deriving a general version of BL model, which integrates investor's subjective beliefs while accounting for the imperfection of information among all investors.
\subsection{Deterministic Implied Market Equilibrium}
The market equilibrium model of equation (\ref{eq2}) extends the perfect information market CAPM to the incomplete information markets described through a vector of shadow-costs of information, highlighting the fact that each investor knows only about a subset of securities in the market. Recalling this equation, we define the expected return for any security $k$, among $n$ securities, in the imperfect information market with a total number $N$ of investors as
\begin{equation}\label{eq1}
\mathbb{E}\bigl[\tilde{R}_k\bigr]=r_f+\lambda_k+\beta_k{\Bigl(\mathbb{E}\bigl[\tilde{R}_M\bigr]-r_f-\lambda_{M}\Bigr)},\;\text{for}\;k=1,\dots,n,
\end{equation}
where $\mathbb{E}\bigl[\tilde{R}_M\bigr]:=\sum_{k=1}^nx_k\mathbb{E}\bigl[\tilde{R}_k\bigl]$ is the market portfolio return, for $x_k$ the market portfolio weight corresponding to security $k$, $\lambda_k:=\sum_{j=1}^N\lambda_k^j/N$ is the aggregate-average shadow-cost per investor of not knowing about security $k$, and $\lambda_M:=\sum_{k=1}^nx_k\lambda_k$ denotes the weighted-average shadow-cost of all securities. The risk drivers in these markets are then given, not only by the random vector of excess returns
\begin{equation*}
\tilde{R}:=\bigl[\tilde{R}_1-r_f,\dots,\tilde{R}_n-r_f\bigr]^\top,
\end{equation*}
but also by the additional random vector of shadow-costs
\begin{equation}\label{eq9_1}
    \tilde\lambda:=[\tilde\lambda_1,\dots,\tilde\lambda_n]^\top,
\end{equation}
whose expected value is given by the vector of shadow-costs
\begin{equation}\label{eq9}
    \lambda:=[\lambda_1,\dots,\lambda_n]^\top.
\end{equation}
In the considered imperfect information markets, some available securities have a non-complete investor-base, i.e., not all investors are informed about these securities. Thus, the distribution of the information about $\bar{R}_k$, $b_k$ and $\sigma_k$ of equation (\ref{eq01}) across investors is not equal, and each investor knows only the parameters related to a subset of securities.  For instance, an investor $j$ knows about the subset of securities $J_j=\{1,\dots,n_j\}$. The resulted positive shadow-cost $\lambda_k$, for security $k$, is computed as the mean of all $\lambda_k^j$ related to investors $j$ that ignore security $k$. As mentioned in the introduction Section \ref{Sec1}, the BL model suggested the use of the CAPM-equilibrium $\pi^c=\delta\Sigma w_M^c$ as the neutral starting point for their equilibrium model (\ref{equ4}). The resulting posterior equilibrium incorporates the investor's beliefs and guides, through a mean-variance optimizer, the prior implied equilibrium market portfolio of equation (\ref{eq04}) towards new weights vector that reflects the given subjective views. Similarly, we shall derive first the implied imperfect information market equilibrium as a vector expression of market data and shadow-costs.
\par We first rewrite equation (\ref{eq1}), by replacing $\tilde{R}_M$ in the $\beta_k$-coefficient by $\sum_{i=1}^{n}x_i\tilde{R}_i$, as
\begin{equation*}
    \mathbb{E}\bigl[\tilde{R}_k\bigr]-r_f=\frac{1}{\text{var}\bigl(\tilde{R}_M\bigr)}\sum_{i=1}^nx_i\text{cov}\bigl(\tilde{R}_k,\tilde{R}_i\bigr){\Bigl(\mathbb{E}\bigl[\tilde{R}_M\bigr]-r_f-\lambda_{M}\Bigr)}+\lambda_k.
\end{equation*}
Thus, combining this equation for all securities $k$, we get the vector formula of the incomplete information market equilibrium model as
\begin{equation}\label{eq3}
    \pi=\biggl(\delta-\frac{\lambda_M}{\sigma_M^2}\biggr)\Sigma w_M+\lambda,
\end{equation}
where $\pi$ represents the vector of expected excess returns as
$$\pi:=\Bigl[\mathbb{E}\bigl[\tilde{R}_1\bigr]-r_f, \dots,\mathbb{E}\bigl[\tilde{R}_n\bigr]-r_f\Bigr]^\top,$$
$\delta$ is the market factor, i.e., the risk-aversion coefficient of equation (\ref{eq04}), with $\sigma_M^2:=\text{var}\bigl(\tilde{R}_M\bigr)$, $\Sigma:=\text{var}\bigl(\tilde{R}+r_f\textbf{1}_{\mathbb{R}^n}\bigr)=\text{var}\bigl(\tilde{R}\bigr)\in\mathbb R^{n\times n}$ the variance-covariance matrix of returns vector $\bigl[\tilde{R}_1,\dots,\tilde{R}_n\bigr]^\top$, and $w_M:=[x_1,\dots,x_n]^\top$ the market portfolio weights vector at equilibrium, for $\textbf{1}_{\mathbb{R}^n}$ the elementary vector of ones in $\mathbb{R}^n$. We should mention that the weighted-average shadow-cost $\lambda_M:={w_M}^\top\lambda$ is computed using the market weights vector $w_M$ and the shadow-costs $\lambda_k^j$ of all investors $j=1,\dots,N$ and any security $k$, these shadow-costs are given by the vector $\lambda$.
\begin{notation}\label{Notation1}
    Let $n$ be the fixed integer number of risky securities in the market. The random variable of $i$-th security's return is denoted by $\tilde{R}_i$, and has a mean $\bar{R}_i:=\mathbb{E}[\tilde{R}_i]$, for $i=1,\dots,n$. The mean $\mathbb{E}[\tilde{R}]$ of the random vector of all risky securities' expected excess returns is denoted by $\pi$ for the incomplete information market, and $\pi^c$ for the complete information market with $\lambda=\mathbf{0}_{\mathbb{R}^n}$. The vector of shadow-costs $\lambda$ is denoted as the expectation of the random vector $\tilde{\lambda}$, i.e., $\mathbb{E}[\tilde\lambda]=:\lambda$, where the variance of this vector is denoted as $\text{Var}[\tilde\lambda]=:\Lambda$. The capital market is such that the weighted-average shadow-costs scalar and equilibrium portfolio vector are denoted by $\lambda_M$ and $w_M$, respectively, while $w_M^c$ denotes the perfect information market equilibrium portfolio. \qed
\end{notation}
We now give the aim of the Section.
\begin{aim}
    The aim is twofold:
    \begin{itemize}
        \item To quantify the value and sensitivity, with respect to shadow-costs and the market portfolio, of the additional excess returns when transitioning from a complete to an incomplete information market;
        \item To derive explicit expressions for the optimal portfolios for both market equilibrium and an investor who is maximizing a quadratic utility function. \qed
    \end{itemize}
\end{aim}
\subsection{Extra Excess Returns Sensitivities}
\par An analysis of the expected excess returns related to the incomplete information market is presented in this section. We first rewrite the equilibrium of equation (\ref{eq3}) as
\begin{equation}\label{q6}
    \pi=\pi^c+\underset{\text{Extra Excess Returns}}{\underbrace{\lambda-\pi^\lambda}},\;\text{where}\;\pi^c:=\delta\Sigma w_M\;\text{and}\;\pi^\lambda:=\frac{\lambda_M}{\sigma_M^2}\Sigma w_M,
\end{equation}
the excess returns vector $\pi$ resulted in being adjusted value of the implied equilibrium excess returns vector $\pi^c$ of equation (\ref{eq04}) related to the complete information market modeled by the CAPM. Thus, moving from the complete to the incomplete information market structure, the implied equilibrium excess returns for the investor is rewarded by the additional term
\begin{equation}\label{d}
\begin{aligned}
    \lambda-\pi^\lambda&=\lambda-\frac{1}{\sigma_M^2}\bigl({w_M}^\top\lambda\bigr)\Sigma w_M,\\
&=\lambda-\lambda_M\beta,
\end{aligned}
\end{equation}
where the vector $\beta:=[\beta_1,\cdots,\beta_n]^\top$ defines the systematic risk related to the market's securities. Given then the vector $\lambda$ of shadow-costs, the implied market equilibrium weights vector $w_M$, and the $\beta$-coefficient one can quantify the extra excess returns an investor might expect to receive from each security in the market of incomplete information. In addition to the CAPM return, each asset $k$ earns its respective shadow-cost $\lambda_k$ penalized by a scaled value of its sensitivity to the complete information market systematic risk vector $\beta:=\Sigma w_M/\sigma_M^2$ of all securities, where the scaling factor is given by the weighted-average shadow-cost $\lambda_M$. Hence, defining the new risk-aversion coefficient $\delta_\lambda:=\lambda_M/\sigma_M^2$, we can express the implied excess returns vector as $\pi=(\delta-\delta_\lambda)\Sigma w_M+\lambda$, or, to remain consistent with the initial equation (\ref{eq2}), as
$$\pi=\Bigl(\mathbb{E}\bigl[\tilde{R}_M\bigr]-r_f-\lambda_M\Bigr)\beta+\lambda.$$
\begin{corollary}[Extra Excess Returns]
The extra excess returns of an asset $k$ is positive when this asset is negatively correlated to the market, or, when it is positively correlated to the market and the two following items hold:
\begin{itemize}
\item The investor-base of this asset, i.e., the number of investors that know about asset $k$ and are interested in allocating it to their portfolios, is small. In other words, this asset's shadow-cost $\lambda_k$ is significantly positive;
\item The market has higher volatility allowing $\delta_\lambda$ to decrease dramatically, this is true since $\lambda_M$ is bounded due to the fact that $x_k\lambda_k$ is minimal for any $k$. In fact, a large market-weighted asset $k$ is related to lower shadow-cost $\lambda_k$, while higher shadow-cost $\lambda_k$ corresponds to lower market weight $x_k$. \qed
\end{itemize}
\end{corollary}

\par A sensitivity analysis of the extra excess returns can now be carried out. The term $\lambda-\pi^\lambda$ is differentiable with respect to shadow-costs vector $\lambda$ and has the following gradient
\begin{equation}\label{nablalambda}
    \begin{aligned}
    \nabla_{\lambda}\bigl(\lambda-\pi^\lambda\bigr):&=\frac{\partial\bigl(\lambda-\pi^\lambda\bigr)}{\partial\lambda}\\
    &=\mathbf{1}_{\mathbb{R}^n}-\frac{1}{\sigma_M^2}D_{w_M}\Sigma w_M\\
    &=\mathbf{1}_{\mathbb{R}^n}-D_{w_M}\beta,
\end{aligned}
\end{equation}
where $D_{w_M}$ denotes the diagonal matrix of diagonal elements being the vector $w_M:=[x_1,\dots,x_n]^\top$ and zeros in off-diagonal elements
\[
D_{w_M} := \begin{pmatrix}
x_1 & 0 & 0 & \cdots & 0 \\
0 & x_2 & 0 & \cdots & 0 \\
0 & 0 & x_3 & \cdots & 0 \\
\vdots & \vdots & \vdots & \ddots & \vdots \\
0 & 0 & 0 & \cdots & x_n
\end{pmatrix}.
\]
Hence, exploring this sensitivity for positive market weight $x_k$, $k=1,\dots,n$, assets with negative $\beta$-coefficient and those with lower market weight and smaller positive $\beta$ tend to experience higher extra excess returns in high-volatility markets, as more investors ignore these assets, i.e., as their respective shadow-cost increases.
\par Additionally, the sensitivity of the extra excess returns to the market portfolio's weights vector $w_M$ is given by the gradient
\begin{equation}\label{nablax}
    \begin{aligned}
    \nabla_{w_M}\bigl(\lambda-\pi^\lambda\bigr):&=\frac{\partial\bigl(\lambda-\pi^\lambda\bigr)}{\partial w_M}\\
    &=-\frac{1}{\sigma_M^2}D_\lambda\Sigma w_M-\frac{\lambda_M}{\sigma_M^2}D_\Sigma\mathbf{1}_{\mathbb{R}^n}\\
    &=-D_\lambda\beta-\delta_\lambda D_\Sigma\mathbf{1}_{\mathbb{R}^n}.
\end{aligned}
\end{equation}
Thus, fixing the shadow-costs vector, assets positively correlated with the market receives lower extra excess return as their respective market weights increase. Conversely, the extra excess return for negatively correlated assets vary based on their respective systematic risk and shadow-costs.
\par The aforementioned results on the sensitivities align with the analysis provided in Merton \cite{Mert87} and are summarized, for each asset $k$ whose market weight $x_k$ is positive, as follows:
\begin{theorem}[Sensitivities of Extra Excess Returns]
The sensitivities of the extra excess returns $\lambda-\pi^\lambda$ of each asset $k$, with nonzero systematic risk $\beta_k$, are derived from equations (\ref{nablalambda}) and (\ref{nablax}), and described by
    \begin{itemize}
        \item \textbf{Sensitivity to Shadow-Cost $\boldsymbol{\lambda_k}$}: In the incomplete information market, the extra excess return of any asset $k$ increases with its shadow-cost $\lambda_k$ if, and only if, the asset is negatively correlated with the market, or its market portfolio weight $x_k$ is less than the inverse of its systematic risk $\beta_k$ when positively correlated with the market. In other words, as more investors remain unaware of security $k$, i.e., $\lambda_k$ increases, its extra excess return increases either when $\beta_k<0$, or when $x_k<1/\beta_k$ for $\beta_k>0$;
        \item \textbf{Sensitivity to Market Weight $\boldsymbol{x_k}$}: In the incomplete information market, each asset $k$ positively correlated with the market experiences a decrease in extra excess return as its market weight $x_k$ increases. Conversely, an asset that is negatively correlated with the market enjoys a higher extra excess return as $x_k$ increases, provided $|\beta_k|>\delta_\lambda\sigma_k^2/\lambda_k$. \qed
    \end{itemize}
\end{theorem}
Next we state a straightforward result by distinguishing sensitivities by the systematic risk.
\begin{corollary}[Distinguishing Sensitivities by $\beta_k$]
    In the incomplete information market, the $k$-th asset expected excess return $\pi_k$ is the sum of $\pi_k^c=\delta\Sigma_kw_M$, the excess return in the complete information market, and $\lambda_k-\pi_k^\lambda=\lambda_k-\lambda_M\beta_k$ its extra excess return. Moreover, we have
    \begin{itemize}
        \item \textbf{Case when $\boldsymbol{\beta_k>0}$}: For a fixed market weight $x_k$, if more investors become interested in asset $k$, i.e., its shadow-cost $\lambda_k$ decreases, then the asset's extra excess return decreases when $x_k<1/\beta_k$, and increases when $x_k>1/\beta_k$. While, when $\lambda_k$ is fixed and $x_k$ increases, then the asset's extra excess return decreases;
        \item \textbf{Case when $\boldsymbol{\beta_k<0}$}: For a fixed market weight $x_k$, if more investors neglect asset $k$, i.e., $\lambda_k$ increases, then the asset's extra excess return increases. While, for a fixed shadow-cost $\lambda_k$ and increasing $x_k$, the extra excess return increases when $|\beta_k|>\delta_\lambda\sigma_k^2/\lambda_k$;
        \item \textbf{Case when $\boldsymbol{\beta_k=0}$}: An asset with null systematic risk $\beta_k$ has an extra excess return given by its respective shadow-cost $\lambda_k$. \qed
    \end{itemize}
\end{corollary}
The part $\pi^\lambda$ in equation (\ref{q6}) is defined as the product of $\lambda_M$ and $\beta$, which both depend on the market portfolio weights vector $w_M$. Hence, the implied expected excess returns vector $\pi$ is given by a quadratic expression of the market portfolio weights vector $w_M$ at equilibrium. The derivation of this vector requires then further development.
\subsection{Market Equilibrium Portfolio}
    \par To derive the market equilibrium portfolios' weights vector $w_M$ for the incomplete information market, we first rewrite the implied equilibrium of equation (\ref{eq3}) as
\begin{equation*}
    \pi=\delta\Sigma w_M-\frac{1}{\sigma_M^2}({w_M}^\top\lambda)\Sigma w_M+\lambda,
\end{equation*}
from which we get
\begin{equation*}
    (\delta\Sigma)^{-1}(\pi-\lambda)=w_M-\frac{1}{\delta\sigma_M^2}({w_M}^\top\lambda)w_M,
\end{equation*}
or, equivalently,
\begin{equation}\label{eq6}
    (\delta\Sigma)^{-1}(\pi-\lambda)=\Bigl(1-\frac{\lambda_M}{\delta\sigma_M^2}\Bigr)w_M.
\end{equation}
The equilibrium weights vector on the right-hand side of ($\ref{eq6}$) is scaled by the scalar-valued coefficient $1-\delta_\lambda/\delta$, for $\delta_\lambda:=\lambda_M/\sigma_M^2$. One might then expect to derive the expression for the weights vector $w_M$ corresponding to the market with incomplete information as
    \begin{equation}\label{eq0022}
    {w_M}:=(\delta-\delta_\lambda)^{-1}\Sigma^{-1}(\pi-\lambda),
    \end{equation}
    thing that remains consistent with the implied equilibrium of equation (\ref{q6}). However, the scalar-valued coefficient also depends on the unknown weights vector and the incomplete information market equilibrium portfolio can not be directly characterized and further development is required. Taking into account equality (\ref{q6}) for a single security $k$, equation ($\ref{eq6}$) is equivalent to the expression
    \begin{align}
    \pi_k&=\pi_k^c+\lambda_k-\pi_k^{\lambda}\nonumber\\
        &=\delta\Sigma_k\Bigl(w_M-\frac{1}{\delta\sigma_M^2}\Bigl(\sum_{i=1}^{n}x_{i}\lambda_{i}\Bigr)w_M\Bigr)+\lambda_k\\
    &=\delta\Sigma_k\biggl(w_M-\frac{1}{\delta\sigma_M^2}\Bigl(\lambda\circ w_M^{2}+w_M\circ\Bigl[\sum_{i=1,i\neq1}^{n}x_{i}\lambda_{i},\dots,\sum_{i=1,i\neq n}^{n}x_{i}\lambda_{i}\Bigr]^\top\Bigr)\biggr)+\lambda_k,
\end{align}
where $\Sigma_k$ represents the $k$-th row of $\Sigma$, $v_k$ denotes the $k$-th element of vector $v$, $w_M^2:=[x_1^2,\dots,x_n^2]^\top$, and the symbol $\circ$ denotes the element-wise multiplication.
\par Hence, the optimal weights vector $w_M$ is given by the solution to the system of non-linear equations
\begin{equation}\label{eq8}
        (\delta\Sigma)^{-1}(\pi-\lambda)=w_M-\frac{1}{\delta\sigma_M^2}\Bigl(D_{\lambda}w_M^2+D_{w_M}\bigl(M_{\lambda}-D_{\lambda}\bigr)w_M\Bigr),
\end{equation}
where $D_v$ is the diagonal matrix of diagonal elements being the vector $v$ and zeros in off-diagonal elements,
\[
M_{\lambda} := \begin{pmatrix}
\lambda_{1} & \cdots & \lambda_{n} \\
\vdots & \ddots & \vdots \\
\lambda_{1} & \cdots & \lambda_{n}
\end{pmatrix},\;\text{and}\;
M_{\lambda}-D_{\lambda} := \begin{pmatrix}
0 & \lambda_{2} & \lambda_{3} & \cdots & \lambda_{n-1} & \lambda_{n} \\
\lambda_{1} & 0 & \lambda_{3} & \cdots & \lambda_{n-1} & \lambda_{n} \\
\lambda_{1} & \lambda_{2} & 0 & \cdots & \lambda_{n-1} & \lambda_{n} \\
\vdots & \vdots & \vdots & \ddots & \vdots & \vdots \\
\lambda_{1} & \lambda_{2} & \lambda_{3} & \cdots & 0 & \lambda_{n} \\
\lambda_{1} & \lambda_{2} & \lambda_{3} & \cdots & \lambda_{n-1} & 0
\end{pmatrix}.
\]
Considering now the resulted expression
$$\pi^{\lambda}=\frac{1}{\sigma_M^2}\Sigma\Bigl(D_{\lambda}w_M^2+D_{w_M}(M_{\lambda}-D_{\lambda})w_M\Bigr),$$
to express the implied equilibrium equation $\pi-\lambda=\delta\Sigma w_M-\pi^\lambda$, for $\pi^\lambda$ being a non-linear expression of the market equilibrium portfolio. Therefore, the market weights vector related to the implied equilibrium excess returns vector $\pi$ of equation (\ref{eq3}) is characterized by
\begin{proposition}[Incomplete Information Market Equilibrium Portfolio]
    An incomplete information market of $n$ assets has an equilibrium portfolio weights vector described by the solution of the system of non-linear equations $F(W)=\boldsymbol{0}_{\mathbb{R}^n}$, where
    \begin{equation}\label{eqq}
    \begin{aligned}
        F:\;&\mathbb{R}^n\rightarrow\mathbb{R}^n\\
        &W\rightarrow W-\frac{1}{\delta\sigma_M^2}\Bigl(D_{\lambda}W^2+D_{W}(M_{\lambda}-D_{\lambda})W\Bigr)-(\delta\Sigma)^{-1}(\pi-\lambda),
    \end{aligned}
    \end{equation}
    for the excess returns vector $\pi$ of incomplete information market. \qed
    \begin{remark}
        A solution for this equation can be given by the iterative scheme of Newton-Raphson
    \begin{equation*}
        W^{k+1}=W^{k}-JF^{-1}(W^{k})F(W^{k}),
    \end{equation*}
    where $W^{k}$ is an initial guess and $JF(W)$ denotes the Jacobian matrix of $F$ evaluated at point $W$. \qed
    \end{remark}
    \end{proposition}

    \par We first used the reverse optimization method and started from the complete information market equilibrium portfolio weights vector $w_M$ and known model parameters $\delta,\sigma_M,\Sigma$ and $\lambda$, to extract the neutral starting point incomplete information market implied equilibrium excess returns vector $\pi$ of equation (\ref{eq3}). This vector accounts for the incompleteness of information and gives the system of equations (\ref{eq8}) as a characterization of the optimal portfolio in a market with incomplete information. Such a market equilibrium portfolio can also be derived through a risk-adjusted return utility optimization problem associated with the incomplete information market.
\begin{corollary}
    The implied equilibrium expected excess returns vector $\pi$ in the incomplete information market is characterized by $\pi=(\delta-\delta_\lambda)\Sigma w_M+\lambda$, and has the related market equilibrium portfolio described by the solution of the system of non-linear equations (\ref{eqq}). \qed
\end{corollary}
\subsection{Risk-Adjusted Return Utility Optimization}
Considering that each investor in the perfect information market aims at allocating optimal portfolios by performing the standard unconstrained mean-variance quadratic utility maximization. Hence, by solving the problem of finding the optimal portfolio weights vector $w^*$ satisfying
\begin{equation}\label{eeq24}
    w^*=\argmax_{w}U_\delta(w),\;\text{where}\;U_\delta(w):=w^\top\pi^c-\frac{1}{2}\delta w^\top\Sigma w.
\end{equation}
The first-order condition is applied to set the gradient of $U_\delta$ to zero, $$\nabla_wU_\delta(w):=\frac{\partial U_\delta(w)}{\partial w}=0,$$ in order to derive the optimal portfolio at equilibrium. This produces the CAPM-equilibrium $w_M^c=(\delta\Sigma)^{-1}\pi^c$ of equation (\ref{eq04}) as value for $w^*$. An adjusted formulation of the above optimization problem for the incomplete information market is given by the above utility function with the additional parameter $\lambda$, while replacing $\pi^c$ by the implied market equilibrium $\pi$ of equation (\ref{eq3}). We then redefine the above utility function (\ref{eeq24}) as
\begin{equation}\label{eq19}
    U_{\delta,\lambda}(w):=\underset{\text{Risk-Adjusted Return in Complete Information Market}}{\underbrace{w^\top\pi^c-\frac{1}{2}\delta w^\top\Sigma w}}+\underset{\text{Information-Adjusted Premium}}{\underbrace{\Bigl(w^\top\lambda-\frac{\lambda_M}{\sigma_M^2}w^\top\Sigma w\Bigr)}},
\end{equation}
where the first term is classic and related to CAPM, while the second allows for integrating the incomplete information asset-specific and market-wide premiums in the portfolio selection process, the selected assets will then reflect the investor's willingness to incorporate the incompleteness of information through the presence of the additional information-adjusted premium. Considering the fact that the investor is looking for the optimal allocation that corresponds to the market equilibrium weights vector $w_M:=\argmax_{w}U_{\delta,\lambda}(w)$, the fact that $U_{\delta,\lambda}$ is quadratic and $U_{\delta,\lambda}(w_M)=U_{\delta}(w_M)$, only the left-hand side can characterize his/her ability to perform the risk-adjusted optimization. In contrast, the right-hand side part is used to control the incompleteness of information in the market through the first inner product term ${w}^\top\lambda$ and the second weighted-average shadow-cost term $\lambda_M$. Indeed, the information-adjusted premium is null, since the term ${w_M}^\top\Sigma w_M/\sigma_M^2$ reduces to one, but helps to incorporate the imperfection of information to the market. Calling now the first-order condition, $\nabla_wU_\delta(w)=0$, we write
\begin{equation*}
    \pi^c-\delta\Sigma w+\frac{\lambda_M}{\sigma_M^2}\Sigma w-\frac{\lambda_M}{\sigma_M^2}\Sigma w+\lambda-\frac{1}{\sigma_M^2}({w_M}^\top\Sigma w_M)\lambda=0.
\end{equation*}
Then, exploring equation (\ref{q6}) and multiplying both sides by $(\delta\Sigma)^{-1}$, we get
\begin{equation*}
    (\delta\Sigma)^{-1}\pi-w+\frac{\delta_\lambda}{\delta}w-(\delta\Sigma)^{-1}\lambda=0,\;\text{where}\;\delta_\lambda:=\frac{\lambda_M}{\sigma_M^2}.
\end{equation*}
Finally, we deduce the optimal market portfolio allocation $w_M$ solution to the above equation and to the utility maximization problem (\ref{eeq24}), which adjusts the complete information market portfolio to shadow-costs vector as
$$w_M=(\delta-\delta_\lambda)^{-1}\Sigma^{-1}(\pi-\lambda).$$
This allocation is given as the solution to the system of non-linear equations (\ref{eqq}).
\par The optimal solution for the investor can be found by directly applying the first order condition, $\nabla_wU_{\delta,\lambda}(w)=0$, to the problem (\ref{eq19}) then deduce that
\begin{equation*}
    \pi^c+\lambda-(\delta+2\delta_\lambda)\Sigma w=0.
\end{equation*}
Then the optimal allocation $w^*$ that accounts for the imperfection of information is deduced as a solution to this equation. The corollary below summarizes these results.
\begin{corollary}[Modified Risk-Aversion Coefficient and Optimal Portfolios]
    In the incomplete information market, the unconstrained risk-adjusted return portfolio optimization problem has the utility function of the form
\begin{equation*}
    U_{\delta,\lambda}(w):=w^\top(\pi^c+\lambda)-\frac{1}{2}(\delta+2\delta_{\lambda}) w^\top\Sigma w,\;\text{for}\;\delta_\lambda:=\frac{{w_M}^\top\lambda}{\sigma_M^2}
\end{equation*}
with the modified risk-aversion coefficient $\delta+2\delta_{\lambda}$ and modified return objective $\pi^c+\lambda$, where $\pi^c$ and $\pi$ are the implied equilibrium in complete and incomplete information markets, respectively. The maximization to find optimal market portfolio at equilibrium gives
$$w_M=(\delta-\delta_\lambda)^{-1}\Sigma^{-1}(\pi-\lambda).$$
Moreover, the maximization over all possible allocations gives the investor's optimal portfolio
$$w^*=(\delta+2\delta_\lambda)^{-1}\Sigma^{-1}(\pi^c+\lambda).$$ \qed
\end{corollary}
\subsection{Solution Scheme Summary}
Scheme \ref{scheme:1} below summarizes the comprehensive study on the equilibrium in imperfect information markets based on Merton's model of equilibrium. Our analysis is the starting point for integrating investor's subjective beliefs in these markets.
\begin{scheme}[H]
  \centering
\begin{tikzpicture}[x=0.75pt,y=0.75pt,yscale=-1,xscale=1]

\draw    (285,35) -- (320,35) ;
\draw [shift={(320,35)}, rotate = 180] [color={rgb, 255:red, 0; green, 0; blue, 0 }  ][line width=0.75]    (10.93,-3.29) .. controls (6.95,-1.4) and (3.31,-0.3) .. (0,0) .. controls (3.31,0.3) and (6.95,1.4) .. (10.93,3.29)   ;
\draw    (455,70) -- (455,90) ;
\draw [shift={(455,90)}, rotate = 270] [color={rgb, 255:red, 0; green, 0; blue, 0 }  ][line width=0.75]    (10.93,-3.29) .. controls (6.95,-1.4) and (3.31,-0.3) .. (0,0) .. controls (3.31,0.3) and (6.95,1.4) .. (10.93,3.29)   ;

\draw [fill=gray!10]  (70,0) -- (285,0) -- (285,70) -- (70,70) -- cycle ;
\draw (100,5) node [anchor=north west][inner sep=0.75pt]  [font=\small] [align=left] {\textbf{Sharpe-Lintner-Mossin}};
\draw (90,25) node [anchor=north west][inner sep=0.75pt]  [font=\small] [align=left] {Perfect Market Equilibrium};
\draw (140,45) node [anchor=north west][inner sep=0.75pt]  [font=\small] [align=left] {$\pi^c=\delta\Sigma w_M^c$};
\draw [fill=gray!10]   (320,0) -- (585,0) -- (585,70) -- (320,70) -- cycle ;
\draw   (400,5) node [anchor=north west][inner sep=0.75pt]  [font=\small] [align=left]{\textbf{Merton's Model}};
\draw (365,25) node [anchor=north west][inner sep=0.75pt]  [font=\small] [align=left]{Shadow-Costs of Information};
\draw   (400,45) node [anchor=north west][inner sep=0.75pt]  [font=\small] [align=left] {$\lambda:=[\lambda_1,\dots\lambda_n]^\top$};
\draw[fill=green!10]  (70,150) -- (285,150) -- (285,90) -- (70,90) -- cycle ;
\draw (105,100) node [anchor=north west][inner sep=0.75pt]  [font=\small] [align=left] {Asset's $k$ Excess Return};
\draw (115,120) node [anchor=north west][inner sep=0.75pt]  [font=\small] [align=left] {$\pi_k=\pi_k^c+\lambda_k-\pi_k^\lambda$};
\draw[]    (70,195) -- (285,195) -- (285,240) -- (70,240) -- cycle ;
\draw (120,200) node [anchor=north west][inner sep=0.75pt]  [font=\small] [align=left] {Extra excess return};
\draw (105,218) node [anchor=north west][inner sep=0.75pt]  [font=\small] [align=left] {$\lambda_k-\pi_k^\lambda=\lambda_k-\lambda_M\beta_k$};

\draw[fill=gray!10]  (70,245) -- (285,245) -- (285,275) -- (70,275) -- cycle ;
\draw (145,253) node [anchor=north west][inner sep=0.75pt]  [font=\small] [align=left] {If $\beta_k>0$};

\draw[fill=gray!10]  (70,455) -- (285,455) -- (285,485) -- (70,485) -- cycle ;
\draw (145,460) node [anchor=north west][inner sep=0.75pt]  [font=\small] [align=left] {If $\beta_k<0$};

\draw[fill=gray!1]  (70,490) -- (285,490) -- (285,520) -- (70,520) -- cycle ;
\draw (105,495) node [anchor=north west][inner sep=0.75pt]  [font=\small] [align=left] {If $\lambda_k\uparrow\;(x_k\;\text{unchanged})$};

\draw[fill=gray!1]  (70,525) -- (285,525) -- (285,555) -- (70,555) -- cycle ;
\draw (145,530) node [anchor=north west][inner sep=0.75pt]  [font=\small] [align=left] {$\lambda_k-\pi_k^\lambda\uparrow$};

\draw[fill=gray!1]  (70,560) -- (285,560) -- (285,590) -- (70,590) -- cycle ;
\draw (105,565) node [anchor=north west][inner sep=0.75pt]  [font=\small] [align=left] {If $x_k\uparrow\;(\lambda_k\;\text{unchanged})$};

\draw[fill=gray!1]  (70,595) -- (285,595) -- (285,625) -- (70,625) -- cycle ;
\draw (78,600) node [anchor=north west][inner sep=0.75pt]  [font=\small] [align=left] {$\lambda_k-\pi_k^\lambda\uparrow\;\text{when}\;|\beta_k|>\delta_\lambda\sigma_k^2/\lambda_k$};


\draw [color={rgb, 255:red, 0; green, 0; blue, 0 }  ,draw opacity=1 ] (70,280) -- (285,280) -- (285,310) -- (70,310) -- cycle ;

\draw (105,288) node [anchor=north west][inner sep=0.75pt]  [font=\small] [align=left] {If $\lambda_k\downarrow\;(x_k\;\text{unchanged})$};

\draw [color={rgb, 255:red, 0; green, 0; blue, 0 }  ,draw opacity=1 ] (70,315) -- (285,315) -- (285,345) -- (70,345) -- cycle ;

\draw (90,318) node [anchor=north west][inner sep=0.75pt]  [font=\small] [align=left] {$\lambda_k-\pi_k^\lambda\downarrow\;\text{when}\;x_k<1/\beta_k$};

\draw [color={rgb, 255:red, 0; green, 0; blue, 0 }  ,draw opacity=1 ] (70,350) -- (285,350) -- (285,380) -- (70,380) -- cycle ;

\draw (90,353) node [anchor=north west][inner sep=0.75pt]  [font=\small] [align=left] {$\lambda_k-\pi_k^\lambda\uparrow\;\text{when}\;x_k>1/\beta_k$};

\draw [color={rgb, 255:red, 0; green, 0; blue, 0 }  ,draw opacity=1 ] (70,385) -- (285,385) -- (285,415) -- (70,415) -- cycle ;

\draw (105,390) node [anchor=north west][inner sep=0.75pt]  [font=\small] [align=left] {If $x_k\uparrow\;(\lambda_k\;\text{unchanged})$};

\draw [color={rgb, 255:red, 0; green, 0; blue, 0 }  ,draw opacity=1 ] (70,420) -- (285,420) -- (285,450) -- (70,450) -- cycle ;
\draw (145,423) node [anchor=north west][inner sep=0.75pt]  [font=\small] [align=left] {$\lambda_k-\pi_k^\lambda\uparrow$};

\draw  [fill=green!10] (320,90) -- (585,90) -- (585,150) -- (320,150) -- cycle ;
\draw (360,95) node [anchor=north west][inner sep=0.75pt]  [font=\small] [align=left] {Imperfect Market Equilibrium};
\draw (330,115) node [anchor=north west][inner sep=0.75pt]  [font=\small] [align=left] {$\pi=(\delta-\delta_\lambda)\Sigma w_M+\lambda$, where $\delta_{\lambda}:=\frac{\lambda_M}{\sigma_M^2}$};
\draw[dashed]  [color={rgb, 255:red, 0; green, 0; blue, 0 }  ,draw opacity=1 ] (320,200) -- (435,200) -- (435,245) -- (320,245) -- cycle ;
\draw (330,205) node [anchor=north west][inner sep=0.75pt]  [font=\small] [align=left] {Mean-Variance};
\draw (350,225) node [anchor=north west][inner sep=0.75pt]  [font=\small] [align=left] {Optimizer};

\draw[fill=gray!10] (340,390) -- (565,390) -- (565,440) -- (340,440) -- cycle ;
\draw (360,400) node [anchor=north west][inner sep=0.75pt, ]  [font=\small] [align=left] {$\argmax_{w}\Bigl\{w^\top\hat\pi-\frac{1}{2}\hat\delta w^\top\Sigma w\Bigr\}$};

\draw  (320,290) -- (435,290) -- (435,350) -- (320,350) -- cycle ;
\draw (335,300) node [anchor=north west][inner sep=0.75pt, ]  [font=\small] [align=left] {$\hat\pi:=\pi^c+\lambda$};
\draw (335,320) node [anchor=north west][inner sep=0.75pt, ]  [font=\small] [align=left] {$\hat\delta:=\delta+2\delta_{\lambda}$};

\draw[dotted]  [color={rgb, 255:red, 0; green, 0; blue, 0 }  ,draw opacity=1 ] (20,-10) -- (600,-10) -- (600,165) -- (20,165) -- cycle ;

\draw[dotted]  [color={rgb, 255:red, 0; green, 0; blue, 0 }  ,draw opacity=1 ] (305,180) -- (600,180) -- (600,640) -- (305,640) -- cycle ;
\draw[dotted]  [color={rgb, 255:red, 0; green, 0; blue, 0 }  ,draw opacity=1 ] (20,180) -- (295,180) -- (295,640) -- (20,640) -- cycle ;
\draw[dashed]  [color={rgb, 255:red, 0; green, 0; blue, 0 }  ,draw opacity=1 ] (440,200) -- (585,200) -- (585,245) -- (440,245) -- cycle ;
\draw (455,205) node [anchor=north west][inner sep=0.75pt]  [font=\small] [align=left] {Non-Linear System};
\draw (490,225) node [anchor=north west][inner sep=0.75pt]  [font=\small] [align=left] {Solver};
\draw [color={rgb, 255:red, 0; green, 0; blue, 0 }  ,draw opacity=1 ] (440,290) -- (585,290) -- (585,350) -- (440,350) -- cycle ;
\draw (455,300) node [anchor=north west][inner sep=0.75pt]  [font=\small] [align=left] {$W$ Solution of the};
\draw (450,320) node [anchor=north west][inner sep=0.75pt]  [font=\small] [align=left] {System $F(W)=\boldsymbol{0}_{\mathbb{R}^n}$};

\draw[dotted]  (320,480) -- (585,480) -- (585,610) -- (320,610) -- cycle ;

\draw[dotted,fill=green!10]  (340,550) -- (550,550) -- (550,590) -- (340,590) -- cycle ;
\draw[dotted,fill=green!10]  (340,540) -- (550,540) -- (550,500) -- (340,500) -- cycle ;

\draw (355,510) node [anchor=north west][inner sep=0.75pt]  [font=\small] [align=left] {$w_M=(\delta-\delta_\lambda)^{-1}\Sigma^{-1}(\pi-\lambda)$};

\draw (400,560) node [anchor=north west][inner sep=0.75pt]  [font=\small] [align=left] {$w^*=\bigl(\hat\delta\Sigma\bigr)^{-1}\hat\pi$};


\draw []   (285,120) -- (320,120) ;
\draw [shift={(285,120)}, rotate = 360] [color={rgb, 255:red, 0; green, 0; blue, 0 }  ][line width=0.75]    (10.93,-3.29) .. controls (6.95,-1.4) and (3.31,-0.3) .. (0,0) .. controls (3.31,0.3) and (6.95,1.4) .. (10.93,3.29)   ;

\draw []   (170,150) -- (170,180) ;
\draw [shift={(170,180)}, rotate = 270] [color={rgb, 255:red, 0; green, 0; blue, 0 }  ][line width=0.75]    (10.93,-3.29) .. controls (6.95,-1.4) and (3.31,-0.3) .. (0,0) .. controls (3.31,0.3) and (6.95,1.4) .. (10.93,3.29)   ;

\draw []   (455,150) -- (455,180) ;
\draw [shift={(455,180)}, rotate = 270] [color={rgb, 255:red, 0; green, 0; blue, 0 }  ][line width=0.75]    (10.93,-3.29) .. controls (6.95,-1.4) and (3.31,-0.3) .. (0,0) .. controls (3.31,0.3) and (6.95,1.4) .. (10.93,3.29)   ;

\draw []   (510,245) -- (510,290) ;
\draw [shift={(510,290)}, rotate = 270] [color={rgb, 255:red, 0; green, 0; blue, 0 }  ][line width=0.75]    (10.93,-3.29) .. controls (6.95,-1.4) and (3.31,-0.3) .. (0,0) .. controls (3.31,0.3) and (6.95,1.4) .. (10.93,3.29)   ;

\draw []   (592,320) -- (592,520) ;
\draw []   (585,320) -- (592,320) ;
\draw []   (550,520) -- (592,520) ;

\draw [shift={(550,520)}, rotate = 360] [color={rgb, 255:red, 0; green, 0; blue, 0 }  ][line width=0.75]    (10.93,-3.29) .. controls (6.95,-1.4) and (3.31,-0.3) .. (0,0) .. controls (3.31,0.3) and (6.95,1.4) .. (10.93,3.29)   ;
\draw []   (312,225) -- (312,415) ;
\draw []   (312,415) -- (340,415) ;
\draw []   (312,225) -- (320,225) ;
\draw [shift={(340,415)}, rotate = -180] [color={rgb, 255:red, 0; green, 0; blue, 0 }  ][line width=0.75]    (10.93,-3.29) .. controls (6.95,-1.4) and (3.31,-0.3) .. (0,0) .. controls (3.31,0.3) and (6.95,1.4) .. (10.93,3.29)   ;

\draw []   (460,440) -- (460,480) ;
\draw [shift={(460,480)}, rotate = 270] [color={rgb, 255:red, 0; green, 0; blue, 0 }  ][line width=0.75]    (10.93,-3.29) .. controls (6.95,-1.4) and (3.31,-0.3) .. (0,0) .. controls (3.31,0.3) and (6.95,1.4) .. (10.93,3.29)   ;
\draw []   (380,350) -- (380,390) ;
\draw [shift={(380,390)}, rotate = 270] [color={rgb, 255:red, 0; green, 0; blue, 0 }  ][line width=0.75]    (10.93,-3.29) .. controls (6.95,-1.4) and (3.31,-0.3) .. (0,0) .. controls (3.31,0.3) and (6.95,1.4) .. (10.93,3.29)   ;
\draw []   (30,220) -- (70,220) ;
\draw []   (30,255) -- (70,255) ;
\draw [shift={(70,255)}, rotate = 180] [color={rgb, 255:red, 0; green, 0; blue, 0 }  ][line width=0.75]    (10.93,-3.29) .. controls (6.95,-1.4) and (3.31,-0.3) .. (0,0) .. controls (3.31,0.3) and (6.95,1.4) .. (10.93,3.29)   ;
\draw []   (30,465) -- (70,465) ;
\draw []   (30,220) -- (30,465) ;
\draw [shift={(70,465)}, rotate = 180] [color={rgb, 255:red, 0; green, 0; blue, 0 }  ][line width=0.75]    (10.93,-3.29) .. controls (6.95,-1.4) and (3.31,-0.3) .. (0,0) .. controls (3.31,0.3) and (6.95,1.4) .. (10.93,3.29)   ;


\draw []   (45,265) -- (45,395) ;
\draw []   (45,265) -- (70,265) ;
\draw []   (45,290) -- (70,290) ;
\draw [shift={(70,290)}, rotate = 180] [color={rgb, 255:red, 0; green, 0; blue, 0 }  ][line width=0.75]    (10.93,-3.29) .. controls (6.95,-1.4) and (3.31,-0.3) .. (0,0) .. controls (3.31,0.3) and (6.95,1.4) .. (10.93,3.29)   ;

\draw []   (45,395) -- (70,395) ;
\draw [shift={(70,395)}, rotate = 180] [color={rgb, 255:red, 0; green, 0; blue, 0 }  ][line width=0.75]    (10.93,-3.29) .. controls (6.95,-1.4) and (3.31,-0.3) .. (0,0) .. controls (3.31,0.3) and (6.95,1.4) .. (10.93,3.29)   ;


\draw []   (45,475) -- (70,475) ;
\draw []   (45,500) -- (70,500) ;
\draw [shift={(70,500)}, rotate = 180] [color={rgb, 255:red, 0; green, 0; blue, 0 }  ][line width=0.75]    (10.93,-3.29) .. controls (6.95,-1.4) and (3.31,-0.3) .. (0,0) .. controls (3.31,0.3) and (6.95,1.4) .. (10.93,3.29)   ;

\draw []   (45,475) -- (45,570) ;
\draw []   (45,570) -- (70,570) ;
\draw [shift={(70,570)}, rotate = 180] [color={rgb, 255:red, 0; green, 0; blue, 0 }  ][line width=0.75]    (10.93,-3.29) .. controls (6.95,-1.4) and (3.31,-0.3) .. (0,0) .. controls (3.31,0.3) and (6.95,1.4) .. (10.93,3.29)   ;
\draw []   (55,300) -- (55,365) ;
\draw []   (55,300) -- (70,300) ;
\draw []   (55,330) -- (70,330) ;
\draw [shift={(70,330)}, rotate = 180] [color={rgb, 255:red, 0; green, 0; blue, 0 }  ][line width=0.75]    (10.93,-3.29) .. controls (6.95,-1.4) and (3.31,-0.3) .. (0,0) .. controls (3.31,0.3) and (6.95,1.4) .. (10.93,3.29)   ;
\draw []   (55,365) -- (70,365) ;
\draw [shift={(70,365)}, rotate = 180] [color={rgb, 255:red, 0; green, 0; blue, 0 }  ][line width=0.75]    (10.93,-3.29) .. controls (6.95,-1.4) and (3.31,-0.3) .. (0,0) .. controls (3.31,0.3) and (6.95,1.4) .. (10.93,3.29)   ;
\draw []   (55,405) -- (55,435) ;
\draw []   (55,405) -- (70,405) ;
\draw []   (55,435) -- (70,435) ;
\draw [shift={(70,435)}, rotate = 180] [color={rgb, 255:red, 0; green, 0; blue, 0 }  ][line width=0.75]    (10.93,-3.29) .. controls (6.95,-1.4) and (3.31,-0.3) .. (0,0) .. controls (3.31,0.3) and (6.95,1.4) .. (10.93,3.29)   ;
\draw []   (55,510) -- (55,540) ;
\draw []   (55,510) -- (70,510) ;
\draw []   (55,540) -- (70,540) ;
\draw [shift={(70,540)}, rotate = 180] [color={rgb, 255:red, 0; green, 0; blue, 0 }  ][line width=0.75]    (10.93,-3.29) .. controls (6.95,-1.4) and (3.31,-0.3) .. (0,0) .. controls (3.31,0.3) and (6.95,1.4) .. (10.93,3.29)   ;

\draw []   (55,580) -- (55,610) ;
\draw []   (55,580) -- (70,580) ;
\draw []   (55,610) -- (70,610) ;
\draw [shift={(70,610)}, rotate = 180] [color={rgb, 255:red, 0; green, 0; blue, 0 }  ][line width=0.75]    (10.93,-3.29) .. controls (6.95,-1.4) and (3.31,-0.3) .. (0,0) .. controls (3.31,0.3) and (6.95,1.4) .. (10.93,3.29)   ;

\end{tikzpicture}
\caption{Scheme for Extra Excess Returns Sensitivities and Implied Equilibrium in Imperfect Information Market with Unconstrained Mean-Variance Quadratic Utility Maximization.}
  \label{scheme:1}
\end{scheme}
\newpage
\section{Subjective Views in Imperfect Information Market}\label{sec.3}
Similar to the original papers by Black and Litterman \cite{blaLit90,blaLit91,blaLit92}, this Section explains how investors' subjective views are incorporated into markets with incomplete information as described in Section \ref{sec2}. First, we derive the reference model for excess returns conditioning on the shadow-costs. Then, we rigorously outline the market structure, providing relevant assumptions and definitions. Next, we adjust the BL formulation to incorporate investors' beliefs while accounting for information imperfection.
\subsection{Incomplete Information Market Structure}
\subsubsection{Market Reference Model}
We consider capital markets where the random vector of all risky securities' excess returns is defined by the risk premium vector $\tilde{R}:=[\tilde{R}_1,\dots,\tilde{R}_n]^\top-r_f\mathbf{1}_{\mathbb{R}^n}$, has a multivariate normal distribution denoted as
\begin{equation}\label{eq_26}
    \tilde{R}\sim\mathcal{N}(\tilde{\mu};\Sigma),\;\text{for}\;\tilde{\mu}\sim\mathcal{N}(\pi;\Sigma_{\tilde{\mu}}:=\tau^\prime\Sigma),
\end{equation}
where $\tilde{\mu}$ is also a random normal vector that determines the behavior of the mean of $\tilde{R}$, for $\tau^\prime$ a non-negative real number as in BL model \cite{blaLit90,blaLit91,blaLit92}. For a value of $\tau^\prime$ equals zero, the risk-premium mean value become deterministic and given by $\pi$, the implied equilibrium of equation (\ref{eq3}) related to the incomplete information market. Thus, the reference model of the considered market where the information is incomplete can be such that
\begin{equation}\label{eq30}
    \tilde{R}\sim\mathcal{N}(\pi;\tau\Sigma),\;\text{for}\;\tau:=1+\tau^\prime.
\end{equation}
Since $\mathbb{E}\bigl[\tilde{R}|\tilde{\mu}\bigr]=\tilde{\mu}$ and $\text{Var}[\tilde{R}|\tilde{\mu}]=\Sigma$, the resulted distribution in (\ref{eq30}) is due to the Law of Total Expectation
$$\mathbb{E}[\tilde{R}]=\mathbb{E}\Bigl[\mathbb{E}\bigl[\tilde{R}|\tilde{\mu}\bigr]\Bigr]=\mathbb{E}[\tilde{\mu}]=\pi,$$
and to the Law of Total Variance
$$\text{Var}[\tilde{R}]=\text{Var}\bigl[\mathbb{E}[\tilde{R}|\tilde{\mu}]\bigr]+\mathbb{E}\bigl[\text{Var}[\tilde{R}|\tilde{\mu}]\bigr]=\Sigma_{\tilde{\mu}}+\Sigma=\tau\Sigma.$$
However, in our formulation, the random vector of shadow-costs $\tilde{\lambda}$ defined in Notation \ref{Notation1} is distributed with deterministic mean vector $\lambda$ and covariance matrix $\Lambda$, generally $\tilde{\lambda}\sim\mathcal{N}(\lambda;\Lambda)$. The resulted conditional distribution $\tilde{R}^{\tilde{\lambda}}\sim\mathcal{P}\bigl(\tilde{R}|\tilde{\lambda}\bigr)$, $\mathcal{P}$ denotes the probability distribution, has a random expectation. Such an expectation characterizes well the market with shadow-costs $\tilde{\lambda}$, has the mean $\pi$, but a variance that is different from $\tau\Sigma$. In fact, from the Law of Total Expectation, we can write
\begin{equation}\label{e}
\mathbb{E}\Bigl[\mathbb{E}\bigl[\tilde{R}\bigl|\tilde{\lambda}\bigr]\Bigr]=\mathbb{E}[\tilde{R}]=\pi,
\end{equation}
which also holds true since, given the random vector of information $\tilde{\lambda}$ and the equilibrium of equation (\ref{eq3}), the random vector $\mathbb{E}\bigl[\tilde{R}\bigl|\tilde{\lambda}\bigr]$ gives the equilibrium random vector $\tilde{R}$ as a linear expression of the random vector $\tilde{\lambda}$ of the imperfect information market whose expectation is $\pi$. Additionally, from the Law of Total Variance, we also write
\begin{equation}\label{v}
\text{Var}\Bigl[\mathbb{E}\bigl[\tilde{R}\bigl|\tilde{\lambda}\bigl]\Bigr]=\text{Var}(\tilde{R})-\mathbb{E}\Bigl[\text{Var}\bigl[\tilde{R}\bigl|\tilde{\lambda}\bigl]\Bigr],\;\text{for}\;\text{Var}(\tilde{R})=\tau\Sigma,
\end{equation}
this term reflects the explained variation on excess returns vector $\tilde{R}$ after accounting for the realizations of the shadow-costs vector $\tilde{\lambda}$, while the second term on the right-hand side of equation (\ref{v}) captures the unexplained part of total variance of $\tilde{R}$ that is due to the variability of $\tilde{R}$ given different levels of $\tilde{\lambda}$. Accounting now for the cross-covariance matrix $\Sigma_{\tilde{R},\tilde{\lambda}}$, between $\tilde{R}$ and $\tilde{\lambda}$, and letting these two variables be joint normal dependent distributions, that is we have
\begin{equation*}
\begin{pmatrix}
\tilde{R} \\
\tilde{\lambda}
\end{pmatrix}
\sim\mathcal{N}
\Biggl(
\begin{pmatrix}
\pi \\
\lambda
\end{pmatrix};
\begin{pmatrix}
\tau\Sigma & \Sigma_{\tilde{R},\tilde{\lambda}} \\
\Sigma_{\tilde{\lambda},\tilde{R}} & \Lambda
\end{pmatrix}
\Biggr),\;\text{where}\;\Sigma_{\tilde{\lambda},\tilde{R}}={\Sigma_{\tilde{R},\tilde{\lambda}}}^\top.
\end{equation*}
Since the incompleteness of information affects the implied equilibrium, it is obvious to assume that
\begin{assumption}
    We assume that the random vector of shadow-costs $\tilde{\lambda}$ is dependent of $\tilde{R}$, the random vector of excess returns in the incomplete information market. \qed
\end{assumption}
This assumption yields a non-null cross-covariance matrix, and the expression
\begin{equation*}
\mathbb{E}\Bigl[\text{Var}\bigl[\tilde{R}\bigl|\tilde{\lambda}\bigl]\Bigr]=\tau\Sigma-\Sigma_{\tilde{R},\tilde{\lambda}}\Lambda^{-1}\Sigma_{\tilde{\lambda},\tilde{R}},
\end{equation*}
which represents the average remaining uncertainty of excess returns after accounting for the information coming from the shadow-costs. Hence, the asset pricing model we shall derive next takes the incomplete information market structure such that the implied equilibrium of excess returns is given by the random conditional expectation of excess returns $\tilde{R}$ given the observed shadow-costs $\tilde{\lambda}$. Based on the above equations (\ref{e}) and (\ref{v}), this equilibrium has the properties
$$\mathbb{E}\Bigl[\mathbb{E}\bigl[\tilde{R}\bigl|\tilde{\lambda}\bigr]\Bigr]=\pi,\;\text{Var}\Bigl[\mathbb{E}\bigl[\tilde{R}\bigl|\tilde{\lambda}\bigl]\Bigr]=\Sigma_{\tilde{R},\tilde{\lambda}}\Lambda^{-1}\Sigma_{\tilde{\lambda},\tilde{R}}.$$
The resulted incomplete information market equilibrium allows investors to take $\pi$ as the neutral starting point, which will tilt toward their views and drive $\pi$ to a new posterior equilibrium expected value with a level of uncertainty given by the variability of this equilibrium.
\begin{proposition}[Market Reference Model]\label{prop1}
From Merton's equilibrium model of incomplete information, the reference prior distribution of implied equilibrium excess returns for imperfect information markets is multivariate normal given by the random conditional expectation $\mathbb{E}\bigl[\tilde{R}\bigl|\tilde{\lambda}\bigr]$ as
\begin{equation}\label{Rlambda}
\tilde{R}^{\tilde{\lambda}}\sim\mathcal{N}(\pi;\Sigma_{\tilde{R},\tilde{\lambda}}\Lambda^{-1}\Sigma_{\tilde{\lambda},\tilde{R}}),
\end{equation}
for $\Sigma_{\tilde{R},\tilde{\lambda}}$ being the cross-covariance matrix between $\tilde{R}$ and $\tilde{\lambda}$. \qed
\end{proposition}
We might use this model of Proposition \ref{prop1} to derive our asset pricing model, or, alternatively, consider the approach where the random mean vector and covariance matrix of $\tilde{R}$ given $\tilde{\lambda}$ are directly taken into account without limiting on the mean and variance of the conditional expectation.
\begin{remark}[Alternative Market Reference Model]\label{remk2}
Alternatively, one might model the market through the random conditional expectation $$\mathbb{E}\bigl[\tilde{R}\bigl|\tilde{\lambda}\bigr]=\pi+\Sigma_{\tilde{R},\tilde{\lambda}}\Lambda^{-1}\bigl(\tilde{\lambda}-\lambda\bigr),$$
which adjusts the implied excess returns $\pi$ by incorporating the information about the observed value of shadow-costs, and the deterministic conditional variance $$\text{Var}\bigl[\tilde{R}\bigl|\tilde{\lambda}\bigr]=\tau\Sigma-\Sigma_{\tilde{R},\tilde{\lambda}}\Lambda^{-1}\Sigma_{\tilde{\lambda},\tilde{R}}.$$
Therefore, letting the reference model be such that
$$\tilde{R}^{\tilde{\lambda}}\sim\mathcal{N}\Bigl(\mathbb{E}\bigl[\tilde{R}\bigl|\tilde{\lambda}\bigr];\text{Var}\bigl[\tilde{R}\bigl|\tilde{\lambda}\bigr]\Bigr).$$
More in precise, by recalling once again the Law of Total Expectation and Law of Total Variance, we get back the initial model of equation (\ref{eq30}) as $\tilde{R}^{\tilde{\lambda}}\sim\mathcal{N}(\pi;\tau\Sigma)$.
\par The two reference models can be merged into a single distribution given by
\begin{equation}\label{eq32}
\tilde{R}^{\tilde{\lambda}}\sim\mathcal{N}\Bigl(\pi;\Sigma_\gamma:=\gamma\tau\Sigma+(1-\gamma)\Sigma_{\tilde{R},\tilde{\lambda}}\Lambda^{-1}\Sigma_{\tilde{\lambda},\tilde{R}}\Bigr),\;\text{for}\;\gamma\in\{0,1\},
\end{equation}
where $\gamma=0$ corresponds to the model based on Conditional Expectation, while $\gamma=1$ is related to the alternative model of $\tilde{\lambda}$-Conditioned Excess Returns vector. \qed
\end{remark}
\begin{proposition}[Optimal Portfolio]
    Given the market reference model with random shadow-costs, an investor who optimizes the unconstrained mean-variance quadratic utility function has to choose the optimal allocation
    $$w_M^\gamma=(\delta\Sigma_\gamma)^{-1}\pi.$$ \qed
\end{proposition}
\begin{remark}[Shadow-Costs with Random Expectation]\label{rk2}
    Similarly to (\ref{eq30}), one might consider that the mean of $\tilde\lambda$ is not deterministic and is multivariate normal. Thus, assuming $$\tilde\lambda\sim\mathcal{N}(\tilde{\lambda}_1;\Lambda),\;\text{for}\;\tilde{\lambda}_1\sim\mathcal{N}(\lambda_1;\tau^{\prime}_1\Lambda)\;\text{and}\;\tau^{\prime}_1>0,$$
    we get $\tilde{\lambda}\sim\mathcal{N}(\lambda_1;\tau_1\Lambda),\;\text{for}\;\tau_1=1+\tau_1^\prime$. The market reference model in (\ref{eq32}) generalized to $\tilde{R}^{\tilde\lambda}\sim\mathcal{N}(\hat{\pi};\hat{\Sigma}_\gamma)$,
    where $\hat{\pi}$ is given by means of equation (\ref{eq3}) as
    \begin{equation*}
    \hat{\pi}=\biggl(\delta-\frac{\lambda_M}{\sigma_M^2}\biggr)\Sigma w_M+\lambda_1,\;\text{where}\;\lambda_M={w_M}^\top\lambda_1,
    \end{equation*}
    and $\hat{\Sigma}_\gamma$ is computed by replacing $\Lambda$ by $\tau_1\Lambda$ in the expression of $\Sigma_\gamma$. \qed
\end{remark}
\subsubsection{Market Structure and Assumptions}
Once the market reference model has been specified, we next give assumptions on the market structure and investor's behavior. We let the market be such that the following assumption holds:
\begin{assumption}[Reference Market Structure]\label{assumption1}
    We assume that the considered capital market follows the reference model outlined in the previous Section, and that
    \begin{enumerate}
        \item The analysis is on a single period of investment;
        \item The market is not necessarily efficient due to the adopted assumption of incomplete information among investors. Therefore, the information is imperfect in the market as well as the competition;
        \item There are no market frictions regarding transaction-costs, short sale, regulatory constraints or other market imperfections;
        \item The risk is measured by the variability of returns using the standard deviation metric;
        \item The statistical distribution of the excess returns random vector that accounts for the imperfection of information and subjective views is used to characterize the market equilibrium;
        \item The investor behavior in this market is such that Assumption \ref{assumption2}, given hereafter, holds true. \qed
    \end{enumerate}
\end{assumption}
Since we are dealing with realistic modeling of capital markets where information is incomplete and investor holds anticipations, we need to adopt the following assumption on the investor's behavior.
\begin{assumption}[Investor Behavior]\label{assumption2}
    In this study, we assume that the investor considers the market with the structure of Assumption \ref{assumption1}, and that
    \begin{enumerate}
        \item The investor is rational, risk-averse and aims to maximize the quadratic utility function related to the risk-adjusted return optimization problem under suitable constraints;
        \item Each investor $j$ does not know about all $n$ risky securities, he/she is only interested in investing in $n_j<n$ risky securities about which the information is acquired. That is, the shadow-cost $\lambda_k^j=0$ for any $k\in J_j=\{1,\dots,n_j\}$, and the allocation is null for assets in $J_j^c=\{n_j+1,\dots,n\}$;
        \item Each investor $j$ can express anticipations about relative and absolute future performances, at the end of the investment period, of the securities on his/her information set $J_j$. \qed
    \end{enumerate}
\end{assumption}
Finally, we define the two main characteristics of the considered capital markets by quantifying the notions of information cost and equilibrium.
\begin{definition}[Information Cost]
In information theory, information cost refers to any type of cost deployed to acquire, transform, and make any type of information useful to make decisions. It can refer to financial costs as well as subjective costs. In our context of portfolio optimization, the information cost is related to the cost an investor has to pay to be aware of a security in the market, and to allocate it in his/her portfolio. Based on Merton's model, the information cost is given by the information-premium $\lambda-\lambda_M\beta$, that is, an investor $j_1$ who is aware about security $k$ has the cost of information given by the extra excess return $\lambda_k-\lambda_M\beta_k$ for this security $k$, where $\lambda_k:=(1/N)\sum_{j=1}^{N}\lambda^{j}_k$ with $\lambda_k^{j_1}=0$. \qed
\end{definition}
\begin{definition}[Market Equilibrium]
    We define both the equilibrium in markets with incomplete information and the equilibrium when subjective views are held in these markets, as
    \begin{enumerate}
        \item In the incomplete information market, we characterize the equilibrium of the securities through the conditional distribution $\tilde R^{\tilde\lambda}\sim\mathcal{P}(\tilde{R}|\tilde{\lambda})$ of the random excess returns vector $\tilde{R}$ given the shadow-costs vector $\tilde{\lambda}$, or, alternatively, by the distribution of the random conditional expectation of excess returns $\mathbb{E}\bigl[\tilde{R}|\tilde{\lambda}\bigr]$. Then, the optimal allocation maximizes the investor's utility function according to one of these distributions in the imperfect information market;
        \item Similarly, in the incomplete information market with subjective views, we derive the equilibrium of the securities through the posterior conditional distribution $\tilde{R}^{\tilde{\lambda}}_{\tilde{\nu}}\sim\mathcal{P}(\tilde{R}|\tilde{\lambda};\tilde{\nu})$ of excess returns vector $\tilde{R}$, given the distributions of the vectors of shadow-costs $\tilde{\lambda}$ and subjective views $\tilde{\nu}$.
    \end{enumerate}
    Therefore, for the two-period investment model in imperfect information markets with views, we define the general market equilibrium through the distribution of the vector of risky securities' risk premiums at the end of the period given by the mean vector $\pi^*=\mathbb{E}\bigl[\tilde{R}_{\tilde{\nu}}^{\tilde{\lambda}}\bigr]$, and the corresponding variance matrix $\Sigma^*=\text{Var}\bigl[\tilde{R}_{\tilde{\nu}}^{\tilde{\lambda}}\bigr]$. \qed
\end{definition}
\subsection{Subjective Views Incorporation}
Experts in capital markets usually build a strong track-record experience and are often asked by quantitative modelers for their opinion on the outcome of the markets. Any belief on the market performance can be used as a perturbation of an initial model predictions, and can have a degree of certainty given by a probability density function. From the introduction Section \ref{Sec1}, we recall that the optimal allocation expression is characterized by
\begin{equation}\label{eq33}
\alpha(\theta)=\argmax_{\alpha\in\mathcal{C}_\theta}\bigl\{\mathcal{S}_\theta(\alpha)\bigr\},
\end{equation}
which obviously is sensitive to the market parameters $\theta$, where its corresponding Bayesian allocation form is
\begin{equation}\label{eq11_}
\alpha_B[i_T,e_C]=\argmax_{\alpha\in\mathcal{C}}\int_{\mathcal{M}}U(\alpha^\top m)f_p(m;i_T,e_C)dm,
\end{equation}
for $m$ the market vector of expected excess returns, whose probability support is $\mathcal{M}$. A slightly change in the parameters $\theta$ gives rise to a significantly different allocation. The sample-based allocation modeling gives best statistical estimators for the parameters based on historical data while minimizing estimation risk. Alternatively, in the Bayesian allocation approach of equation (\ref{eq11_}) and BL model, the distribution of the market parameters is shrunk towards the investor's prior in order to limit the sensitivity of the optimal allocation to $\theta$. For our case, we let the probability support for the market parameters be given by the joint-distribution of the shadow-costs of information and the subjective beliefs of investor defined hereafter, i.e., $\Theta:=\bigl[\tilde\lambda,\tilde{\nu}\bigr]^\top$. Therefore, for any investor $j$ who is informed about, and are interested in investing in, the subset $J_j=\bigl\{1,\dots,n_j\bigr\}$, $n_j<n$, of assets, our aim is to find an explicit density function $f_p$ that accounts for the imperfection of information in the market and incorporates his/her beliefs about some of, or all, the assets $1,\dots,n_j$. Hence, explicitly describing the corresponding distribution $f_p$ for the market vector $m$ by representing the investor's information $i_T$ and experience $e_C$ by the distributions $\tilde\lambda$ and $\tilde{\nu}$, respectively.
\par In the present study, similar to BL model, a view on the market can be absolute, related to a single security, or relative, associated with two or more securities. Each view form a portfolio and the combination of all views is related to a Pick-Matrix $P$ and a Pick-Vector of subjective beliefs $q$. We describe how investor's views are defined to form the conditional distribution $\tilde{\nu}$ of beliefs for the Bayesian inference approach we shall explore to derive our equilibrium model for the incomplete information market with subjective views.
\begin{definition}[View-Portfolio, Pick-Matrix and Pick-Vector]\label{def3}
A view portfolio $P_l$, on $n$ risky securities, is a row of $n$ elements with few non-null values ranging between $-1$ and $1$ and many zeros. Each view portfolio can express a relative or an absolute anticipation on the securities' future performances. The vision related to this portfolio is filled in the $l$-th row of a pick-matrix $P\in\mathcal M_{v\times n}(\mathbb{R})$, and is quantified by $q_l$, the real valued $l$-th element of a pick-vector $q:=[q_1,\dots,q_v]^\top\in\mathbb R^v$. A characterization of investor's subjective beliefs is given by the equation $P\mathbb{E}\bigl[\tilde{R}^{\tilde{\lambda}}\bigr]=q$, for $\tilde{R}^{\tilde{\lambda}}:=\bigl[\tilde{R}_1^{\tilde{\lambda}}-r_f,\dots,\tilde{R}_n^{\tilde{\lambda}}-r_f\bigr]^\top$ the $\tilde{\lambda}$-conditioned excess returns random vector related to the incomplete information market. This equation formulates $v$ view portfolios filled in the rows of $P$, and quantified by the elements of $q$. Additionally, the sum of weights in a view portfolio is zero for relative views and one for absolute views. \qed
    \end{definition}
    \begin{example}
Let us consider a view-portfolio of three subjective views, $l=1,\dots,v=3$ for an investor $j$ with information set $J_j$, that is defined by the following pick-matrix $P$ and pick-vector $q$:
\[
P=\left(\underset{n\;Securities}{\underbrace{
{\begin{array}{cccccccccc|c}
0 & 1 & 0 & 0 & 0 & 0 & \cdots & 0 & 0 & 0 & \twogroup \\
0 & -1 & 0 & 1 & 0 & 0 & \cdots & 0 & 0 & 0 & \twogroup \\
0 & 0.5 & -0.2 & -0.8 & 0.4 & 0 & \cdots & 0 & 0.1 & 0 & \oneGroup\\
\end{array} } }}\right),\;q=\left(
{\begin{array}{c}
5 \%\\
3 \%\\
4 \%
\end{array} } \right).
\]
Here, due to the incompleteness of information in the market, the right block of matrix $P$ is filled with zeros since we assumed that investor $j$ is only interested in investing in assets $1,\dots,n_j$ within his/her set of information $J_j$, while assets $n_{j}+1,\dots,n$ are ignored and not allocated in his/her portfolio. Therefore, no views are related to assets $n_{j}+1,\dots,n$. Each investor has his/her own order of assets in the matrix $P$, in the sense that assets that are ignored by investor $j$, i.e., $n_{j}+1,\dots,n$, are not necessarily ignored by another investor $j^\prime$ and belong to his/her left block of the view-matrix.
\par This example deals with the following three subjective beliefs:
\begin{itemize}
    \item[\textbf{(i)}] \textbf{Absolute View $1$}: Asset $2$ has an absolute excess return of $q_1=5\%$;
    \item[\textbf{(ii)}] \textbf{Relative View $2$}: Asset $4$ will outperform Asset $2$ by $q_2=3\%$;
    \item[\textbf{(iii)}] \textbf{Multi-Assets Relative View $3$}: Assets $2$, $5$ and $n_j-1$ will outperform Assets $3$ and $4$ by $q_3=4\%$. Here, the weights add up to $1$ for the outperforming assets, and to $-1$ for the under-performing assets, where the weightings are not necessarily equal. \qed
\end{itemize}
\end{example}
    \begin{notation}[Random Pick-Vector]
    Each investor is assumed to hold $v$ subjective views $l=1,\dots,v$, where each view $l$ is quantified by a random pick-variable $\tilde{q}_l$. The random pick-vector $\tilde{\nu}:=\bigl[\tilde{q}_1,\dots,\tilde{q}_v\bigr]^\top$, whose expectation is the pick-vector $\mathbb{E}[\tilde{\nu}]=q$, denotes the collection of these beliefs and has a variance matrix $\text{Var}(\tilde{\nu})=\Omega$. This matrix measures the certainty of the views, and is diagonal when all views are independent. \qed
\end{notation}
This distribution of views is related to the securities in the market of imperfect information, and can be seen as $\tilde{\nu}:=P\tilde{R}^{\tilde{\lambda}}$. More in precise, it is crucial to mention that a more general expression of the above equation is given as $P\bigl(\pi+\varepsilon^\pi\bigr)\sim q+\varepsilon^q$, such that the random noise vector
\begin{equation}\label{noises}
\begin{pmatrix}
\varepsilon^\pi \\
\varepsilon^q
\end{pmatrix}
\sim\mathcal{N}\Biggl(\textbf{0}_{\mathbb{R}^{n+v}};
\begin{pmatrix}
\Sigma_\gamma & \textbf{0}_{\mathbb{R}^{n\times v}} \\
\textbf{0}_{\mathbb{R}^{v\times n}} & \Omega
\end{pmatrix}\Biggr),
\end{equation}
is related to $\varepsilon^\pi$ and $\varepsilon^q$, two normally distributed un-observable independent random noises with zero mean each. The views are built from an expert's opinion as a prediction of the expectation $q$ of the random vector $\tilde{\nu}$, which can be viewed as a perturbation of the initial market expected excess returns $\pi$ of the random vector $\tilde{R}^{\tilde{\lambda}}$ coming from all view portfolios in the pick-matrix $P$, or, more generally, as statements on the reference model (\ref{eq32}). We assume that an investor my have zero or many views, and that it is possible for the views to conflict since they can merge due to the mixing process of confidence in the prior implied equilibrium and views. Moreover, the view's confidence matrix $\Omega$ describes the uncertainty of the expert and might have the expression
$$\Omega=\Bigl(\frac{1}{c}-1\Bigr)P\Sigma P^\top,$$
Attilio \cite[Chapter $9$]{Attilio05}. The scalar $c$ helps the expert to adjust his/her absolute confidence in the views:
\begin{itemize}
    \item\textbf{Case when $\boldsymbol{c\rightarrow 0}$}: Expert's views have no impact, and this case gives rise to an infinitely disperse distribution $\tilde{\nu}$;
    \item\textbf{Case when $\boldsymbol{c\rightarrow 1}$}: An infinitely peaked distribution $\tilde{\nu}$ resulted from this case. That is, the expert is trusted completely over the prior market model distribution $\tilde{R}^{\tilde{\lambda}}$;
    \item\textbf{Case $\boldsymbol{c=1/2}$}: The investor is as trusted as the prior market model.
    \end{itemize}
    \par In the BL model \cite{blaLit90,blaLit91,blaLit92}, the excess returns vector $\tilde{R}$ in the complete information market is given by the market reference model (\ref{equ5}) and yields the equilibrium of the equation (\ref{equ4}). When the mean of the reference model is deterministic, i.e., $\tilde{\mu}=\pi$ and $\tau^\prime=0$, the posterior equilibrium is the same of equation (\ref{equ4}) for $\tau=1$. The reference model of $\tilde{R}^{\tilde{\lambda}}$ and the random vector $\tilde{\nu}$ are then crucial in describing the distribution of excess returns vector at equilibrium. Meaning that, a non-informative view-vector with $\Omega\rightarrow\infty$ returns the prior equilibrium $\tilde{R}\sim\mathcal{N}(\pi^c;\tau\Sigma)$ for BL or $\tilde{R}^{\tilde{\lambda}}\sim\mathcal{N}(\pi;\Sigma_\gamma)$ for our model, while a view-vector with confidence level $\Omega$ gives the posterior distribution (\ref{equ4}) or the one we shall derive in the Section below.
    \par At this point the BL quantitative model integrates the subjective views to compute a posterior distribution of the complete information market conditioned on the expert's anticipations. Such anticipations are useful for scenario analysis of investment decision, and, based on Attilio \cite[Chapter $9$]{Attilio05}, can be given, in a default setting when only qualitative subjective views are hold by the expert in the complete information market, as
    \begin{equation}\label{equ38}
        q_k=(P\pi^c)_k+\eta_k\sqrt{(P\Sigma P^\top)_{k,k}},\;k=1,\dots,v,
    \end{equation}
for any view-portfolio $k$, where $\eta_k\in\{-\beta,-\alpha,+\alpha,+\beta\}$ for "very bearish", "bearish", "bullish" and "very bullish" views, respectively, with typical choice $\alpha=1$ and $\beta=2$. For our modeling, we have to replace $\pi^c$ by $\pi$ in the above equation to quantify the qualitative views.
\par This approach of incorporating subjective views in the considered markets of Section \ref{sec2} with incomplete information, drives optimal allocations of the form (\ref{eq33}) to be expressed as the Bayesian optimal allocation
\begin{align*}
    \alpha_B[i_T,e_C]&=\argmax_{\alpha\in C_{\{\tilde{\lambda},\tilde{\nu}\}}}\Bigl\{S_{\{\tilde{\lambda},\tilde{\nu}\}}(\alpha)\Bigr\}\\
    &=\argmax_{\alpha\in C_{\{\tilde{\lambda},\tilde{\nu}\}}}\Bigl\{\mathbb{E}\Bigl[U\bigl(\Psi_\alpha^{\{\tilde{\lambda},\tilde{\nu}\}}\bigr)\Bigr]\Bigr\},
\end{align*}
for which the parameter $\theta$ encompassing market information parameter $i_T$ and investor's experience $e_C$ is encapsulated into the random vector $\{\tilde{\lambda},\tilde{\nu}\}$ of shadow-costs and subjective views. For further developments, we now need to consider that the distribution $\tilde{\nu}$ satisfies the following:
\begin{assumption}\label{assumption4}
    Assume that the random vector $\tilde{\nu}$ of subjective views is independent of the random vector $\tilde{\lambda}$ of shadow-costs, given the excess returns vector $\tilde{R}$ in the complete information market. That is, we assume
    $$\mathcal{P}\bigl(\tilde{\lambda},\tilde{\nu}\bigl|\tilde{R}\bigr)=\mathcal{P}\bigl(\tilde{\lambda}\bigl|\tilde{R}\bigr)\mathcal{P}\bigl(\tilde{\nu}\bigl|\tilde{R}\bigr).$$ \qed
\end{assumption}
This assumption is not restrictive, in fact, the subjective beliefs do not come from the shadow-costs, especially when the implied equilibrium is observed. The Section to follow deals with the second and main aim of our study.
\begin{aim}
    Incorporation of expert's subjective views, measured by the random view-vector $\tilde{\nu}$, in markets with incomplete information characterized by the vector $\tilde{\lambda}$ of shadow-costs and the random implied equilibrium vector $\tilde{R}^{\tilde{\lambda}}$. Thus, providing an explicit posterior distribution $f_p$ in the Bayesian allocation framework (\ref{eq11_}) for the random vector of excess returns $\tilde{R}_{\tilde{\nu}}^{\tilde{\lambda}}\sim\mathcal{P}\bigl(\tilde{R}|\tilde{\lambda};\tilde{\nu}\bigr)$. \qed
\end{aim}
\section{Equilibrium Model and Optimal Portfolio Derivation}\label{sec.4}
In this Section, we derive the closed-form formula for the posterior distribution $\tilde{R}^{\tilde\lambda}_{\tilde\nu}$ of the random vector of equilibrium excess returns in markets with incomplete information, which incorporates investor's subjective views. The Bayesian approach of statistical inference is used to infer the posterior distribution based on the given implied imperfect information market equilibrium prior distribution $\tilde{R}^{\tilde{\lambda}}$, and the likelihood function that models the distribution $\tilde{\nu}$ of the investor's beliefs in these markets given the equilibrium. We first give a scheme for the Bayesian approach.
\subsection{Bayesian Approach Scheme}
Bayesian approach of statistical inference is a powerful statistical tool that helps to make inferences by expressing the posterior distribution $\mathcal{P}(\theta|x)$ of a random parameter $\theta$, given data observations $x$. This approach consists in exploring the Bayes' rule to update the distribution of beliefs $\mathcal{P}(\theta)$ about the parameter in response to the distribution of evidence $\mathcal{P}(x|\theta)$ and write the posterior distribution of the parameter. Using the definition of the marginal distribution $\mathcal{P}(x)$ in the denominator, the posterior writes as
\begin{equation*}
    \mathcal{P}(\theta|x)=\frac{\mathcal{P}(x|\theta)\mathcal{P}(\theta)}{\int_{\Theta}\mathcal{P}(x|\theta)\mathcal{P}(\theta)d\theta},
\end{equation*}
for $\Theta$ being the probability support of $\theta$, thus, one only needs to fix the two distributions $\mathcal{P}(x|\theta)$ and $\mathcal{P}(\theta)$. As in the BL model \cite{blaLit90,blaLit91,blaLit92}, the Bayesian approach framework is suitable for capturing subjective views of investors in imperfect information markets, leading to a fair valuation of the securities in these markets where expert's anticipations are incorporated within the valuation. Based on the results concerning the imperfect information market, and the formulation of the views, we adopt assumptions on the prior implied market equilibrium, the shadow-costs of information and the subjective views in incomplete information markets we introduced in Sections \ref{sec2} and \ref{sec.3}.
\begin{assumption}[Bayesian Approach Setting]\label{Assump05}
Let us assume that
\begin{itemize}
\item \textbf{Shadow-Costs of Information $\boldsymbol{{\tilde{\lambda}}}$}: The distribution of the random vector of imperfect information market's shadow-costs is such that $\mathcal{P}\bigl({\tilde\lambda}\bigr)\sim\mathcal{N}(\lambda;\Lambda)$, where $\lambda$ is the deterministic mean vector of the shadow-costs related to Merton's model, and $\Lambda$ represents the diagonal covariance matrix of $\tilde{\lambda}$;
\item \textbf{Implied Market Equilibrium $\boldsymbol{\tilde{R}^{\tilde{\lambda}}}$ and $\boldsymbol{\tilde{R}}$}: The distribution of the implied equilibrium in the market with incomplete information is given by $\mathcal{P}\bigl(\tilde R\bigl|\tilde{\lambda}\bigr)\sim\mathcal{N}(\pi;\Sigma_\gamma)$, where $\gamma=0$ for the Conditional Expectation Reference Model, and $\gamma=1$ for the $\tilde{\lambda}$-Conditioned Excess Returns Reference Model. In the complete information market it is given by $\mathcal{P}\bigl(\tilde R\bigr)\sim\mathcal{N}(\pi^c;\tau\Sigma)$, for $\tau>0$;
\item \textbf{Subjective Beliefs of Investor $\boldsymbol{\tilde{\nu}}$}:
The distribution of the random vector $\tilde{\nu}$ of views is such that $\mathcal{P}\bigl(\tilde{\nu}\bigr)\sim\mathcal{N}(q;\Omega)$, for $q$ the pick-vector and $\Omega$ the covariance matrix of $\tilde{\nu}$. \qed
\end{itemize}
\end{assumption}
Later we explore Bayes' rule to derive the distribution $\mathcal{P}\bigl(\tilde{R}\bigl|\tilde{\lambda},\tilde{\nu}\bigr)$ of the implied equilibrium excess returns vector $\tilde{R}^{\tilde\lambda}_{\tilde\nu}$. First, we give Scheme \ref{scheme:1} to illustrate the equilibrium model, where the random vectors $\varepsilon^\pi$ and $\varepsilon^q$ are the un-observable noises (\ref{noises}).

\begin{scheme}
\centering
\begin{tikzpicture}[x=0.75pt,y=0.75pt,yscale=-1,xscale=1]
\draw  [color={rgb, 255:red, 0; green, 0; blue, 0 }  ,draw opacity=1, dotted ] (0,0) -- (190,0) -- (190,165) -- (0,165) -- cycle ;
\draw (5,4) node [anchor=north west][inner sep=0.75pt]  [font=\small] [align=left] {Complete Market Equilibrium};
\draw (20,24) node [anchor=north west][inner sep=0.75pt]  [font=\small] [align=left] {\textbf{Sharpe-Lintner-Mossin}};
\draw[fill=gray!10]   (15,115) -- (180,115) -- (180,150) -- (15,150) -- cycle ;
\draw (60,124) node [anchor=north west][inner sep=0.75pt]  [font=\small] [align=left] {$\boldsymbol{\pi^c=\delta\Sigma w_M^c}$};
\draw (16,50) node [anchor=north west][inner sep=0.75pt]  [font=\small] [align=left] {$\delta:=\frac{\mathbb{E}[R_m]-r_f}{\sigma_M^2}$};
\draw (112,60) node [anchor=north west][inner sep=0.75pt]  [font=\small] [align=left] {$w_M^c$};
\draw (158,60) node [anchor=north west][inner sep=0.75pt]  [font=\small] [align=left] {$\Sigma$};
\draw [dotted]   (195,0) -- (400,0) -- (400,165) -- (195,165) -- cycle ;
\draw[fill=gray!10] (210.33,115) -- (382,115) -- (382,150.5) -- (210,150.5) -- cycle ;
\draw (215,122) node [anchor=north west][inner sep=0.75pt]  [font=\small] [align=left] {$\boldsymbol{\pi=\bigl(\delta-\delta_\lambda\bigr)\Sigma w_M+\lambda}$};
\draw (200,4) node [anchor=north west][inner sep=0.75pt]  [font=\small] [align=left] {Incomplete Market Equilibrium};
48\draw (245,24) node [anchor=north west][inner sep=0.75pt]  [font=\small] [align=left] {\textbf{Merton Model}};
\draw (198,60) node [anchor=north west][inner sep=0.75pt]  [font=\small] [align=left] {$\lambda_M$};

\draw (248,58) node [anchor=north west][inner sep=0.75pt]  [font=\small] [align=left] {$\delta_\lambda:=\frac{\lambda_M}{\sigma_M^2}$};

\draw (425,62) node [anchor=north west][inner sep=0.75pt]  [font=\small] [align=left] {$\tau$};
\draw (318,60) node [anchor=north west][inner sep=0.75pt]  [font=\small] [align=left] {$\Lambda:=\text{Var}(\lambda)$};
\draw [dotted]  (405,0) -- (590,0) -- (590,165) -- (405,165) -- cycle ;
\draw (420,4) node [anchor=north west][inner sep=0.75pt]  [font=\small] [align=left] {{Investors' subjective views}};
\draw (428,24) node [anchor=north west][inner sep=0.75pt]  [font=\small] [align=left] {\textbf{Black-Literman Model}};
\draw (450,60) node [anchor=north west][inner sep=0.75pt]  [font=\small] [align=left] {$P$};
\draw (480,62) node [anchor=north west][inner sep=0.75pt]  [font=\small] [align=left] {$q$};
\draw (500,60) node [anchor=north west][inner sep=0.75pt]  [font=\small] [align=left] {$\text{Var}(q):=\Omega$};
\draw[fill=gray!10]   (420,115) -- (580,115) -- (580,150) -- (420,150) -- cycle ;
\draw (430,124) node [anchor=north west][inner sep=0.75pt]  [font=\small] [align=left] {$\boldsymbol{P(\pi+\varepsilon^\pi)=q+\varepsilon^q}$};
\draw[dotted, fill=green!10]   (35,200) -- (238,200) -- (238,275) -- (35,275) -- cycle ;
\draw (40,215) node [anchor=north west, rotate=0][inner sep=0.75pt]  [font=\small] [align=left] {\textbf{Complete Information Market}};
\draw (75,240) node [anchor=north west, rotate=0][inner sep=0.75pt]  [font=\small] [align=left] {$\boldsymbol{w_M^c=(\delta\Sigma)^{-1}\pi^c}$};

\draw [fill=gray!10]  (300,305) -- (490,305) -- (490,380) -- (300,380) -- cycle ;
\draw (310,313) node [anchor=north west][inner sep=0.75pt]  [font=\small] [align=left] {Prior Distribution of Implied};
\draw (330,333) node [anchor=north west][inner sep=0.75pt]  [font=\small] [align=left] {\textbf{Market equilibrium}};
\draw (340,355) node [anchor=north west][inner sep=0.75pt]  [font=\small] [align=left] {$\boldsymbol{\tilde{R}^{\tilde{\lambda}}\sim\mathcal{N}\bigl(\pi;\Sigma_\gamma\bigr)}$};
\draw [fill=gray!10]  (495,305) -- (580,305) -- (580,380) -- (495,380) -- cycle ;
\draw (505,325) node [anchor=north west][inner sep=0.75pt]  [font=\small] [align=left] {Likelihood};

\draw (510,345) node [anchor=north west][inner sep=0.75pt]  [font=\small] [align=left] {$\boldsymbol{\mathcal{P}\bigl(\tilde{\nu}\bigl|\tilde{R}\bigr)}$};
\draw [fill=gray!10]  (245,200) -- (418,200) -- (418,275) -- (245,275) -- cycle ;
\draw (257,208) node [anchor=north west][inner sep=0.75pt]  [font=\small] [align=left] {Distribution of Merton's};
\draw (282,228) node [anchor=north west][inner sep=0.75pt]  [font=\small] [align=left] {\textbf{Shadow-Costs}};
\draw (282,250) node [anchor=north west][inner sep=0.75pt]  [font=\small] [align=left] {$\boldsymbol{\tilde{\lambda}\sim\mathcal{N}(\lambda;\Lambda)}$};
\draw [fill=gray!10]  (425,200) -- (580,200) -- (580,275) -- (425,275) -- cycle ;
\draw (445,208) node [anchor=north west][inner sep=0.75pt]  [font=\small] [align=left]
{
Distribution of BL 
};
\draw (450,228) node [anchor=north west][inner sep=0.75pt]  [font=\small] [align=left]
{
\textbf{Subjective views}
};
\draw (455,252) node [anchor=north west][inner sep=0.75pt]  [font=\small] [align=left] {$\boldsymbol{\tilde{\nu}\sim\mathcal{N}(q;\Omega)}$};
\draw[dotted, fill=green!10]   (35,305) -- (270,305) -- (270,380) -- (35,380) -- cycle ;
\draw (50,318) node [anchor=north west, rotate=0][inner sep=0.75pt]  [font=\small] [align=left] {\textbf{Incomplete Information Market}};
\draw (45,343) node [anchor=north west, rotate=0][inner sep=0.75pt]  [font=\small] [align=left] {$\boldsymbol{w_M^\lambda:=(\delta-\delta_\lambda)^{-1}\Sigma^{-1}(\pi-\lambda)}$};
\draw[dotted, fill=green!10]   (35,445) -- (270,445) -- (270,515) -- (35,515) -- cycle ;
\draw (42,455) node [anchor=north west, rotate=0][inner sep=0.75pt]  [font=\small] [align=left] {\textbf{Subjective Beliefs of the BL Model}};
\draw (60,476) node [anchor=north west, rotate=0][inner sep=0.75pt]  [font=\small] [align=left] {$\boldsymbol{w_M^\nu=\Bigl[\delta\text{Var}\bigl[\tilde{R}_{\tilde{\nu}}\bigr]\Bigr]^{-1}\mathbb{E}\bigl[\tilde{R}_{\tilde{\nu}}\bigr]}$};

\draw[fill=gray!10]   (280,445) -- (580,445) -- (580,515) -- (280,515) -- cycle ;
\draw (300,455) node [anchor=north west][inner sep=0.75pt]  [font=\small] [align=left] {Posterior Distribution in Market with Views};
\draw (375,485) node [anchor=north west][inner sep=0.75pt]   [align=left] {$\boldsymbol{\tilde{R}^{\tilde{\lambda}}_{\tilde{\nu}}\sim\mathcal{N}(\pi^*;\Sigma^*)}$};
\draw[dotted, fill=green!10]   (35,670) -- (350,670) -- (350,610) -- (35,610) -- cycle ;
\draw (45,618) node [anchor=north west][inner sep=0.75pt]  [font=\small] [align=left] {Optimal Portfolio in Imperfect market with views};
\draw (45,635) node [anchor=north west][inner sep=0.75pt]  [font=\small] [align=left] {\large{$\boldsymbol{w_M^{\lambda,\nu}=\bigl[\delta\Sigma^*_{1,\dots,n_j\times 1,\dots,n_j}\bigr]^{-1}\pi^*_{1,\dots,n_j}}$}};
\draw[dashed]   (370,565) -- (560,565) -- (560,590) -- (370,590) -- cycle ;
\draw (380,570) node [anchor=north west][inner sep=0.75pt]   [align=left] {Mean-Variance Optimizer};
\draw   (385,610) -- (547,610) -- (547,670) -- (385,670) -- cycle ;
\draw (390,620) node [anchor=north west][inner sep=0.75pt]   [align=left] {Investor's $j$ with Set $J_j$};
\draw (400,645) node [anchor=north west][inner sep=0.75pt]   [align=left] {$\boldsymbol{J_j=\{1,\dots,n_j\}}$};

\draw[dashed]   (420,400) -- (560,400) -- (560,425) -- (420,425) -- cycle ;
\draw (425,405) node [anchor=north west][inner sep=0.75pt]   [align=left] {Bayesian Approach};
\draw[dotted, fill=green!10]   (135,395) -- (270,395) -- (270,430) -- (135,430) -- cycle ;
\draw (140,404) node [anchor=north west][inner sep=0.75pt]   [align=left] {$\boldsymbol{w_M^\gamma=(\delta\Sigma_\gamma)^{-1}\pi}$};
\draw  [color={rgb, 255:red, 0; green, 0; blue, 0 }  ,draw opacity=1, dotted ] (0,178) -- (590,178) -- (590,535) -- (0,535) -- cycle ;
\draw  [color={rgb, 255:red, 0; green, 0; blue, 0 }  ,draw opacity=1, dotted ] (0,550) -- (590,550) -- (590,685) -- (0,685) -- cycle ;
\draw    (310,150) -- (310,198.5) ;
\draw [shift={(310,200.5)}, rotate = 270] [color={rgb, 255:red, 0; green, 0; blue, 0 }  ][line width=0.75]    (10.93,-3.29) .. controls (6.95,-1.4) and (3.31,-0.3) .. (0,0) .. controls (3.31,0.3) and (6.95,1.4) .. (10.93,3.29)   ;
\draw    (510,150) -- (510,197) ;
\draw [shift={(510,200.67)}, rotate = 271.59] [color={rgb, 255:red, 0; green, 0; blue, 0 }  ][line width=0.75]    (10.93,-3.29) .. controls (6.95,-1.4) and (3.31,-0.3) .. (0,0) .. controls (3.31,0.3) and (6.95,1.4) .. (10.93,3.29)   ;
\draw[dotted]    (120,275) -- (120,305) ;
\draw[dotted] [shift={(120,305)}, rotate = 271.59] [color={rgb, 255:red, 0; green, 0; blue, 0 }  ][line width=0.75]    (10.93,-3.29) .. controls (6.95,-1.4) and (3.31,-0.3) .. (0,0) .. controls (3.31,0.3) and (6.95,1.4) .. (10.93,3.29)   ;
\draw    (290,275) -- (290,340) ;
\draw    (270,340) -- (290,340) ;
\draw [shift={(270,340)}, rotate = 360] [color={rgb, 255:red, 0; green, 0; blue, 0 }  ][line width=0.75]    (10.93,-3.29) .. controls (6.95,-1.4) and (3.31,-0.3) .. (0,0) .. controls (3.31,0.3) and (6.95,1.4) .. (10.93,3.29)   ;
\draw[dotted]    (20,250) -- (35,250) ;
\draw[dotted]    (20,250) -- (20,480) ;
\draw[dotted]    (20,480) -- (35,480) ;
\draw[dotted] [shift={(35,480)}, rotate = -180] [color={rgb, 255:red, 0; green, 0; blue, 0 }  ][line width=0.75]    (10.93,-3.29) .. controls (6.95,-1.4) and (3.31,-0.3) .. (0,0) .. controls (3.31,0.3) and (6.95,1.4) .. (10.93,3.29)   ;
\draw[dotted]    (10,230) -- (35,230) ;
\draw[dotted]    (10,230) -- (10,640) ;
\draw[dotted]    (10,640) -- (35,640) ;
\draw[dotted] [shift={(35,640)}, rotate = 180] [color={rgb, 255:red, 0; green, 0; blue, 0 }  ][line width=0.75]    (10.93,-3.29) .. controls (6.95,-1.4) and (3.31,-0.3) .. (0,0) .. controls (3.31,0.3) and (6.95,1.4) .. (10.93,3.29)   ;
\draw    (350,275) -- (350,305) ;
\draw [shift={(350,305)}, rotate = 271.59] [color={rgb, 255:red, 0; green, 0; blue, 0 }  ][line width=0.75]    (10.93,-3.29) .. controls (6.95,-1.4) and (3.31,-0.3) .. (0,0) .. controls (3.31,0.3) and (6.95,1.4) .. (10.93,3.29)   ;
\draw    (540,275) -- (540,305) ;
\draw [shift={(540,305)}, rotate = 271.59] [color={rgb, 255:red, 0; green, 0; blue, 0 }  ][line width=0.75]    (10.93,-3.29) .. controls (6.95,-1.4) and (3.31,-0.3) .. (0,0) .. controls (3.31,0.3) and (6.95,1.4) .. (10.93,3.29)   ;
\draw    (540,380) -- (540,400) ;
\draw [shift={(540,400)}, rotate = 271.59] [color={rgb, 255:red, 0; green, 0; blue, 0 }  ][line width=0.75]    (10.93,-3.29) .. controls (6.95,-1.4) and (3.31,-0.3) .. (0,0) .. controls (3.31,0.3) and (6.95,1.4) .. (10.93,3.29)   ;
\draw    (450,380) -- (450,400) ;
\draw [shift={(450,400)}, rotate = 270.37] [color={rgb, 255:red, 0; green, 0; blue, 0 }  ][line width=0.75]    (10.93,-3.29) .. controls (6.95,-1.4) and (3.31,-0.3) .. (0,0) .. controls (3.31,0.3) and (6.95,1.4) .. (10.93,3.29)   ;
\draw [dotted]   (375,380) -- (375,410) ;
\draw [dotted]   (270,410) -- (375,410) ;
\draw [shift={(270,410)}, rotate = 360, dotted] [color={rgb, 255:red, 0; green, 0; blue, 0 }  ][line width=0.75]    (10.93,-3.29) .. controls (6.95,-1.4) and (3.31,-0.3) .. (0,0) .. controls (3.31,0.3) and (6.95,1.4) .. (10.93,3.29)   ;
\draw    (490,425) -- (490,445) ;
\draw [shift={(490,445)}, rotate = 270.37] [color={rgb, 255:red, 0; green, 0; blue, 0 }  ][line width=0.75]    (10.93,-3.29) .. controls (6.95,-1.4) and (3.31,-0.3) .. (0,0) .. controls (3.31,0.3) and (6.95,1.4) .. (10.93,3.29)   ;
\draw    (470,515) -- (470,565) ;
\draw [shift={(470,565)}, rotate = 270.37] [color={rgb, 255:red, 0; green, 0; blue, 0 }  ][line width=0.75]    (10.93,-3.29) .. controls (6.95,-1.4) and (3.31,-0.3) .. (0,0) .. controls (3.31,0.3) and (6.95,1.4) .. (10.93,3.29)   ;
\draw    (470,590) -- (470,610) ;
\draw [shift={(470,610)}, rotate = 270.37] [color={rgb, 255:red, 0; green, 0; blue, 0 }  ][line width=0.75]    (10.93,-3.29) .. controls (6.95,-1.4) and (3.31,-0.3) .. (0,0) .. controls (3.31,0.3) and (6.95,1.4) .. (10.93,3.29)   ;
\draw    (350,640) -- (385,640) ;
\draw [shift={(350,640)}, rotate = 360] [color={rgb, 255:red, 0; green, 0; blue, 0 }  ][line width=0.75]    (10.93,-3.29) .. controls (6.95,-1.4) and (3.31,-0.3) .. (0,0) .. controls (3.31,0.3) and (6.95,1.4) .. (10.93,3.29)   ;
%
\draw    (30,78) -- (30,110) ;
\draw [shift={(30,110)}, rotate = 271.59] [color={rgb, 255:red, 0; green, 0; blue, 0 }  ][line width=0.75]    (10.93,-3.29) .. controls (6.95,-1.4) and (3.31,-0.3) .. (0,0) .. controls (3.31,0.3) and (6.95,1.4) .. (10.93,3.29)   ;

\draw    (124,78) -- (124,110) ;
\draw [shift={(124,110)}, rotate = 271.59] [color={rgb, 255:red, 0; green, 0; blue, 0 }  ][line width=0.75]    (10.93,-3.29) .. controls (6.95,-1.4) and (3.31,-0.3) .. (0,0) .. controls (3.31,0.3) and (6.95,1.4) .. (10.93,3.29)   ;
\draw    (165,78) -- (165,110) ;
\draw [shift={(165,110)}, rotate = 271.59] [color={rgb, 255:red, 0; green, 0; blue, 0 }  ][line width=0.75]    (10.93,-3.29) .. controls (6.95,-1.4) and (3.31,-0.3) .. (0,0) .. controls (3.31,0.3) and (6.95,1.4) .. (10.93,3.29);
\draw    (255,78) -- (255,110) ;
\draw [shift={(255,110)}, rotate = 271.59] [color={rgb, 255:red, 0; green, 0; blue, 0 }  ][line width=0.75]    (10.93,-3.29) .. controls (6.95,-1.4) and (3.31,-0.3) .. (0,0) .. controls (3.31,0.3) and (6.95,1.4) .. (10.93,3.29)   ;
\draw    (224,68) -- (242,68) ;
\draw [shift={(242,68)}, rotate = -180] [color={rgb, 255:red, 0; green, 0; blue, 0 }  ][line width=0.75]    (10.93,-3.29) .. controls (6.95,-1.4) and (3.31,-0.3) .. (0,0) .. controls (3.31,0.3) and (6.95,1.4) .. (10.93,3.29)   ;

\draw    (430,78) -- (430,110) ;
\draw [shift={(430,110)}, rotate = 271.59] [color={rgb, 255:red, 0; green, 0; blue, 0 }  ][line width=0.75]    (10.93,-3.29) .. controls (6.95,-1.4) and (3.31,-0.3) .. (0,0) .. controls (3.31,0.3) and (6.95,1.4) .. (10.93,3.29)   ;
\draw    (335,78) -- (335,110) ;
\draw [shift={(335,110)}, rotate = 271.59] [color={rgb, 255:red, 0; green, 0; blue, 0 }  ][line width=0.75]    (10.93,-3.29) .. controls (6.95,-1.4) and (3.31,-0.3) .. (0,0) .. controls (3.31,0.3) and (6.95,1.4) .. (10.93,3.29)   ;

\draw    (457,78) -- (457,110) ;
\draw [shift={(457,110)}, rotate = 271.59] [color={rgb, 255:red, 0; green, 0; blue, 0 }  ][line width=0.75]    (10.93,-3.29) .. controls (6.95,-1.4) and (3.31,-0.3) .. (0,0) .. controls (3.31,0.3) and (6.95,1.4) .. (10.93,3.29)   ;
\draw    (484,78) -- (484,110) ;
\draw [shift={(484,110)}, rotate = 271.59] [color={rgb, 255:red, 0; green, 0; blue, 0 }  ][line width=0.75]    (10.93,-3.29) .. controls (6.95,-1.4) and (3.31,-0.3) .. (0,0) .. controls (3.31,0.3) and (6.95,1.4) .. (10.93,3.29)   ;
\draw    (545,78) -- (545,110) ;
\draw [shift={(545,110)}, rotate = 271.59] [color={rgb, 255:red, 0; green, 0; blue, 0 }  ][line width=0.75]    (10.93,-3.29) .. controls (6.95,-1.4) and (3.31,-0.3) .. (0,0) .. controls (3.31,0.3) and (6.95,1.4) .. (10.93,3.29)   ;
\draw    (180,132) -- (210,132) ;
\draw [shift={(210,132)}, rotate = 178.53] [color={rgb, 255:red, 0; green, 0; blue, 0 }  ][line width=0.75]    (10.93,-3.29) .. controls (6.95,-1.4) and (3.31,-0.3) .. (0,0) .. controls (3.31,0.3) and (6.95,1.4) .. (10.93,3.29)   ;
\draw    (382,132) -- (420,132) ;
\draw [shift={(420,132)}, rotate = 178.53] [color={rgb, 255:red, 0; green, 0; blue, 0 }  ][line width=0.75]    (10.93,-3.29) .. controls (6.95,-1.4) and (3.31,-0.3) .. (0,0) .. controls (3.31,0.3) and (6.95,1.4) .. (10.93,3.29)   ;
\draw [dotted]   (120,150) -- (120,200) ;
\draw [shift={(120,200)}, rotate = 271.59, dotted] [color={rgb, 255:red, 0; green, 0; blue, 0 }  ][line width=0.75]    (10.93,-3.29) .. controls (6.95,-1.4) and (3.31,-0.3) .. (0,0) .. controls (3.31,0.3) and (6.95,1.4) .. (10.93,3.29)   ;

\draw (130,280) node [anchor=north west, rotate=0][inner sep=0.75pt]  [font=\small] [align=left] {\footnotesize{$\mathbb{E}[\tilde{\lambda}]$-Adjustment}};
\draw (25,405) node [anchor=north west, rotate=0][inner sep=0.75pt]  [font=\small] [align=left] {\footnotesize{$\tilde{\nu}$-Adjustment}};
\draw (15,570) node [anchor=north west, rotate=0][inner sep=0.75pt]  [font=\small] [align=left] {\footnotesize{$\{\tilde{\lambda},\tilde{\nu}\}$-Adjustment}};
\draw (290,415) node [anchor=north west, rotate=0][inner sep=0.75pt]  [font=\small] [align=left] {\footnotesize{$\tilde{\lambda}$-Adjustment}};
\end{tikzpicture}
\caption{Scheme for the Equilibrium Model in Markets of Imperfect Information and Subjective Views with Unconstrained Mean-Variance Quadratic Utility Maximization.}
\label{Scheme:2}
\end{scheme}
\newpage
\subsection{Posterior Distribution of the Market Equilibrium}\label{sec4.2}
Now, we explore the Bayesian approach of statistical inference to derive the mean and variance of the distribution $\tilde{R}_{\tilde{\nu}}^{\tilde{\lambda}}$ of the random vector of excess returns given the distributions of the shadow-costs and subjective views. The result is stated as follows:
\begin{theorem}[Posterior Distribution of Equilibrium Excess Returns]\label{thm1}
    The posterior distribution of the random vector $\tilde{R}^{\tilde\lambda}_{\tilde\nu}\sim\mathcal{P}\bigl(\tilde{R}\bigl|\tilde\lambda;\tilde\nu\bigr)$ of excess returns, given the subjective views' random vector $\tilde{\nu}\sim\mathcal{N}(q;\Omega)$, in the imperfect information market with random vector $\tilde{\lambda}\sim\mathcal{N}(\lambda;\Lambda)$ of shadow-costs has a multivariate normal distribution characterized by
    \begin{equation}\label{eq36_}
    \tilde{R}^{\tilde\lambda}_{\tilde\nu}\sim\mathcal{N}\bigl(\pi^*:={\Sigma^*}\pi^\lambda_\nu;\Sigma^*\bigr),
    \end{equation}
    where
$$\pi^\lambda_\nu:=\Sigma_\gamma^{-1}\pi+P^\top\Omega^{-1} q,$$
and
$$\Sigma^*:=\bigl[\Sigma_\gamma^{-1}+P^\top\Omega^{-1} P\bigr]^{-1},$$
for $\pi$ and $\Sigma_\gamma$ are given by equations (\ref{eq3}) and (\ref{eq32}), respectively.
\end{theorem}
\begin{proof}
We start by exploring Bayesian inference framework to compute the posterior equilibrium distribution, given the distributions of shadow-costs $\tilde{\lambda}$ and subjective views $\tilde{\nu}$. This posterior turns out to be the double-conditional distribution we express by means of Bayes' Rule as
\begin{equation}\label{eq26}
    \mathcal{P}\bigl(\tilde{R}\bigl|\tilde\lambda;\tilde\nu\bigr)=\frac{\mathcal{P}\bigl(\tilde\lambda;\tilde\nu\bigl|\tilde{R}\bigr)\mathcal{P}\bigl(\tilde{R}\bigr)}{\mathcal{P}\bigl(\tilde\lambda;\tilde\nu\bigr)},
\end{equation}
where
\begin{itemize}
    \item[$\boldsymbol{(i)}$] $\mathcal{P}\bigl(\tilde{R}\bigr)$ is the "Prior" distribution of investor's belief on the CAPM-equilibrium in the complete information market before observing the shadow-costs and incorporating subjective views;
    \item[$\boldsymbol{(ii)}$] $\mathcal{P}\bigl(\tilde\lambda;\tilde\nu\bigl|\tilde{R}\bigr)$ represents the "Likelihood" of observing the shadow-costs and incorporating the subjective views, given the implied equilibrium;
    \item[$\boldsymbol{(iii)}$] $\mathcal{P}\bigl(\tilde\lambda;\tilde\nu\bigr)$ the "Marginal Likelihood" of observing the imperfection of information among investors as well as incorporating the subjective beliefs.
\end{itemize}
The knowledge about $\boldsymbol{(i)}$ and $\boldsymbol{(ii)}$ is sufficient to characterize the conditional posterior distribution, in fact we can marginalize the distribution in $\boldsymbol{(iii)}$ by the prior in $\boldsymbol{(i)}$ as
\begin{equation*}
\mathcal{P}\bigl(\tilde\lambda;\tilde\nu\bigr)=\int_{\mathcal{R}}\mathcal{P}\bigl(\tilde\lambda;\tilde\nu\bigl|\tilde{R}\bigr)\mathcal{P}\bigl(\tilde{R}\bigr)d\tilde{R},
\end{equation*}
for $\mathcal{R}$ being the probability support of $\tilde{R}$. From Assumption \ref{assumption4}, the Likelihood rewrites as
$$\mathcal{P}\bigl(\tilde\lambda\bigl|\tilde{R}\bigr)\mathcal{P}\bigl(\tilde\nu\bigl|\tilde{R}\bigr),$$
combined with equation (\ref{eq26}) yields
\begin{align*}
\mathcal{P}\bigl(\tilde{R}\bigl|\tilde\lambda;\tilde\nu\bigr)&=\frac{\mathcal{P}\bigl(\tilde\lambda\bigr)\mathcal{P}\bigl(\tilde\nu\bigl|\tilde{R}\bigr)\mathcal{P}\bigl(\tilde{R}\bigl|\tilde\lambda\bigr)}{\mathcal{P}\bigl(\tilde\lambda;\tilde\nu\bigr)}\\
&=\frac{\mathcal{P}\bigl(\tilde\nu\bigl|\tilde{R}\bigr)\mathcal{P}\bigl(\tilde{R}\bigl|\tilde\lambda\bigr)}{\mathcal{P}\bigl(\tilde\nu\bigl|\tilde\lambda\bigr)}.
\end{align*}
Thus, the posterior distribution is the product of the two conditional distributions
\begin{itemize}
    \item \textbf{The Prior $\boldsymbol{\mathcal{P}\bigl(\tilde{R}\bigl|\tilde\lambda\bigr)}$}: Market reference model distribution $\tilde{R}^{\tilde{\lambda}}$ that account for the shadow-costs, it is in fact the posterior distribution of equilibrium for random shadow-costs;
    \item \textbf{The Likelihood $\boldsymbol{\mathcal{P}\bigl(\tilde\nu\bigl|\tilde{R}\bigr)}$}: Distribution of subjective beliefs one obtain after observing the prior CAPM-equilibrium, which turns to be $\mathcal{P}\bigl(P\tilde{R}^{\tilde{\lambda}}\bigr)$.
\end{itemize}
While the denominator cancels to the constant of integration. Therefore, we deduce that the posterior distribution satisfies
\begin{equation*}
\tilde{R}^{\tilde\lambda}_{\tilde\nu}\propto\mathcal{P}\bigl(P\tilde{R}^{\tilde{\lambda}}\bigr)\tilde{R}^{\tilde{\lambda}},
\end{equation*}
where $\propto$ stands for "proportional to". Recalling now Assumption \ref{Assump05}, we write
\begin{equation*}
    \begin{aligned}
    \mathcal{P}\bigl(\tilde{R}\bigl|\tilde\lambda;\tilde\nu\bigr)\propto&\exp\Bigl(-\frac{1}{2}\bigl(P\tilde{R}^{\tilde\lambda}-q\bigr)^\top\Omega^{-1}\bigl(\tilde P\tilde{R}^{\tilde\lambda}-q\bigr)\Bigr)\\
    &\exp\Bigl(-\frac{1}{2}\bigl(\tilde{R}^{\tilde\lambda}-\pi)^\top\Sigma_\gamma^{-1}\bigl(\tilde{R}^{\tilde\lambda}-\pi\bigr)\Bigr),
\end{aligned}
\end{equation*}
next, we develop the right-hand side of this equation to obtain
\begin{equation*}
    \begin{aligned}
    \mathcal{P}\bigl(\tilde{R}\bigl|\tilde\lambda;\tilde\nu\bigr)\propto\;\exp\Bigl(-\frac{1}{2}\Bigl(&\bigl(P\tilde{R}^{\tilde\lambda}\bigr)^\top\Omega^{-1}P\tilde{R}^{\tilde\lambda}-\bigl(P\tilde{R}^{\tilde\lambda}\bigr)^\top\Omega^{-1}q-q^\top\Omega^{-1}P\tilde{R}^{\tilde\lambda}+q^\top\Omega^{-1}q\\
        &+\bigl(\tilde{R}^{\tilde\lambda}\bigr)^\top\Sigma_\gamma^{-1}\tilde{R}^{\tilde\lambda}-\bigl(\tilde{R}^{\tilde\lambda}\bigr)^\top\Sigma_\gamma^{-1}\pi-\pi^\top\Sigma_\gamma^{-1}\tilde{R}^{\tilde\lambda}+\pi^\top\Sigma_\gamma^{-1}\pi\Bigr)\Bigr).
    \end{aligned}
\end{equation*}
Since matrices $\Sigma_\gamma$ and $\Omega$ are symmetric, it follows that
\begin{equation*}
    \begin{aligned}
        \mathcal{P}\bigl(\tilde{R}\bigl|\tilde\lambda;\tilde\nu\bigr)\propto\;\exp\Bigl(-\frac{1}{2}\Bigl(&\bigl(\tilde{R}^{\tilde\lambda}\bigr)^\top P^\top\Omega^{-1}P\tilde{R}^{\tilde\lambda}-2q^\top\Omega^{-1}P\tilde{R}^{\tilde\lambda}+q^\top\Omega^{-1}q\\
        &+\bigl(\tilde{R}^{\tilde\lambda}\bigr)^\top\Sigma_\gamma^{-1}\tilde{R}^{\tilde\lambda}-2\pi^\top\Sigma_\gamma^{-1}\tilde{R}^{\tilde\lambda}+\pi^\top\Sigma_\gamma^{-1}\pi\Bigr)\Bigr),
    \end{aligned}
\end{equation*}
then, we get
\begin{equation*}
    \begin{aligned}
        \mathcal{P}\bigl(\tilde{R}\bigl|\tilde\lambda;\tilde\nu\bigr)\propto\;\exp\Bigl(-\frac{1}{2}\Bigl(&\bigl(\tilde{R}^{\tilde\lambda}\bigr)^\top\Bigl(\underset{H}{\underbrace{P^\top\Omega^{-1}P+\Sigma_\gamma^{-1}}}\Bigr)\tilde{R}^{\tilde\lambda}-2\Bigl(\underset{C^\top}{\underbrace{q^\top\Omega^{-1}P+\pi^\top\Sigma_\gamma^{-1}}}\Bigr)\tilde{R}^{\tilde\lambda}\\
        &+\underset{A}{\underbrace{q^\top\Omega^{-1}q+\pi^\top\Sigma_\gamma^{-1}\pi}}\Bigr)\Bigr),\\
    \end{aligned}
\end{equation*}
therefore,
\begin{equation*}
\mathcal{P}\bigl(\tilde{R}\bigl|\tilde\lambda;\tilde\nu\bigr)\propto\;\exp\Bigl(-\frac{1}{2}\Bigl(\bigl(\tilde{R}^{\tilde\lambda}\bigr)^\top H^\top\tilde{R}^{\tilde\lambda}-2C^\top\tilde{R}^{\tilde\lambda}+A\Bigr)\Bigr).
\end{equation*}
Since the first term in the right-hand side is equal to $\bigl(H\tilde{R}\bigr)^\top H^{-1}H\tilde{R}$, the second term is equal to $2C^\top H^{-1}H\tilde{R}$ and the last term $A$ vanishes due to the integration, we obtain
\begin{equation*}
\mathcal{P}\bigl(\tilde{R}\bigl|\tilde\lambda;\tilde\nu\bigr)\propto\;\exp\bigl(-\frac{1}{2}(H\tilde{R}^{\tilde\lambda}-C)^\top H^{-1}(H\tilde{R}^{\tilde\lambda}-C)-C^\top H^{-1}C\bigr),
\end{equation*}
that is, since the expression $C^\top H^{-1}C$ cancels to the constant of integration,
\begin{equation*}
    \mathcal{P}\bigl(\tilde{R}\bigl|\tilde{\lambda};\tilde{\nu}\bigr)\propto\;\exp\bigl(-\frac{1}{2}(\tilde{R}^{\tilde\lambda}-H^{-1}C)^\top H(\tilde{R}^{\tilde\lambda}-H^{-1}C)\bigr).
\end{equation*}
Finally, the posterior distribution $\tilde{R}^{\tilde\lambda}_{\tilde\nu}$ of the random vector $\tilde{R}$ of excess returns, given the distributions of the imperfection of information $\tilde{\lambda}$ and subjective views $\tilde{\nu}$, is multivariate normal with mean vector $H^{-1}C$ and covariance matrix $H^{-1}$. The proof is then finished.
\end{proof}
\begin{corollary}[Shadow-Costs with Random Expectation]
    Let the expectation of $\tilde\lambda$ be random and given by a multivariate normal distribution as in Remark \ref{rk2}. We write $\tilde{\lambda}\sim\mathcal{N}(\lambda_1;\tau_1\Lambda)$, for $\tau_1>0$, then we deduce from Theorem \ref{thm1} the distribution of $\tilde{R}^{\tilde{\lambda}}_{\tilde{\nu}}$ as
    $$\mathcal{N}\bigl(\hat\pi^*:={\hat\Sigma^*}\hat\pi^\lambda_\nu;\hat\Sigma^*\bigr),$$
    where, for $\hat\pi$ and $\hat{\Sigma}_\gamma$ are given as in Remark \ref{rk2},

    $$\hat\pi^\lambda_\nu:=\hat\Sigma_\gamma^{-1}\hat\pi+P^\top\Omega^{-1} q,$$
and
$$\hat\Sigma^*:=\bigl[\hat\Sigma_\gamma^{-1}+P^\top\Omega^{-1} P\bigr]^{-1}.$$ \qed
\end{corollary}
We now present an alternative approach for the derivation of the equilibrium on which the prior is given by the posterior distribution of the equilibrium related to the incomplete information market.
\begin{proposition}[Bayes' Rule Conditioning on $\tilde{\lambda}$]\label{prop3}
    The equilibrium posterior distribution of Theorem \ref{thm1} assumed the prior to be the implied equilibrium of the complete information market. Equivalently, the expression of the posterior distribution is
    \begin{equation}\label{eq40}
    \mathcal{P}\bigl(\tilde{R}\bigl|\tilde{\lambda};\tilde{\nu}\bigr)=\frac{\overset{\text{Predictive Distribution}}{\overbrace{\mathcal{P}\bigl(\tilde{\nu}\bigl|\tilde{R}\bigr)}}\;\overset{\text{Posterior Distribution}}{\overbrace{\mathcal{P}\bigl(\tilde{R}\bigl|\tilde{\lambda}\bigr)}}}{\underset{\text{Posterior Predictive Distribution}}{\underbrace{\mathcal{P}\bigl(\tilde{\nu}\bigl|\tilde{\lambda}\bigr)}}},
    \end{equation}
    which gives the equilibrium by means of our prior reference model $\tilde{R}^{\tilde{\lambda}}\sim\mathcal{P}\bigl(\tilde{R}|\tilde{\lambda}\bigr)$.
    \end{proposition}
    \begin{proof}
    It suffices to prove that the formula given in equation (\ref{eq40}) is equivalent to the straight forward equation (\ref{eq26}). We express the posterior distribution of $\tilde{R}$, conditioned on the shadow-costs and subjective views distributions, by first conditioning on shadow-costs $\tilde{\lambda}$ and using Bayes' rule as
    \begin{equation*}            \mathcal{P}\bigl(\tilde{R}\bigl|\tilde{\lambda};\tilde{\nu}\bigr)=\frac{\mathcal{P}\bigl(\tilde{\nu}\bigl|\tilde{R};\tilde{\lambda}\bigr)\mathcal{P}\bigl(\tilde{R}\bigl|\tilde{\lambda}\bigr)}{\mathcal{P}\bigl(\tilde{\nu}\bigl|\tilde{\lambda}\bigr)},
    \end{equation*}
    considering then Assumption \ref{assumption2}, we obtain
    \begin{equation*}
    \mathcal{P}\bigl(\tilde{R}\bigl|\tilde{\lambda};\tilde{\nu}\bigr)=\frac{\mathcal{P}\bigl(\tilde{\nu}\bigl|\tilde{R}\bigr)\mathcal{P}\bigl(\tilde{R}\bigl|\tilde{\lambda}\bigr)}{\mathcal{P}\bigl(\tilde{\nu}\bigl|\tilde{\lambda}\bigr)}.
    \end{equation*}
    Now, to get the equivalence, we start by marginalizing the joint likelihood over $\tilde{\lambda}$ to write
    $$\mathcal{P}\bigl(\tilde{\lambda};\tilde{\nu}\bigl|\tilde{R}\bigr)=\mathcal{P}\bigl(\tilde{\nu}\bigl|\tilde{\lambda};\tilde{R}\bigr)\mathcal{P}\bigl(\tilde{\lambda}\bigl|\tilde{R}\bigr).$$
    Then, we rewrite equation (\ref{eq26}) as
    \begin{equation}\label{equ42}
    \mathcal{P}\bigl(\tilde{R}\bigl|\tilde{\lambda};\tilde{\nu}\bigr)=\frac{\mathcal{P}\bigl(\tilde{\nu}\bigl|\tilde{\lambda};\tilde{R}\bigr)\mathcal{P}\bigl(\tilde{\lambda}\bigl|\tilde{R}\bigr)\mathcal{P}(\tilde{R})}{\mathcal{P}(\tilde{\nu};\tilde{\lambda})},
    \end{equation}
    for the marginal likelihood being
\begin{equation*}
    \begin{aligned}
        \mathcal{P}(\tilde{\nu};\tilde{\lambda})&=\int_{\mathcal{R}}\mathcal{P}\bigl(\tilde{\nu}\bigl|\tilde{\lambda};\tilde{R}\bigr)\mathcal{P}\bigl(\tilde{\lambda}\bigl|\tilde{R}\bigr)\mathcal{P}(\tilde{R})d\tilde{R}\\
        &=\int_{\mathcal{R}}\mathcal{P}\bigl(\tilde{\nu}\bigl|\tilde{\lambda};\tilde{R}\bigr)\mathcal{P}\bigl(\tilde{R}\bigl|\tilde{\lambda}\bigr)\mathcal{P}\bigl(\tilde{\lambda}\bigr)d\tilde{R}\\
                &=\mathcal{P}\bigl(\tilde{\nu}\bigl|\tilde{\lambda}\bigr)\mathcal{P}\bigl(\tilde{\lambda}\bigr).
    \end{aligned}
\end{equation*}
Recalling assumption (\ref{Assump05}), the numerator of equation (\ref{equ42}) become $\mathcal{P}\bigl(\tilde{\nu}|\tilde{R}\bigr)\mathcal{P}\bigl(\tilde{R}|\tilde{\lambda}\bigr)\mathcal{P}\bigl(\tilde{\lambda}\bigr)$, then substituting in this equation completes the proof.
\end{proof}
This allows investors to first use the available knowledge about the shadow-costs of information to update their beliefs about the equilibrium without incorporating their subjective views. This step involves computing the posterior distribution $\mathcal{P}\bigl(\tilde{R}|\tilde{\lambda}\bigl)$. Subsequently, investors integrate their subjective views through the likelihood distribution $\mathcal{P}\bigl(\tilde{\nu}|\tilde{R}\bigl)$ of their anticipations, given the implied equilibrium, along with the posterior distribution encompassing the shadow-costs, where the normalization constant is determined by the posterior predictive distribution $\mathcal{P}\bigl(\tilde{\nu}|\tilde{\lambda}\bigl)$. Exploring the result of the above Proposition \ref{prop3}, a similar proof to Theorem \ref{thm1} leads us to the alternative equilibrium below.
\begin{corollary}[Posterior Distribution by Bayes' Rule]
The posterior distribution $\tilde{R}_{\tilde{\nu}}^{\tilde{\lambda}}$ of the random vector of excess returns, given the views in a market with incomplete information, is expressed by the posterior distribution $\tilde{R}^{\tilde{\lambda}}$ of the equilibrium related to the incomplete information market and the distribution $\tilde{\nu}$ of subjective views, and has the multivariate normal distribution (\ref{eq36_}). \qed
\end{corollary}
\subsection{Optimal Portfolio in Incomplete Information Market with Views}
As mentioned in Assumptions \ref{assumption1} and \ref{assumption2}, investor $j$ aims at allocating optimal portfolios in markets with incomplete information while reflecting there subjective beliefs, and chooses to perform the standard "unconstrained mean-variance quadratic utility maximization" problem. Thus, similarly to the optimization problem (\ref{eeq24}), the investor solves the problem of finding the portfolio allocation
\begin{equation}\label{equ49}
w_M=\argmax_{w}\biggl\{w^\top\mathbb{E}\Bigl[{\bigl(\tilde{R}^{\tilde{\lambda}}_{\tilde\nu}\bigr)}^{J_j}\Bigl]-\frac{1}{2}\delta w^\top\text{Var}\Bigl[{\bigl(\tilde{R}^{\tilde{\lambda}}_{\tilde\nu}\bigr)}^{J_j}\Bigl]w\biggr\},
\end{equation}
for the equilibrium excess returns sub-vector $\bigl(\tilde{R}^{\tilde{\lambda}}_{\tilde\nu}\bigr)^{J_j}$ being the posterior distribution of our model's equilibrium related to the securities about which the investor holds information. Any security $k$ that is held on $j$-th investor's information set $J_j$ can be subject of weighting in the allocated portfolio, yielding the equivalent expression for the vector of excess returns
$$\tilde{R}^{\tilde{\lambda}}_{\tilde\nu}:=\biggl[{\bigl(\tilde{R}^{\tilde{\lambda}}_{\tilde\nu}\bigr)}^{J_j},{\bigl(\tilde{R}^{\tilde{\lambda}}_{\tilde\nu}\bigr)}^{J_j^c}\biggr]^\top,$$
where, letting $\tilde{R}_k^{\tilde{\lambda},\tilde{\nu}}$ be the $\{\tilde{\lambda},\tilde{\nu}\}$-conditioned random excess return of asset $k$ in the incomplete information market with subjective views,
$$\bigl(\tilde{R}^{\tilde{\lambda}}_{\tilde\nu}\bigr)^{J_j}:=\Bigl\{\tilde{R}_1^{\tilde{\lambda},\tilde{\nu}}-r_f,\dots,\tilde{R}_{n_j}^{\tilde{\lambda},\tilde{\nu}}-r_f\Bigr\}\;\text{and}\;\bigl(\tilde{R}^{\tilde{\lambda}}_{\tilde\nu}\bigr)^{J_j^c}:=\Bigl\{\tilde{R}_{n_j+1}^{\tilde{\lambda},\tilde{\nu}}-r_f,\dots,\tilde{R}_{n}^{\tilde{\lambda},\tilde{\nu}}-r_f\Bigr\}.$$
The solution to the problem (\ref{equ49}) is then given as follows:
\begin{theorem}[Unconstrained Optimal Portfolio]
    The optimal allocation in the imperfect information market with subjective views and risk-adjusted return utility maximization is described, for investor $j$ with shadow-costs and subjective views distributions $\tilde{\lambda}$ and $\tilde{\nu}$, respectively, as
    \begin{align*}
    w_M^{\lambda,\nu}:&=\delta^{-1}\biggl[\text{Var}\Bigl[{\bigl(\tilde{R}^{\tilde{\lambda}}_{\tilde\nu}\bigr)}^{J_j}\Bigl]\biggr]^{-1}\mathbb{E}\Bigl[{\bigl(\tilde{R}^{\tilde{\lambda}}_{\tilde\nu}\bigr)}^{J_j}\Bigl]\\
        &=\delta^{-1}\bigl[\Sigma^*_{1,\dots,n_j\times1,\dots,n_j}\bigr]^{-1}\pi^*_{1,\dots,n_j},
    \end{align*}
    where $J_j=\{1,\dots,n_j\}$ is the set of securities subject of allocation for investor $j$. \qed
\end{theorem}
\par As an alternative, the investor might choose to solve other optimization problems depending on his/her goal, and these might include the following: 
\begin{itemize}
    \item \textbf{Risk-Constrained Problem}: A more classical approach for portfolio optimization problem is the maximization of the expectation while maintaining the variance under a risk threshold, it is given as
\begin{equation*}
\left\{
\begin{aligned}
    &{w_M^{\text{RiskConstraints}}}=\argmax_{w}\Bigl\{w^\top\mathbb{E}\Bigl[{\bigl(\tilde{R}^{\tilde{\lambda}}_{\tilde\nu}\bigr)}^{J_j}\Bigl]\Bigr\},\\
&\text{Subject to:}\;w^\top\text{Var}\Bigl[{\bigl(\tilde{R}^{\tilde{\lambda}}_{\tilde\nu}\bigr)}^{J_j}\Bigl]w\leq\sigma^2;
\end{aligned}
\right.
\end{equation*}
\item \textbf{Risk- and Budget-Constrained Problem}: In this type of problem we write
\begin{equation*}
\left\{
\begin{aligned}
    &{w_M^{\text{RiskBudgetConstraints}}}=\argmax_{w}\Bigl\{w^\top\mathbb{E}\Bigl[{\bigl(\tilde{R}^{\tilde{\lambda}}_{\tilde\nu}\bigr)}^{J_j}\Bigl]\Bigr\},\\
&\text{Subject to:}\;w^\top\text{Var}\Bigl[{\bigl(\tilde{R}^{\tilde{\lambda}}_{\tilde\nu}\bigr)}^{J_j}\Bigl]w\leq\sigma^2\;\text{and}\;w^\top\textbf{1}=1;
\end{aligned}
\right.
\end{equation*}
\item \textbf{Minimum Variance Problem}: Here we only minimize the variance of the portfolio's return by solving
\begin{equation*}
    {w_M^{\text{MinimumVariance}}}=\argmin_{w}\Bigl\{w^\top\text{Var}\Bigl[{\bigl(\tilde{R}^{\tilde{\lambda}}_{\tilde\nu}\bigr)}^{J_j}\Bigl]w\Bigr\}.
\end{equation*}
\end{itemize}
By solving these alternative problems, one gets the following result:
\begin{proposition}[Constrained and Minimum Variance Optimal Portfolios]
    The optimal portfolio allocation in the imperfect information market with subjective views and risk-constraint is
    $$w_M^{^{\text{RiskConstraints}}}:=\frac{\sigma}{\sqrt{\bigl(w_M^{\lambda,\nu}\bigr)^\top\text{Var}\Bigl[{\bigl(\tilde{R}^{\tilde{\lambda}}_{\tilde\nu}\bigr)}^{J_j}\Bigl] w_M^{\lambda,\nu}}}w_M^{\lambda,\nu},$$
    $$w_M^{\text{RiskBudgetConstraints}}:=aw_M^{\lambda,\nu}+b{w_M^{\text{MinimumVariance}}},$$
    for $w_M^{\lambda,\nu}$ being the solution of the unconstrained problem, $a,b$ weighting coefficients satisfying the risk and budget constraints, and $w_M^{\text{MinimumVariance}}$ the optimal portfolio for the minimum variance portfolio given as
    $$w_M^{\text{MinimumVariance}}=\frac{1}{\boldsymbol{1}_{\mathbb{R}^{n_j}}^\top{\text{Var}\Bigl[{\bigl(\tilde{R}^{\tilde{\lambda}}_{\tilde\nu}\bigr)}^{J_j}\Bigl]}^{-1}\boldsymbol{1}_{\mathbb{R}^{n_j}}}{\text{Var}\Bigl[{\bigl(\tilde{R}^{\tilde{\lambda}}_{\tilde\nu}\bigr)}^{J_j}\Bigl]}^{-1}\boldsymbol{1}_{\mathbb{R}^{n_j}}.$$
    Here, $\text{Var}\Bigl[{\bigl(\tilde{R}^{\tilde{\lambda}}_{\tilde\nu}\bigr)}^{J_j}\Bigl]=\Sigma^*_{1,\dots,n_j\times1,\dots,n_j}$. \qed
\end{proposition}
\section{Numerical Simulations}\label{sec.6}
Let us know present a numerical example to illustrate our findings, we give the results for a five-assets case.
\subsection{Computational Data}
Data about market parameters, shadow-costs of information, cross-covariance matrix and subjective views are given in the following tables:
\begin{table}[H]
    \centering
    \begin{tabular}{cccccc}
        \bottomrule
        \rowcolor{black!10} Asset & 1 & 2 & 3 & 4 & 5 \\
        \toprule
        $\boldsymbol{\Sigma}$ & 0.0500 & 0.0200 & 0.0400 & 0.0300 & 0.0100 \\
         & 0.0200 & 0.0800 & 0.0200 & 0.0200 & 0.0300 \\
         & 0.0400 & 0.0200 & 0.0900 & 0.0100 & 0.0200 \\
         & 0.0300 & 0.0200 & 0.0100 & 0.0700 & 0.0100 \\
         & 0.0100 & 0.0300 & 0.0200 & 0.0100 & 0.0600 \\
        \midrule
         CAPM Excess Returns ($\boldsymbol{\pi^c}$) & 0.0100 & 0.0300 & 0.0150 & 0.0400 & 0.0350 \\
        \bottomrule
        \rowcolor{black!10} Parameter & $\boldsymbol{r_f}$ & $\boldsymbol{\mathbb{E}\bigl[\tilde{R}_M\bigr]}$ & $\boldsymbol{\sigma_M}$ & & \\
        \toprule
        Value & 0.0200 & 0.0400 & 0.0500 & & \\
        \bottomrule
    \end{tabular}
    \caption{Covariance Matrix of Excess Returns, CAPM-Equilibrium and Market Parameters.}
    \label{tab1}
\end{table}
\begin{table}[H]
    \centering
    \begin{tabular}{cccccc}
    \bottomrule
       \rowcolor{black!10} Asset & 1 & 2 & 3 & 4 & 5 \\
       \toprule
       Shadow-Costs ($\boldsymbol{\lambda}$) & 0.0100 & 0.0250 & 0.0200 & 0.0150 & 0.0300\\
       \midrule
       $\boldsymbol{\Lambda}$ & 0.0800 & 0      & 0      & 0      & 0 \\
        & 0      & 0.0120 & 0      & 0      & 0 \\
        & 0      & 0      & 0.0200 & 0      & 0 \\
        & 0      & 0      & 0      & 0.0500 & 0 \\
        & 0      & 0      & 0      & 0      & 0.0100 \\
        \midrule
        $\boldsymbol{\Sigma_{\tilde{R},\tilde{\lambda}}}$ & 0.0100 & 0.0200 & 0.0400 & 0.0300 & 0.0500 \\
         & 0.0200 & 0.0200 & 0.1000 & 0.0240 & 0.0400 \\
         & 0.0400 & 0.1000 & 0.0500 & 0.0130 & 0.0600 \\
         & 0.0300 & 0.0240 & 0.1300 & 0.0600 & 0.0400 \\
         & 0.0500 & 0.0400 & 0.0600 & 0.0400 & 0.0900 \\
        \midrule
        $\boldsymbol{\tau}$ Values & 0.1 & 0.5 & 0.9 & & \\
        \bottomrule
    \end{tabular}
    \caption{Covariance Matrix of Shadow-Costs, Cross-Covariance Matrix, and Scaling Factor $\tau$.}
    \label{tab2}
\end{table}
\begin{table}[H]
    \centering
    \begin{tabular}{ccccccc}
        \bottomrule
        \rowcolor{black!10} Asset & 1 & 2 & 3 & 4 & 5 & \textbf{q} \\
        \toprule
        \textbf{P} & 1  & -1 & 0  & 0  & 0  & 0.0500 \\
        & 0  & 1  & 0 & 0  & 0  & 0.0300 \\
        & -0.2  & 0.1 & -0.8  & 0  & 0.9  & 0.0800 \\
        & 0 & 0 & 0  & 0  & 1  & 0.1000 \\
        \bottomrule
        \rowcolor{black!10} View & 1 & 2 & 3 & 4 & & \\
        \toprule
       	$\boldsymbol{\Omega=\bigl(\frac{1}{c}-1\bigr)P\Sigma P^\top}$ & 0.0900 & -0.0600 & -0.0460 & -0.0200 & & \\
        & -0.0600 & 0.0800 & 0.0150 & 0.0300 & & \\
        & -0.0460 & 0.0150 & 0.0908 & 0.0390 & & \\
        & -0.0200 & 0.0300 & 0.0390 & 0.0600 & & \\
        \bottomrule
        c & 0.01 & 0.5 & 0.99 & & & \\
        \toprule
    \end{tabular}
    \caption{View-Matrix and View-Vector of Subjective Views, with Resulting Uncertainty Matrix for $c=0.5$.}
    \label{tab3}
\end{table}
\subsection{Incomplete Information Market}
\subsubsection{Deterministic Shadow-Costs}
The obtained results, for the deterministic shadow-costs vector $\lambda$, include expected and extra excess returns, and optimal portfolios $w_M^c$ and $w_M^\lambda:=w_M$ in the CAPM complete information market and the incomplete information market, respectively.
    \begin{table}[H]
        \centering
    \begin{tabular}{lcccccc}
        \bottomrule
        \rowcolor{black!10} Asset & & 1 & 2 & 3 & 4 & 5 \\
        \toprule
        \text{CAPM Market} & $\boldsymbol{\pi^c}$ & 0.0100 & 0.0300 & 0.0150 & 0.0400 & 0.0350 \\
        \text{Extra Excess Returns} & $\boldsymbol{\lambda-\pi^\lambda}$ & 0.0100 & 0.0248 & 0.0202 & 0.0141 & 0.0298 \\
        \rowcolor{green!10} \text{Incomplete Market} & $\boldsymbol{\pi}$ & 0.0200 & 0.0548 & 0.0352 & 0.0541 & 0.0648 \\
        \text{CAPM Market} & $\boldsymbol{w_M^c}$ & -0.0587 & 0.0154 & 0.0225 & 0.0813 & 0.0540 \\
        \rowcolor{green!10} \text{Incomplete Market} & $\boldsymbol{w_M^\lambda}$ & -0.0394 & -0.0004 & 0.0025 & 0.0605 & 0.0063 \\
        \text{Investor's Portfolio} & $\boldsymbol{w^*}$ & -0.0727 & 0.0290 & 0.0395 & 0.0951 & 0.0946 \\
        \bottomrule
        \rowcolor{black!10} \text{Portfolio's Return \& Risk} & $\boldsymbol{\mathbb{E}\bigl[\tilde{R}_M^c\bigr]}$ & $\boldsymbol{{\sigma_M^c}}$ & $\boldsymbol{\mathbb{E}\bigl[\tilde{R}_M^\lambda\bigr]}$ & $\boldsymbol{\sigma_M^\lambda}$ & $\boldsymbol{\mathbb{E}\bigl[\tilde{R}_P^*\bigr]}$ & $\boldsymbol{{\sigma_P^*}}$ \\
        \toprule
        Computed Value & 0.0076 & 0.0259 & 0.0035 & 0.0138 & 0.0165 & 0.0388 \\
        \bottomrule
    \end{tabular}
    \caption{Expected Excess Returns, Optimal Portfolio Weights, and Return \& Risk at Equilibrium for Complete and Incomplete Information Market, and Investor's Portfolio.}
    \label{tab4}
\end{table}
\begin{table}[H]
    \centering
        \begin{tabular}{ccc}
        \toprule
        \text{Risk-Aversion Coefficient} & $\boldsymbol{\delta:=\frac{\mathbb{E}[\tilde{R}_M]-r_f}{\sigma_M^2}}$ & 8.0000 \\
        \midrule
        \text{Weighted-Average Shadow-Cost} & $\boldsymbol{\lambda_M:=({w_M^\lambda})^\top\lambda}$ & 7.4175e-04 \\
        \midrule
        \text{Modified Risk-Aversion Coefficient} & $\boldsymbol{\delta_\lambda:=\frac{\lambda_M}{\sigma_M^2}}$ & 0.2967 \\
    \bottomrule
    \end{tabular}
    \caption{Computed Parameters.}
    \label{tab5}
\end{table}
\begin{figure}[H]
    \centering
    \includegraphics[width=0.495\linewidth]{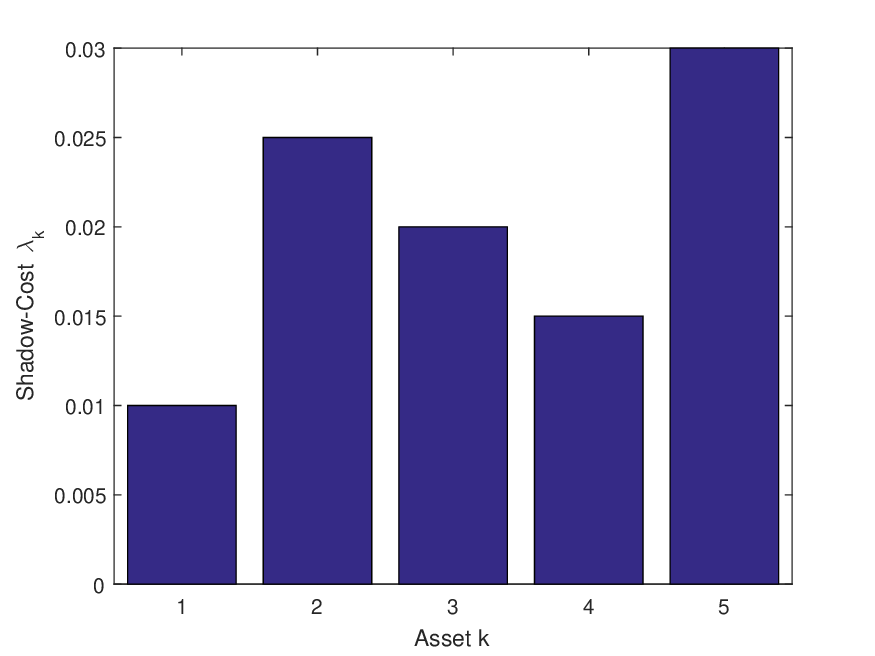}
    \includegraphics[width=0.495\linewidth]{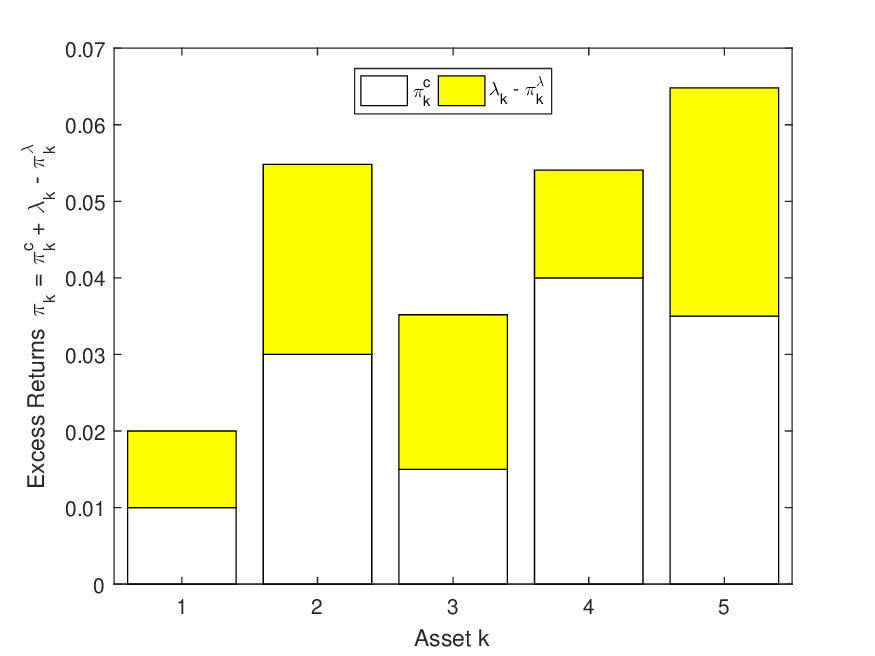}
    \caption{Shadow-Costs Vector $\lambda$ (left) and Expected Excess Returns Vector $\pi=\pi^c+\lambda-\pi^\lambda$ (right).}
    \label{fig1}
\end{figure}
\begin{figure}[H]
    \centering
    \includegraphics[width=0.495\linewidth]{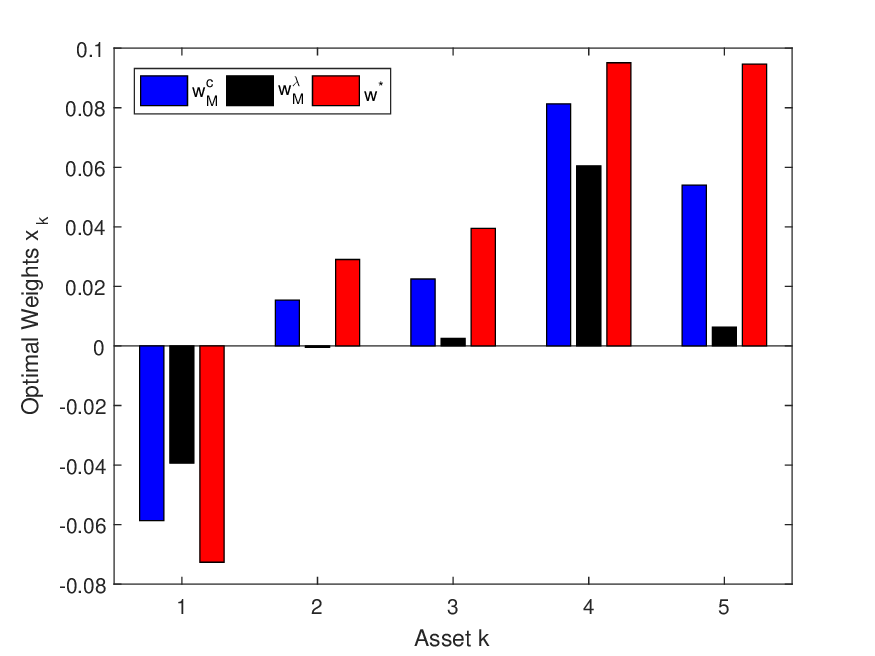}
    \includegraphics[width=0.495\linewidth]{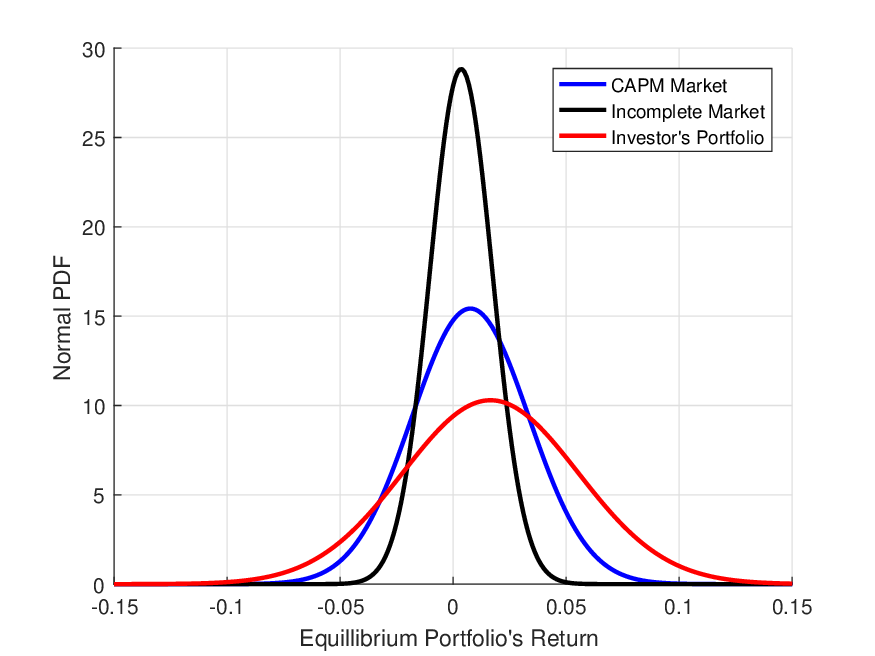}
    \caption{Optimal Portfolio Weights Vector (left) and Return \& Risk for Normally Distributed Market Equilibrium (right) for Complete and Incomplete Information Market, and Investor's Portfolio.}
    \label{fig2}
\end{figure}
\newpage
\subsubsection{Extra Excess Returns}
The following figures present the sensitivities (\ref{nablalambda}) and (\ref{nablax}) of the extra excess return of an asset $k$ to shadow-cost $\lambda_k$ and to optimal market portfolio weight $x_k$.
\begin{figure}[H]

\vspace{\fill}
    \centering
    \includegraphics[width=0.48\linewidth]{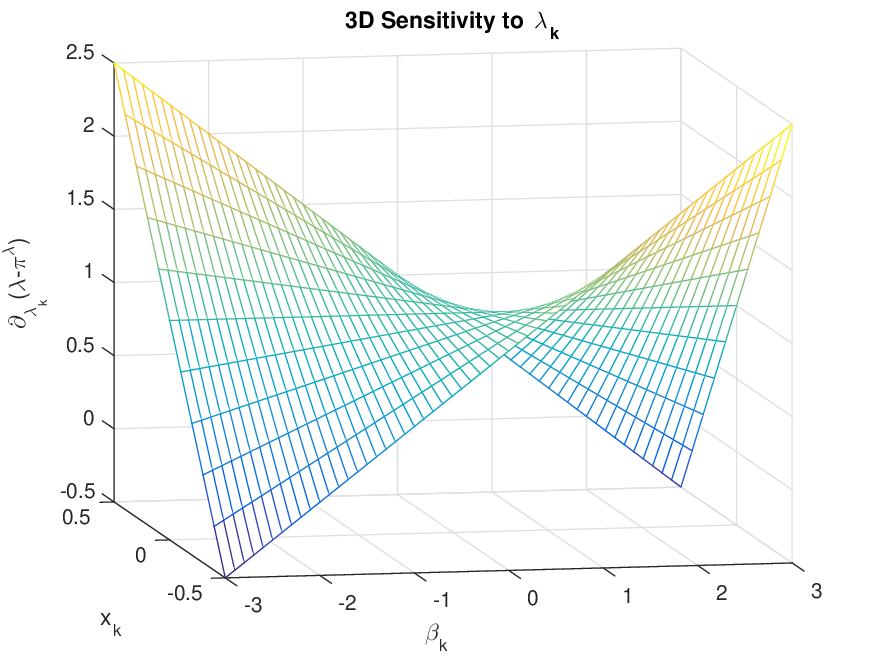}
    \includegraphics[width=0.48\linewidth]{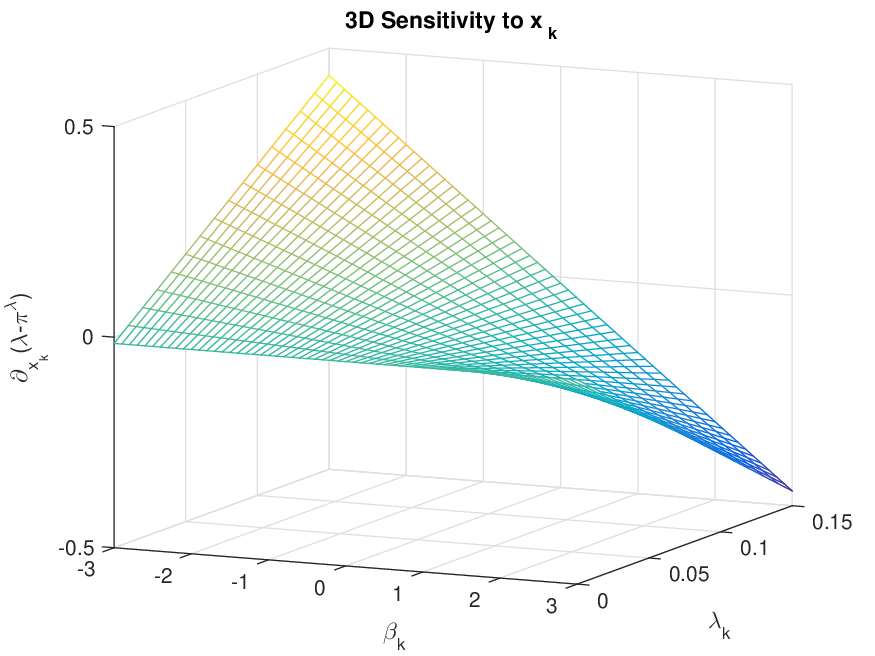}\\
    \includegraphics[width=0.48\linewidth]{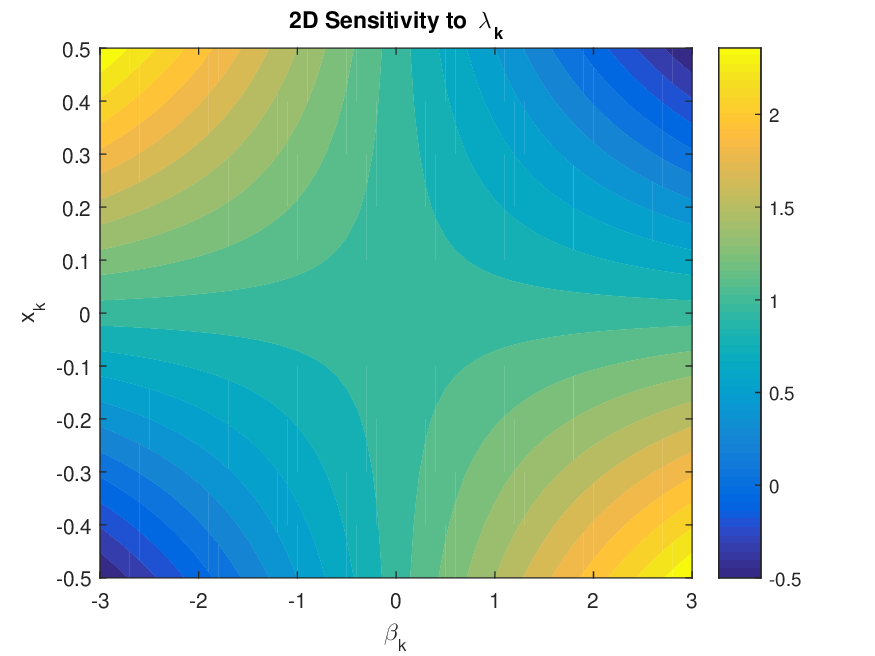}
    \includegraphics[width=0.48\linewidth]{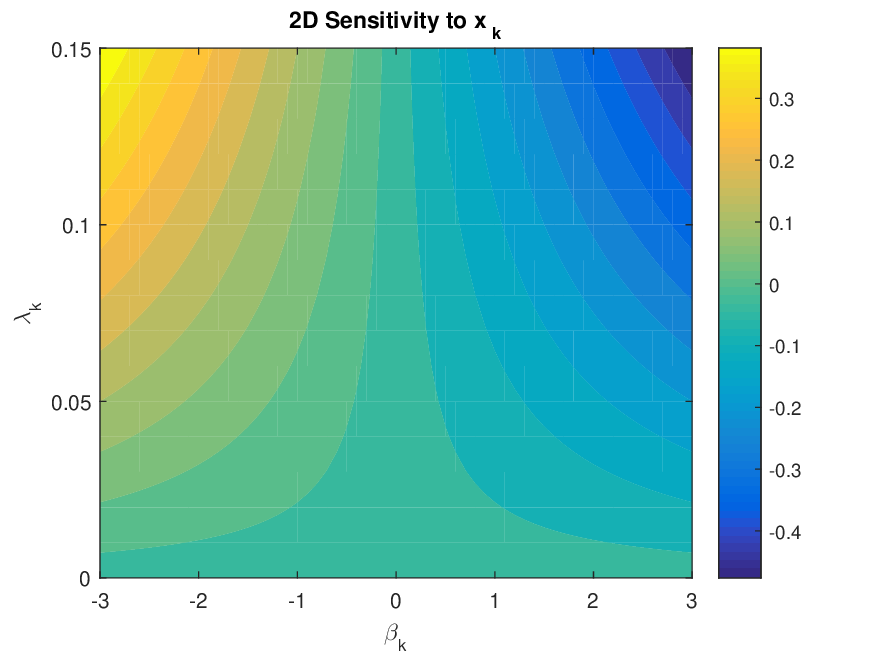}\\
    \includegraphics[width=0.48\linewidth]{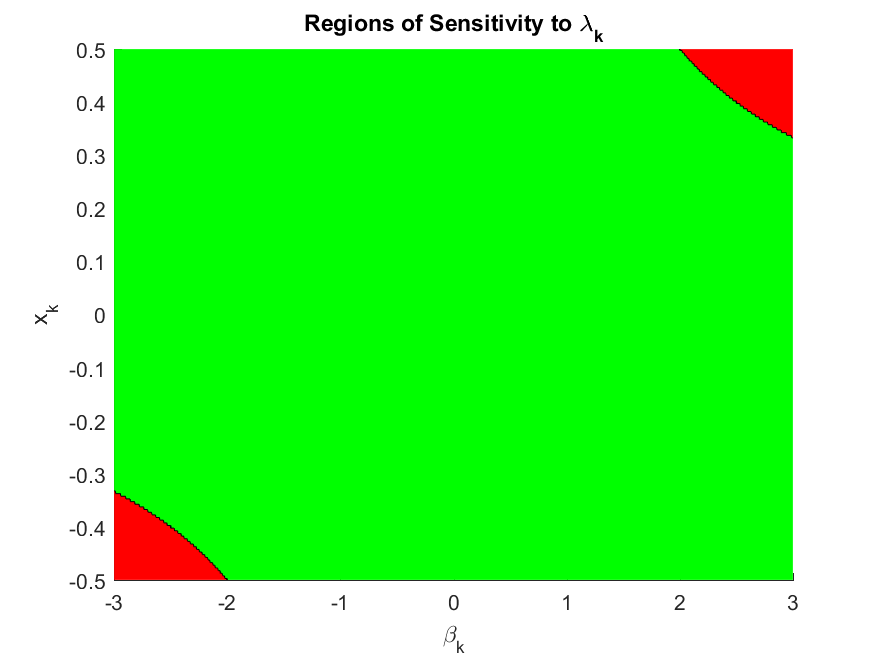}
    \includegraphics[width=0.48\linewidth]{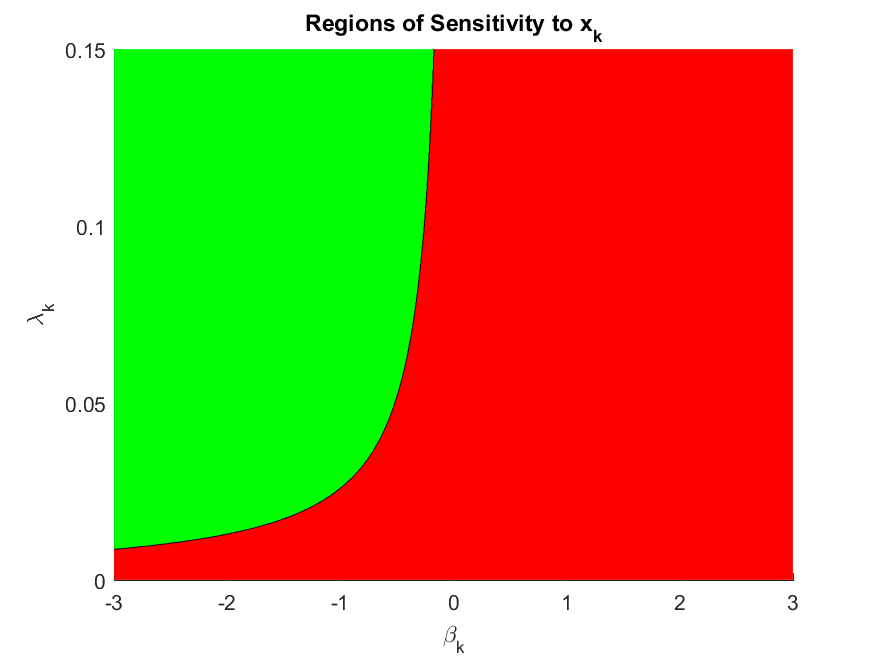}\\
    
    \caption{Sensitivities of the Expected Extra Excess Returns, Red for Negative Sensitivity and Green for Positive, of Asset $k$ to its Shadow-Cost (left) and to Optimal Market Portfolio Weight (right).}

   \label{fig3}

\end{figure}

\subsubsection{Random Shadow-Costs}
Here, we give the results concerning the reference model in (\ref{eq32}) with random shadow-costs vector. We mainly give the equilibrium covariance matrix and optimal portfolio for different levels of shadow-costs certainty.
\begin{figure}[H]
\vspace{\fill}
    \centering
    \includegraphics[width=0.495\linewidth]{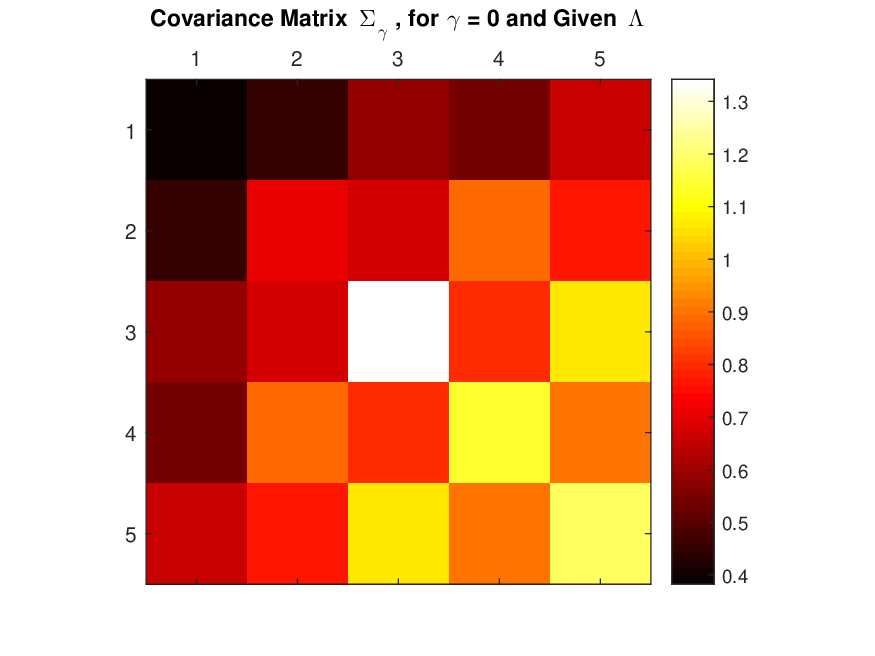}
    \includegraphics[width=0.495\linewidth]{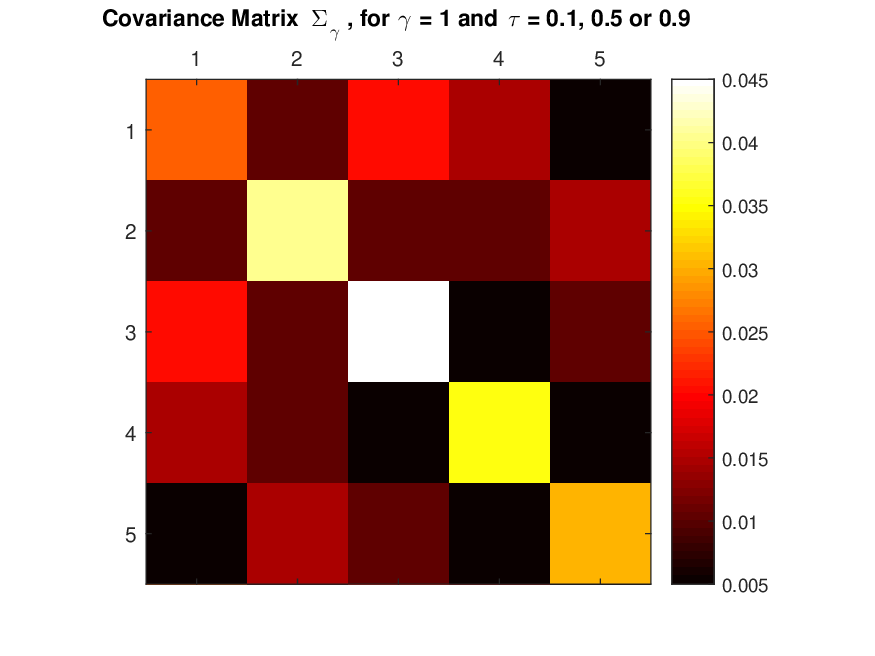}\\ \includegraphics[width=0.495\linewidth]{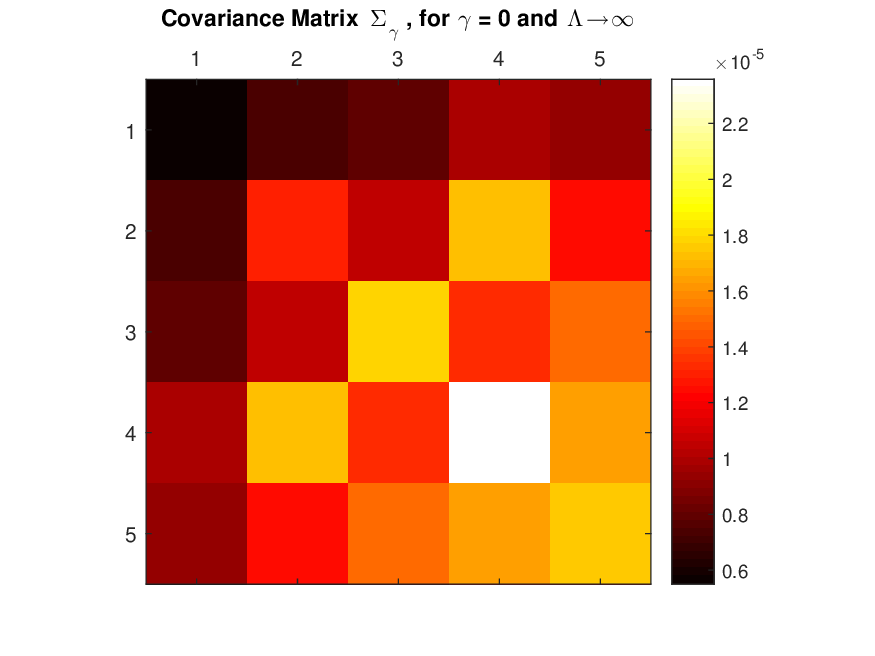}    \includegraphics[width=0.495\linewidth]{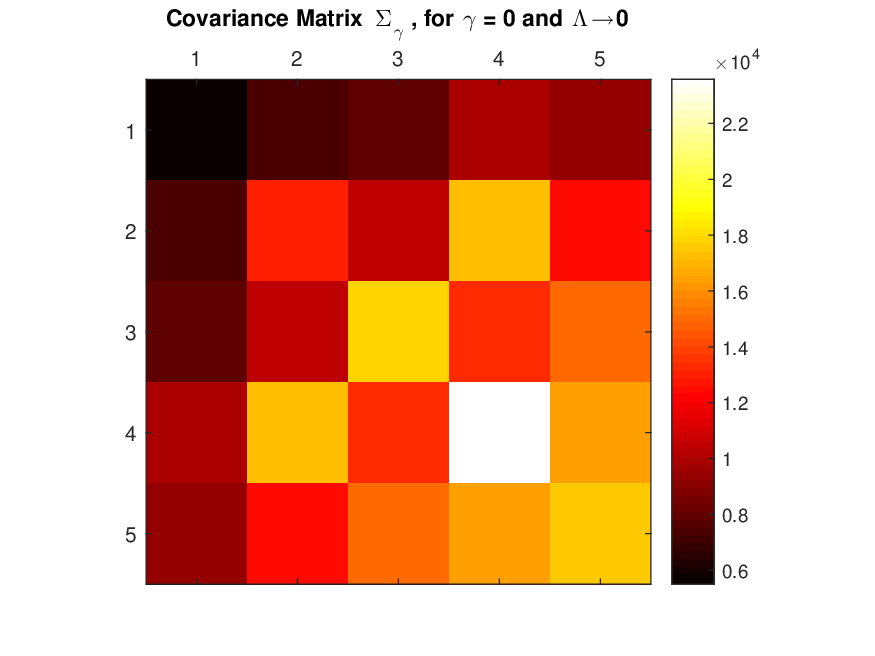}
    \caption{Equilibrium Covariance Matrix of Excess Returns Across Reference Models $\gamma=0,1$ with Flexible Certainty of Shadow-Costs.}

   \label{fig4}

\end{figure}
\begin{table}[H]
    \centering
    \begin{tabular}{lcccccc}
        \bottomrule
        \rowcolor{black!10} Asset & & 1 & 2 & 3 & 4 & 5 \\
        \toprule
        \text{Excess Returns} & $\boldsymbol{\pi}$ & 0.0200 & 0.0548 & 0.0352 & 0.0541 & 0.0648 \\
        \midrule
        \rowcolor{green!10} \text{Model} $\gamma=0$ & $\boldsymbol{w_M^\gamma}$ & -0.4966 & 0.0903 & -0.0260 & -0.0219 & 0.2679 \\
        \rowcolor{green!10} \text{Model} $\gamma=1$ & $\boldsymbol{w_M^\gamma}$ & -0.1532 & 0.0624 & 0.0847 & 0.1999 & 0.2029 \\
        \bottomrule
        \rowcolor{black!10} \text{Return \& Risk} & $\boldsymbol{\mathbb{E}\bigl[\tilde{R}_M^\lambda\bigr]}$ & $\boldsymbol{{\sigma_M^\lambda}}$ & $\boldsymbol{\mathbb{E}\bigl[\tilde{R}_P^{\gamma = 0}\bigr]}$ & $\boldsymbol{\sigma_P^{\gamma = 0}}$ & $\boldsymbol{\mathbb{E}\bigl[\tilde{R}_P^{\gamma = 1}\bigr]}$ & $\boldsymbol{{\sigma_P^{\gamma = 1}}}$ \\
        \toprule
        Value & 0.0035 & 0.0138 & 0.0103 & 0.0359 & 0.0273 & 0.0584 \\
        \bottomrule
    \end{tabular}
    \caption{Expected Excess Returns, Optimal Portfolio Weights, and Return \& Risk at Equilibrium for Incomplete Market with Random Shadow-Costs, given $\Lambda$ and $\tau=0.5$.}
    \label{tab6}
\end{table}

\begin{figure}[H]
    \centering
    \includegraphics[width=0.495\linewidth]{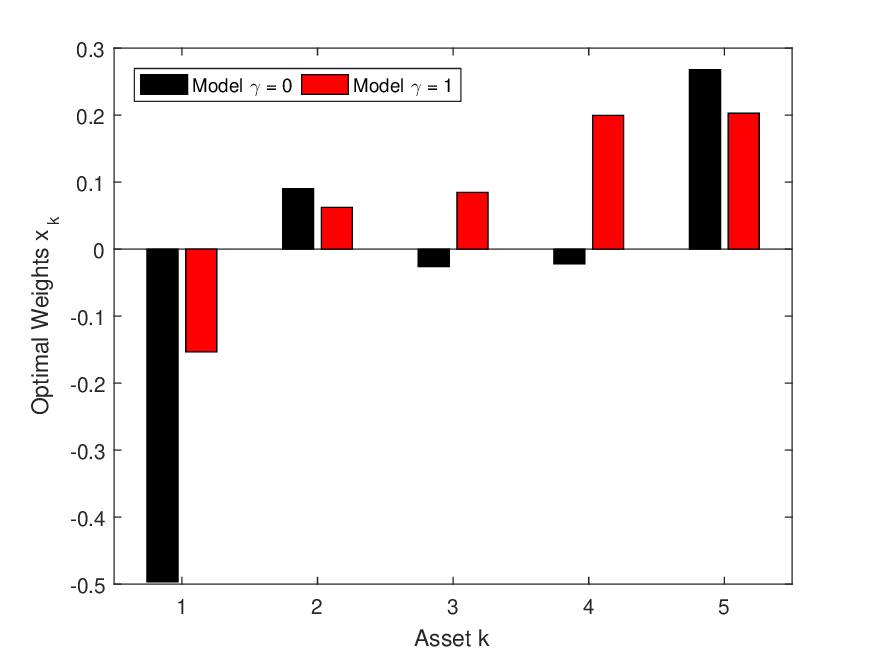}
    \includegraphics[width=0.495\linewidth]{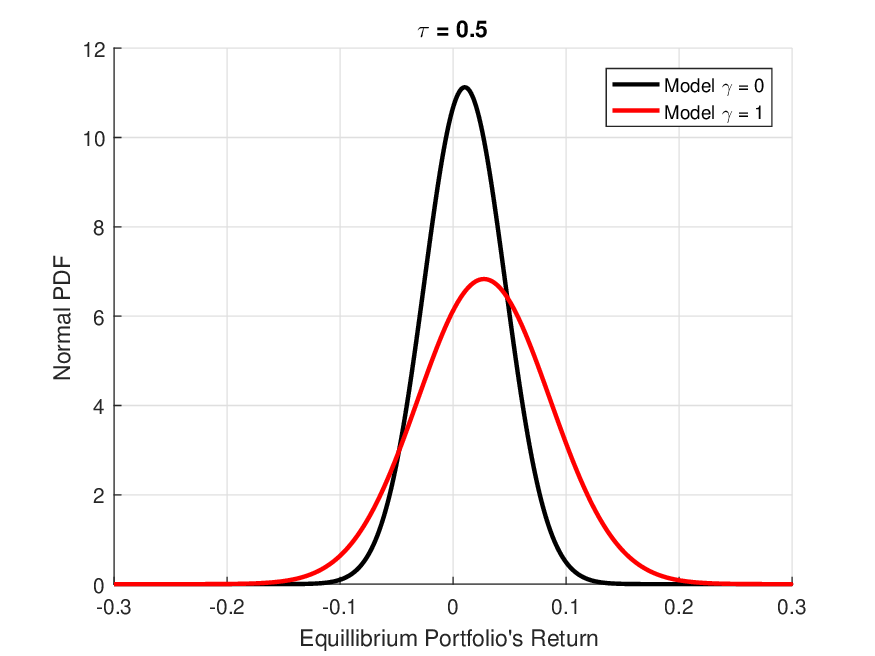}\\
    \includegraphics[width=0.495\linewidth]{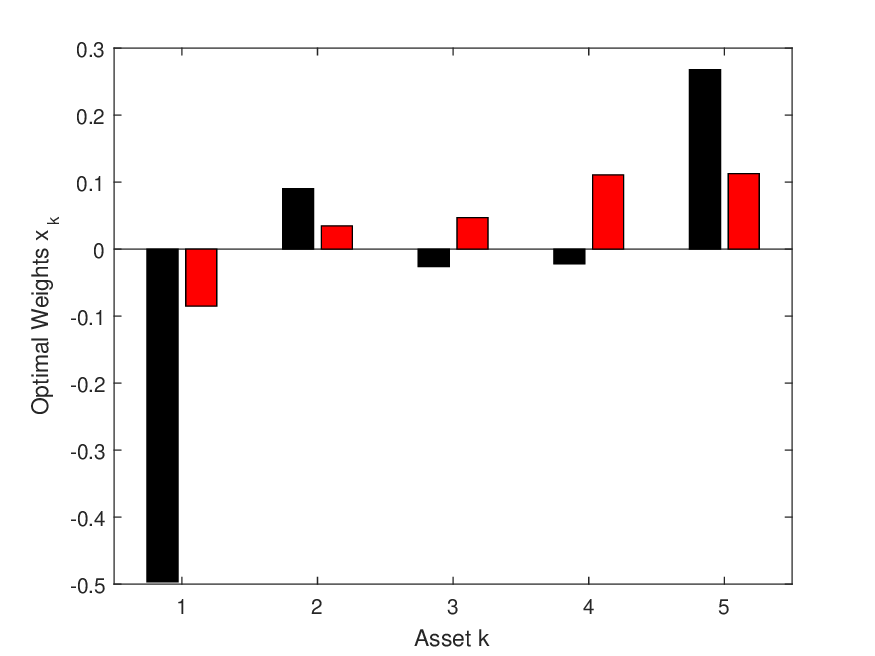}
    \includegraphics[width=0.495\linewidth]{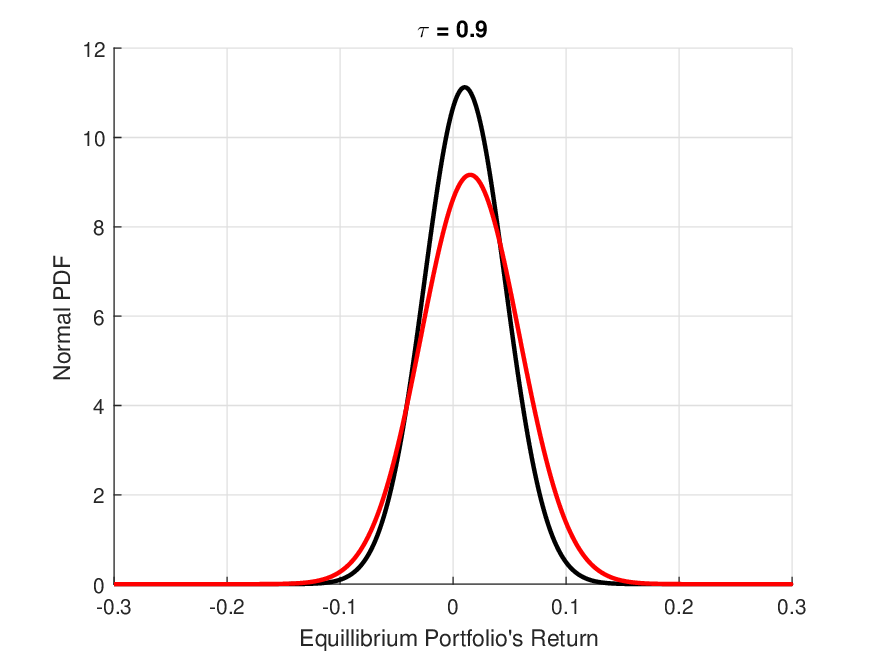}\\ \includegraphics[width=0.495\linewidth]{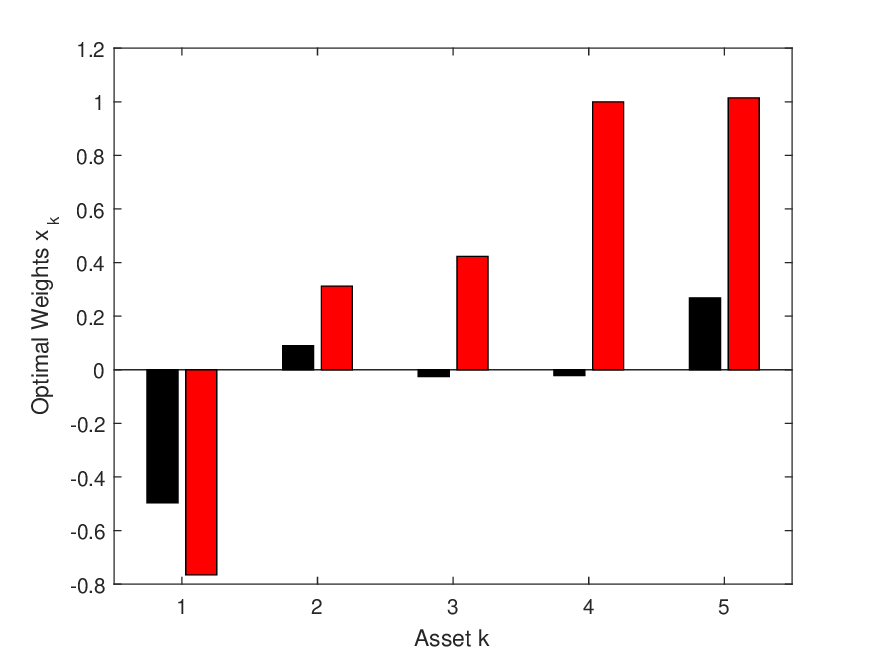}
    \includegraphics[width=0.495\linewidth]{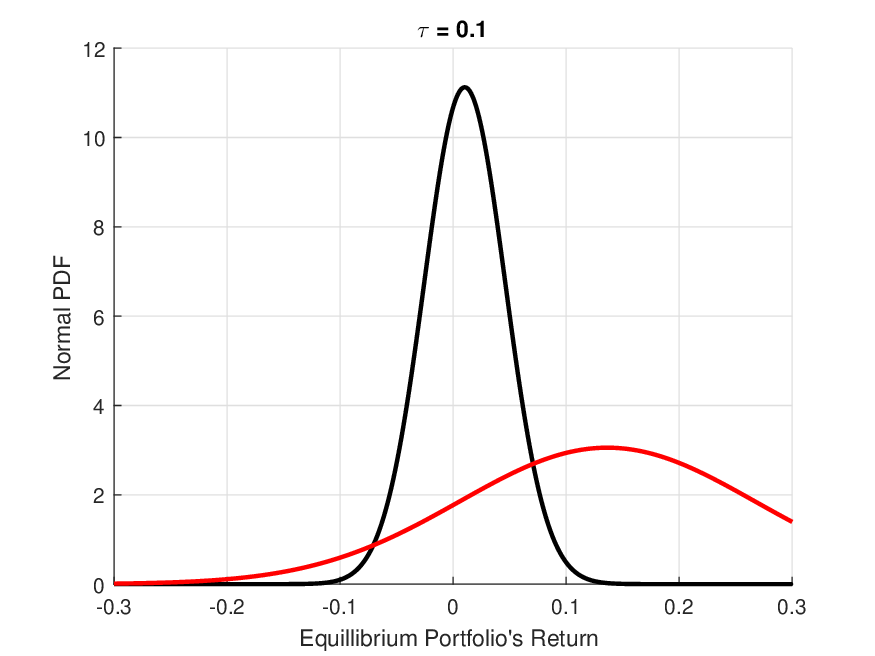}\\        
	\caption{Optimal Portfolio Weights Vector (left) and Return \& Risk for Normally Distributed Market Equilibrium (right) in Incomplete Market with Random Shadow-Costs, Given $\Lambda$ and Flexible $\tau$ Values.}
    \label{fig5}
\end{figure}
\newpage
\subsection{Incomplete Information Market with Views}
Now, assuming that the investor invests in all five assets, the integration of the subjective views of Table \ref{tab3} gives the following results:
\begin{table}[H]
    \centering
    \begin{tabular}{ccccccc}
        \bottomrule
        \rowcolor{black!10} Model & & Asset 1 & Asset 2 & Asset 3 & Asset 4 & Asset 5 \\
        \toprule   
        BL Model & $\mathbb{E}[\tilde{R}_{\tilde{\nu}}]$ & 0.0333 & 0.0300 & 0.0087 & 0.0611 & 0.0567 \\
        \rowcolor{green!10} & $w_M^{\nu}$ & 0.2203 & -0.0558 & -0.1489 & 0.1625 & 0.3679 \\
        Model $\boldsymbol{\gamma=0}$ & $\mathbb{E}[\tilde{R}_{\tilde{\nu}}^{\tilde{\lambda}}]$ & 0.0458 & 0.0479 & -0.0084 & 0.0400 & 0.1069 \\
        \rowcolor{green!10} & $w_M^{\lambda,\nu}$ & -0.6554 & 0.0940 & -0.2459 & -0.0437 & 0.7958 \\
        Model $\boldsymbol{\gamma=1}$ & $\mathbb{E}[\tilde{R}_{\tilde{\nu}}^{\tilde{\lambda}}]$ & 0.0320 & 0.0499 & 0.0274 & 0.0653 & 0.0719 \\
        \rowcolor{green!10} & $w_M^{\lambda,\nu}$ &  0.0314 & 0.0382 & -0.0245 & 0.3997 & 0.6657 \\
        \bottomrule
        \rowcolor{black!10} \text{Return \& Risk} & $\boldsymbol{\mathbb{E}\bigl[\tilde{R}_M^{BL}\bigr]}$ & $\boldsymbol{{\sigma_M^{BL}}}$ & $\boldsymbol{\mathbb{E}\bigl[\tilde{R}_P^{\gamma = 0,\nu}\bigr]}$ & $\boldsymbol{\sigma_P^{\gamma = 0,\nu}}$ & $\boldsymbol{\mathbb{E}\bigl[\tilde{R}_P^{\gamma = 1,\nu}\bigr]}$ & $\boldsymbol{{\sigma_P^{\gamma = 1,\nu}}}$ \\
        \toprule
        Value & 0.0351 & 0.0663 & 0.0599 & 0.0865 & 0.0762 & 0.0976 \\
        \bottomrule
    \end{tabular}
    \caption{Expected Excess Returns, Optimal Portfolio Weights, and Return \& Risk at Equilibrium for Incomplete Market with Subjective Views and $\tau=0.5$.}
    \label{tab7}
\end{table}
\begin{figure}[H]
\vspace{\fill}
    \centering
    \includegraphics[width=0.495\linewidth]{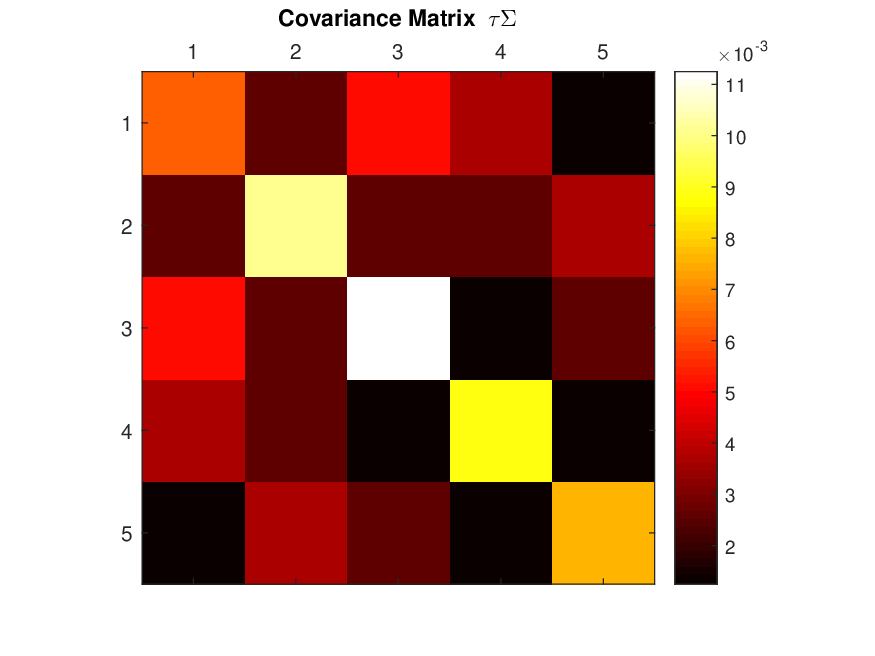}
    \includegraphics[width=0.495\linewidth]{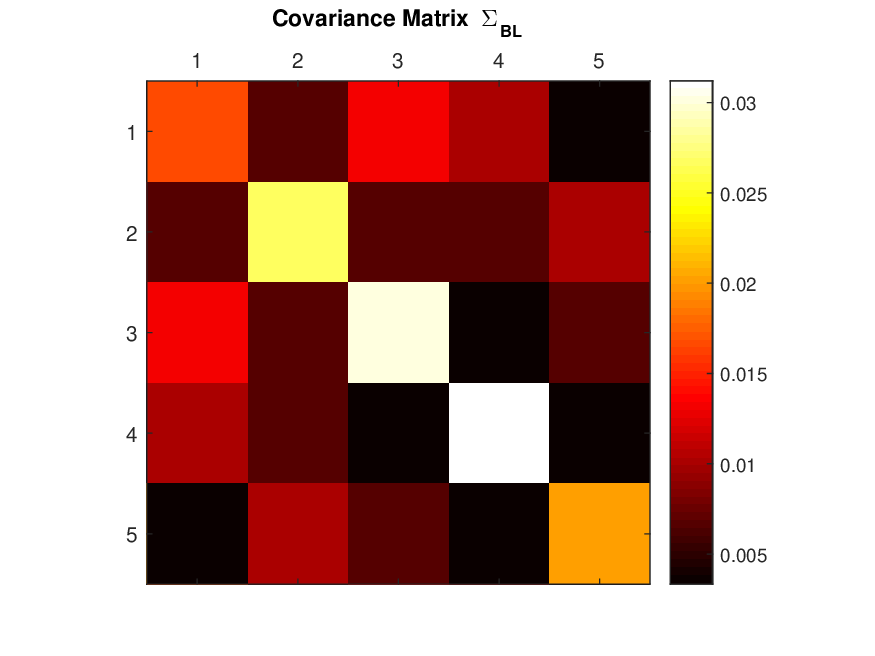}\\ \includegraphics[width=0.495\linewidth]{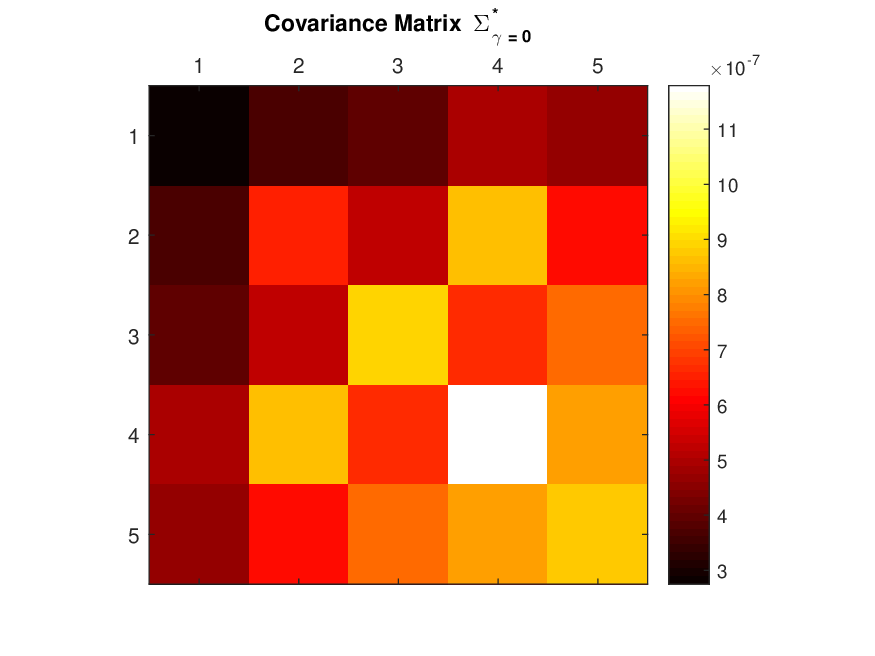}    \includegraphics[width=0.495\linewidth]{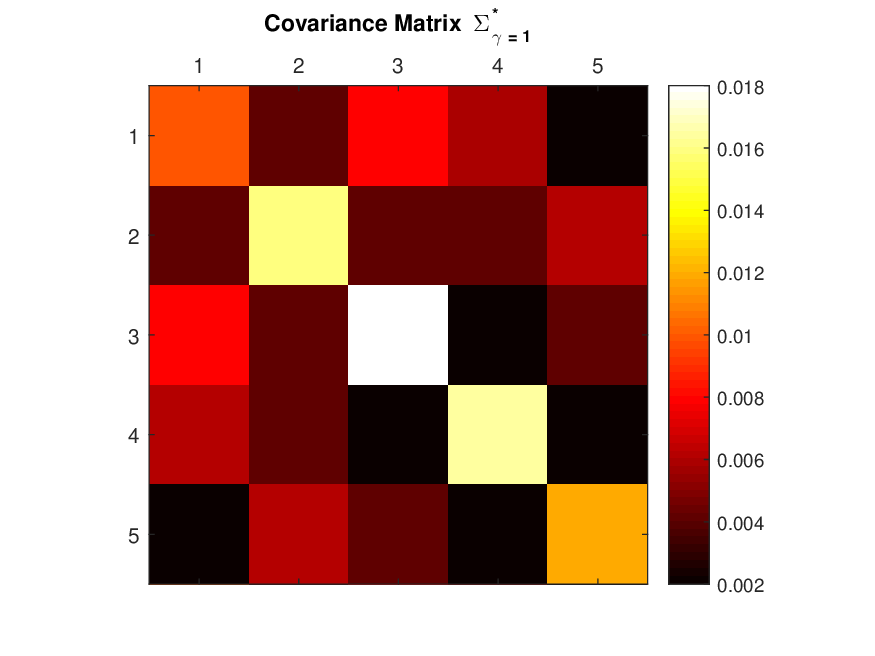}
    \caption{Equilibrium Covariance Matrix of Excess Returns Across Different Models for $\tau=0.5$.}

   \label{fig6}

\end{figure}
\begin{figure}[H]
    \centering
    \includegraphics[width=.49\linewidth]{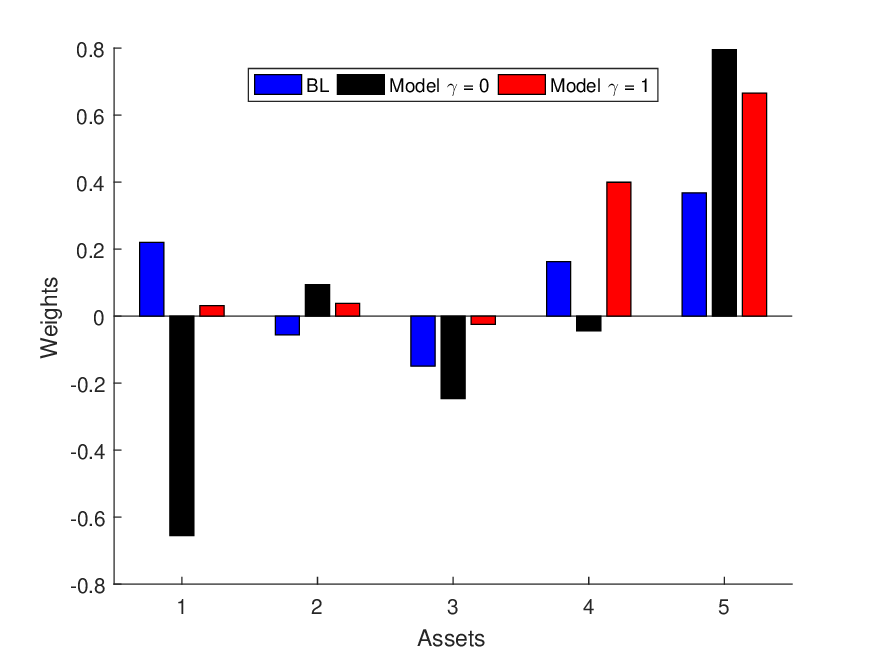}
    \includegraphics[width=.49\linewidth]{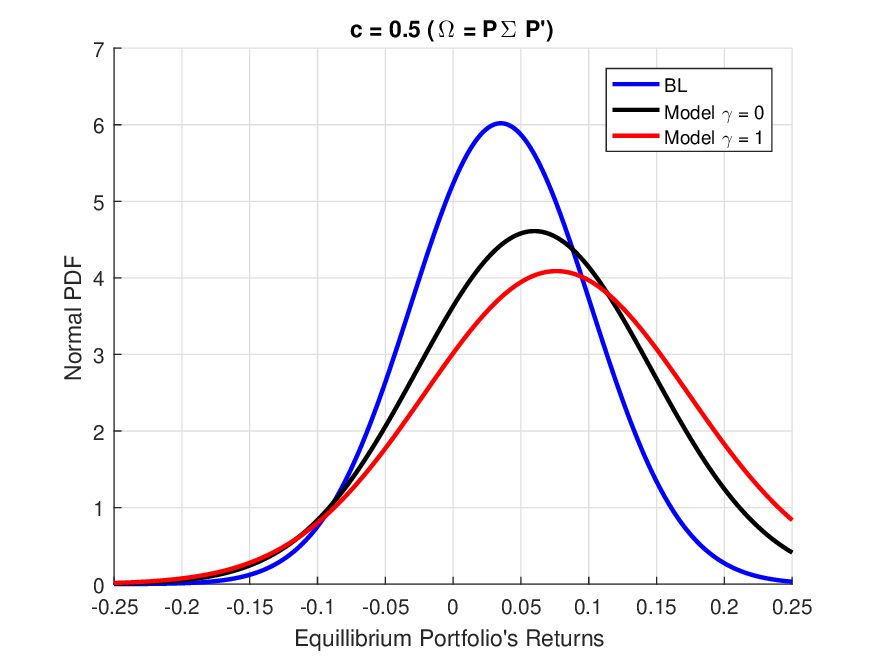}\\
    \includegraphics[width=.49\linewidth]{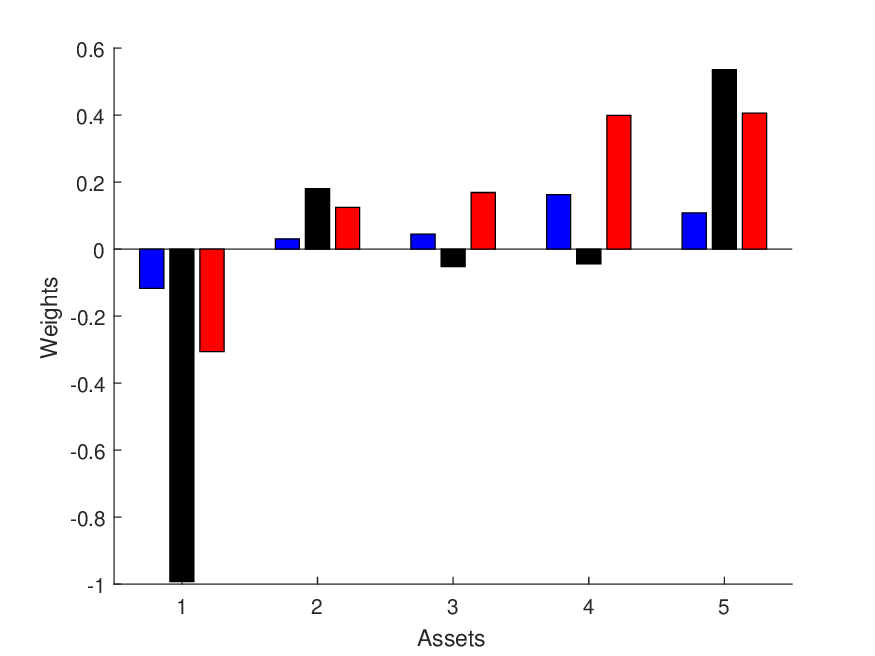}
    \includegraphics[width=.49\linewidth]{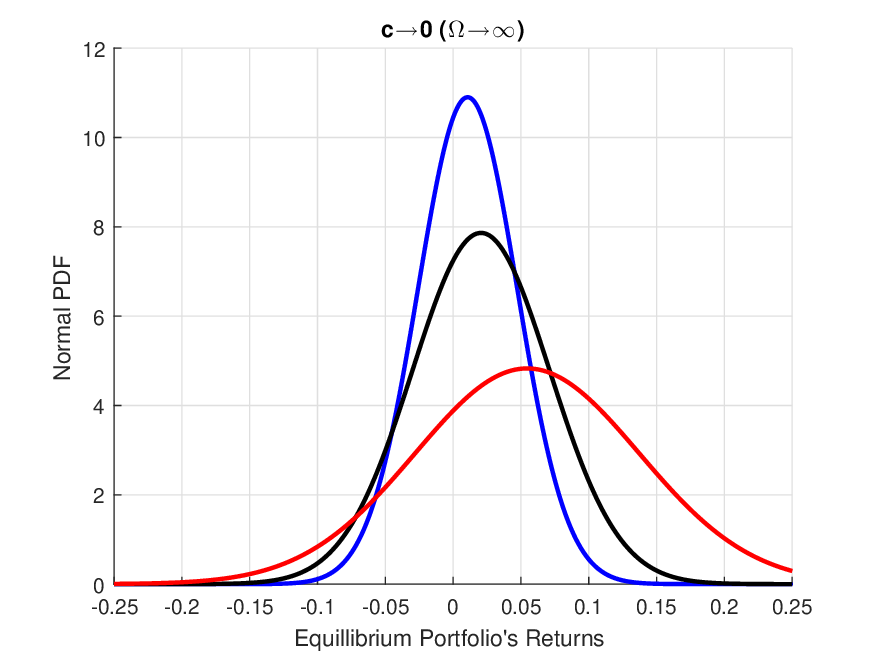}\\
    \includegraphics[width=.49\linewidth]{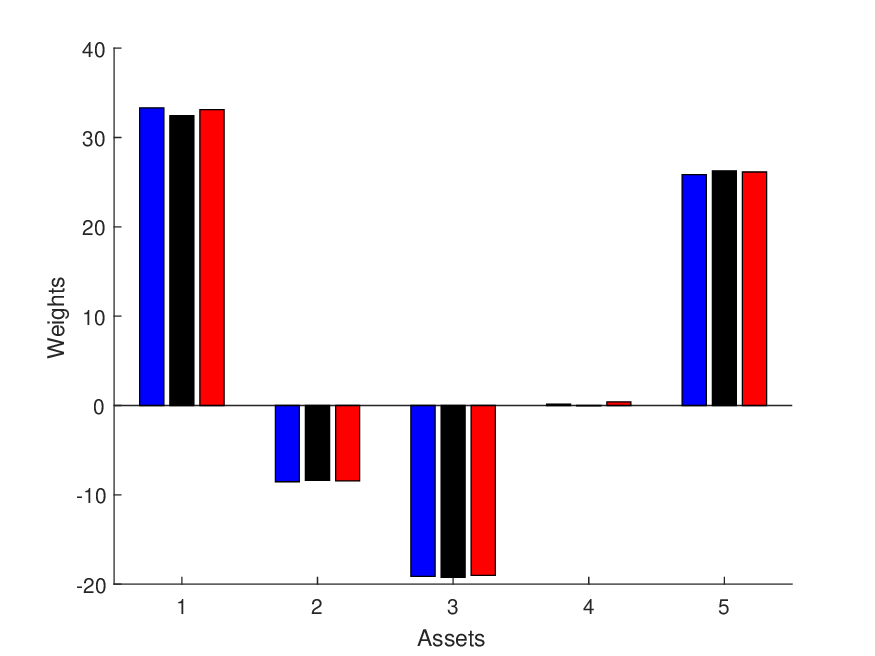}
    \includegraphics[width=.49\linewidth]{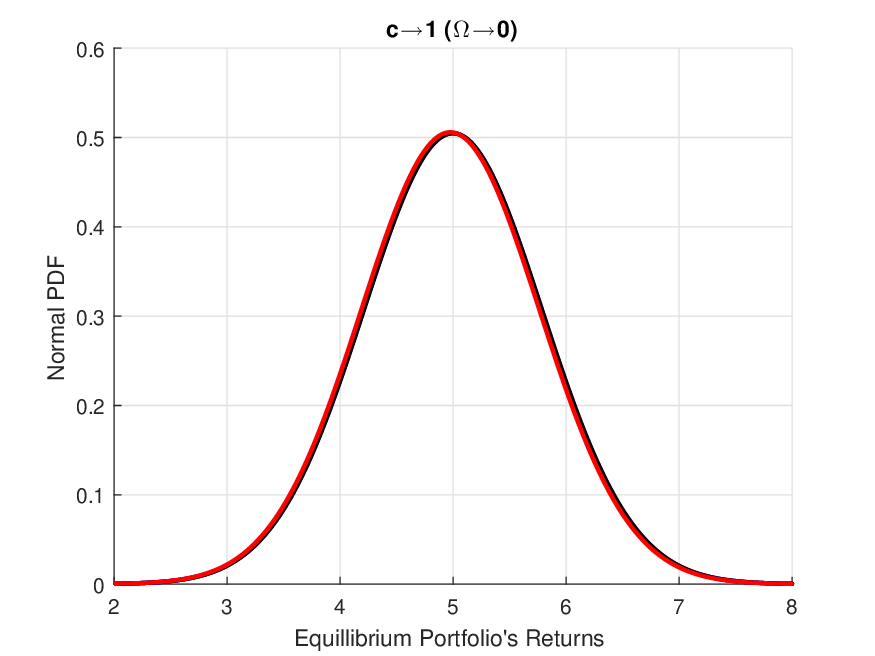}
    \caption{Optimal Portfolio Weights Vector (left) and Return \& Risk for Normally Distributed Market Equilibrium (right) in Incomplete Market with Views, Given $\Lambda$, $\tau=0.5$ and Flexible Certainty of Subjective Views.}	
    \label{fig7}
\end{figure}
\newpage
\section{Conclusion}
This paper developed a capital market equilibrium model that captures investors' subjective views in markets with incomplete information. Each investor can hold a set of beliefs about the relative and/or absolute performance of a subset of securities in the market. The markets under consideration are imperfect, meaning that information about available securities is not equally acquired by all investors. The proposed model enables investment decision-makers to incorporate their beliefs, along with their degree of certainty, using a pick-matrix of views and a related pick-vector, while also quantifying the imperfection of information among the rest of the investors through the so-called vector of shadow-costs of information. First, we characterized the implied equilibrium in markets with incomplete information using Merton's model of incomplete information, we then extracted the associated market portfolio and analyze the sensitivity of the equilibrium excess returns to the shadow-costs of information and the market portfolio. The optimal market portfolio is found to be the solution of a system of non-linear equations, while the investor is allowed to use a modified risk-adjusted return utility maximization problem to determine optimal allocations. Next, we theoretically modeled the structure of the considered markets by rigorously introducing the subjective views of investors within the context of imperfect information markets. The Bayesian inference framework is then explored to derive the asset pricing model in these markets, which gives the posterior distribution of the excess returns conditioned on both the shadow-costs and the subjective beliefs distributions. The resulted model presents a novel sophisticated approach to portfolio allocation in which the prior-allocation approach, that consists in using investor's prior knowledge and experience to allocate assets while the historical information about the market is disregarded, and the sample-based allocation are combined. Such a combination yields the optimal Bayesian allocation $\alpha_{\text{B}}[i_T,e_C]$ which, on the one hand, leverages the sample-based allocation approach by incorporating the incompleteness of information among all investors through a market estimation of the vector of shadow-costs, and, on the other hand, overcomes the biases from the prior allocation approach by reflecting subjective beliefs on a posterior of the market excess returns, which contains the market data reflected in the vector of shadow-costs. More precisely, the Bayesian allocation
\begin{equation*}
    \alpha_B[i_T,e_C]=\argmax_{\alpha\in\mathcal{C}}\int_{\mathcal{R}}U(\alpha^\top r)f_p(r;i_T,e_C)dr,
\end{equation*}
as defined in Section \ref{Sec1}, for the posterior density
\begin{equation*}
    f_p(r;i_T,e_C)=\int_{\Theta}f_{\theta}(r)f(\theta;i_T,e_C)d\theta,
\end{equation*}
can be directly derived from the results of our study by considering the market vector $M_\theta$ and market parameter $\theta$ as the realizations of the distribution of the equilibrium in the imperfect information market and the realizations of the joint-distribution of shadow-costs and subjective beliefs, $\Theta:=[\tilde\lambda,\tilde\nu]^\top$ , respectively. Such an allocation can be extracted from our equilibrium using the derived posterior density
$$f_p(r;i_T,e_C)=\frac{1}{(2\pi)^{n_j/2}\bigl|\boldsymbol{\hat{\Sigma}}\bigr|^{1/2}}\exp\Bigl(-\frac{1}{2}\bigl(r-\boldsymbol{\hat\pi}\bigr)^\top\boldsymbol{\hat{\Sigma}}^{-1}\bigl(r-\boldsymbol{\hat\pi}\bigr)\Bigr),$$
for an investor with information set $J_j$, where $|.|$ denotes the determinant, $$\boldsymbol{\hat{\Sigma}}:=\text{Var}\Bigl[{\bigl(\tilde{R}^{\tilde{\lambda}}_{\tilde\nu}\bigr)}^{J_j}\Bigl]=\Sigma_{1,\dots,n_j\times1,\dots,n_j}^*\text{,}\;\text{and}\;\boldsymbol{\hat\pi}:=\mathbb{E}\Bigl[{\bigl(\tilde{R}^{\tilde{\lambda}}_{\tilde\nu}\bigr)}^{J_j}\Bigl]=\pi_{1,\dots,n_j}^*.$$
Sampling techniques, such as Markov Chain Monte Carlo method, can be used to sample from the posterior equilibrium.
\par Consequently, each investor can reflect his/her subjective beliefs in optimal allocations for portfolios constructed in market with imperfections related to the incompleteness of information among other investors. The derivation and analysis of the posterior predictive distribution $\mathcal{P}(\tilde{\nu}|\tilde{\lambda})$ is of great interest, as it characterizes the future visions of the investor, given the observed distribution of shadow-costs and the model's posterior equilibrium. Moreover, controlling the future subjective beliefs is possible by observing the shadow-costs and analyzing the posterior predictive distribution of views. An extension of this work could involve exploring the updating scheme for the distributions of $\tilde{\lambda}$, $\tilde{\nu}$ and $\tilde{R}$. This would yield a dynamic portfolio optimization model where the equilibrium utilizes an adjusted market parameter $\theta$, computed from the posterior distribution derived from the previous market parameter. This approach allows for one-step-ahead correction of subjective beliefs and the vector of shadow-costs, leading to more accurate and adaptive allocations through the incorporation of new information. Another extension could consider Merton's formulation, where the market returns for an investor $j$, whose investment decision is concerned with a market of $n$ risky securities, a synthetic security indexed by $n+1$ and a risk-free security indexed by $n+2$, are given by:
\begin{equation*}
\tilde{R}^j:=\sum_{k=1}^{n}w_k^j\tilde{R}_k^j+w_{n+1}^j\tilde{R}_{n+1}^j+w_{n+2}^jR=\bar{R}^j+b^j\tilde{Y}+\sigma^j\tilde\varepsilon^j;
\end{equation*}
and then proceed with an investigation of the multi-constrained optimization problem
\begin{subequations}
    \begin{alignat}{2}
            &\max_{b^j,\;w^j}\Bigl\{\mathbb{E}\bigl[\tilde{R}^j\bigr]-\frac{\delta_j}{2}\text{Var}\bigl[\tilde{R}^j\bigr]\Bigr\};\\
&\;\text{Subject to Constraints:}
\left\{
\begin{aligned}\label{eq5b}
    &P^j\mathbb{E}[\tilde{R}]=\mathbb{E}[\tilde{q}^j],\,\text{where}\;\mathbb{E}[\tilde{q}^j]=q^j;\\
    &\text{Var}(\tilde{q}^j)=\Omega^j;\\
&w_k^j=0\;\text{when}\;k\in J_j^c,\;\text{and}\;w_k^j\;\text{is non-negative for}\;k\in J_j,
\end{aligned}
\right.
    \end{alignat}
\end{subequations}
where $w^j:=[w_1^j,\dots,w_{n+2}^j]^\top$, $\delta_j$ is the risk-aversion coefficient, and $J_j:=\{1,\dots,n_j\}$ is the set of risky securities on which the investor $j$ invests. Solving such a problem yields an allocation that reflects the investor's experience, coming from the pick-matrix $P^j$ and the pick-vector $q^j$, as well as the market's information imperfection, characterized by the set $J_j$ of risky securities related to the third constraint.
\bibliographystyle{apa}
\bibliography{myReferences}
\end{document}